\newcounter{myctr}

\documentclass{ws-acs}
\usepackage{url}

\usepackage[T1]{fontenc}
\usepackage[latin9]{inputenc}
\usepackage{color}
\usepackage{array}
\usepackage{multirow}
\usepackage{amsmath}
\usepackage{graphicx}

\makeatletter

\providecommand{\tabularnewline}{\\}

\makeatother

\begin{document}

\makeatletter
\def\@biblabel#1{[#1]}
\makeatother

\markboth{CHALLET, CHICHEPORTICHE, LALLOUACHE,  KASSIBRAKIS}{Statistically validated lead-lag networks and inventory prediction in the foreign exchange
market}

%%%%%%%%%%%%%%%%%%%%% Publisher's Area please ignore %%%%%%%%%%%%%%%
%
\catchline{}{}{}{}{}
%
%%%%%%%%%%%%%%%%%%%%%%%%%%%%%%%%%%%%%%%%%%%%%%%%%%%%%%%%%%%%%%%%%%%%

\title{Statistically validated lead-lag networks\\
 and inventory prediction in the foreign exchange
market
}

\author{\footnotesize DAMIEN CHALLET}

\address{Laboratoire MICS, CentraleSupélec,
Université Paris Saclay, 91190 Gif-sur-Yvette, France\\
and Encelade Capital SA, EPFL Innovation Park, Building C, 1015 Lausanne,
Switzerland
damien.challet@centralesupelec.fr}

\author{R\'EMY CHICHEPORTICHE}

\address{Capital Fund Management, 23, rue de l'Université, 75007 Paris,
France}

\author{MEHDI LALLOUACHE}
\address{BNP Paribas, 20, boulevard des Italiens, 75009 Paris, France}

\author{SERGE KASSIBRAKIS}
\address{ Swissquote Bank SA,  chemin de la Crétaux 33, 1196 Gland, Switzerland}

\maketitle

\begin{history}
\received{(received date)}
\revised{(revised date)}
%\accepted{(Day Month Year)}
%\comby{(xxxxxxxxxx)}
\end{history}

\begin{abstract}
We introduce a method to infer lead-lag networks
of agents' actions in complex systems. These networks open the way
to both microscopic and macroscopic states prediction in such systems.
We apply this method to trader-resolved data in the foreign exchange
market. We show that these networks are remarkably persistent, which
explains why and how order flow prediction is possible from trader-resolved
data. In addition, if traders' actions depend on past prices, the
evolution of the average price paid by traders may also be predictable.
Using random forests, we verify that the predictability of both the
sign of order flow and the direction of average transaction price
is strong for retail investors at an hourly time scale, which is of
great relevance to brokers and order matching engines. Finally, we
argue that the existence of trader lead-lag networks explains in a
self-referential way why a given trader becomes active, which is in
line with the fact that most trading activity has an endogenous origin.
\end{abstract}


\section{Introduction}

Predicting the evolution of complex systems is of
great practical interest but difficult given their nonlinear collective
dynamics and, in practice, {\emph{a priori }unknown
networks of interaction between their elements (see e.g. \cite{johnsonlargechanges,SanMiguel2012}).
Determining the underlying causal
action structure of agents is an important first step to understand
the dynamics of a given system, and also to assess what types of structure
are really informative about future states of complex systems. Assuming
that the actions of some agents trigger the actions of some other
(or same) agents, we propose here a generic method to infer such causal
lead-lag activity networks and to exploit their presence with machine
learning methods so as to predict the dynamics of macroscopic quantities. \textcolor{black}{Although the method is generic, we focus here on financial markets, because the traders' behavior leads to maximally unpredictable
dynamics because all of them try to
learn as much as possible to exploit any potential predictability
\cite{bouchaud2017have}. The fact that dynamics becomes predictable in the most unpredictable complex systems when on accounts for agent lead-lag activity networks demonstrates the potential of our method.}

\textcolor{black}{Since a sizable proportion of financial activity
is endogenous \cite{Filimonov2012,hardiman2013critical}, i.e. triggered
by past activity, we shall illustrate how financial markets become
predictable if the trader lead-lag activity network is persistent
enough. While most of available financial data is fully anonymous,
some financial actors such as brokers have access to trader-resolved
data: to each order is associated a client identification number.
Brokers are interested in managing the risk associated to their temporary
inventory against future price moves or large volume imbalance (buy-sell).
How much predictability either of order imbalance or of average transaction
price change trader-resolved data brings is an open question. Here,
we take the point of view that predictability of complex systems such
as financial markets is due in part to an underlying structure of
systematic lead-lag between the actions of traders. Because of the
large quantity and high quality of data generated by brokers, financial
markets provide an ideal testing ground for the new method we introduce
and to illustrate this point of view.}

\textcolor{black}{This work is relevant to several types of market
participants who try to match buy and sell order flows (which are
submitted asynchronously) by keeping some of them in their inventory. Order crossing may thus occur if incoming
orders can be matched against current inventory or added to it and
made available for future matching, which may happen at many levels.
For example, two strategies of an investor may have opposite opinions
about the same asset at about the same time, in which case internal
crossing saves transaction fees and reduces uncertainty. If an order
needs to be sent to an exchange, it may be matched on its way at crossing
networks, dark pools, or even at an internal matching system within
the exchange (e.g. IMS at NYSE Euronext) \cite{harris2002trading}. }

\textcolor{black}{Managing one's inventory in this situation is of
great practical importance, as keeping an inventory is risky.
A standard approach would be to solve an stochastic optimization problem
and to minimize some cost function (e.g. risk or probability to reach an inventory limit) over a given time horizon,
usually \textcolor{black}{with stochastic processes without any predictability 
both }for the order flow and the future prices. Here, we consider this
situation as an on-line prediction problem instead and show how being
able to identify the source of orders makes it possible to predict
the direction of both the order flow and the volume weighted
average price (VWAP) for retail clients. }

\textcolor{black}{Accordingly, we focus here on matching engines that
are able to identify the source of the order (strategy, investor,
broker, etc.). We link order flow and VWAP direction prediction with
the existence and persistence of trader lead-lag networks which encode
how the activity of some traders (e.g. buy) systematically leads on that
of other traders (e.g. sell). Our first contribution is to introduce
an unsupervised method to infer such networks which is generic enough
to be applied at all levels of order crossing and in non-financial
contexts in which the similarity of choices of items by customers
may be used to predict future actions. The method consists in first clustering
traders into groups according to their synchronicity with the method
of \cite{Michele2012}, \textcolor{black}{determining the aggregate state (buy/sell/neutral) of each group and then  applying the same method to lagged group states }to detect systematic lead-lag between groups. While synchronicity
is likely due to the use of the same strategies or of the same source
of information, or both, lead-lag networks are likely caused by the
different reaction speed of the respective strategies (e.g. two moving
averages with different parameters); an alternative explanation is
that some traders react with different delays to common information
\cite{boudoukh1994tale,jegadeesh1995overreaction}. }

\textcolor{black}{If these trader lead-lag networks are sufficiently
persistent, some quantities become predictable. First, order flow
is predictable if the state of groups (buy/sell/neutral) is partially
causal. In addition, \textcolor{black}{proper } inventory management \textcolor{black}{requires to predict price directions as well}. Our second main contribution is
thus to assess the predictability of both order flow and price direction.
Only the simplest prediction scenario is studied: we try to predict
the sign of each quantity from the global state (buy, sell, neutral)
of each trader group and their lagged values. In other words, we reduce
prediction to a classification problem from discrete variables and
use standard methods of machine learning.}  \textcolor{black}{This setup is both crude and robust to outliers. The point is not to provide a finely tuned method to manage inventory for brokers, but to provide evidence that the persistence of lead-lag trader networks  yields successful predictions.}

\subsection{Literature review}

\textcolor{black}{Our contribution is related to several areas of
finance pertaining to order flow segmentation, predictability, and
inventory control by an informed market maker.}

\textcolor{black}{The first broad related area is market making in
the presence of predictable order flow. In the context of market microstructure, the role of meta-orders in the long memory of the sign of market orders
is well documented \cite{farmersign,BouchaudFarmerLillo}. The order
flow of individual traders is known to be anti-correlated with previous
daily or weekly price returns \cite{grinblatt2000investment,kaniel2008individual}.
A related topic, although less directly relevant to our contribution,
is the predictive power of order flow on other quantities than itself
such as price returns \cite{kelley2013wise}. Whereas market makers
were assumed to be fighting against informed traders in the early
literature, the long memory of market order signs leads to a new paradigm
of optimal market making \cite{avellaneda2008high}. Internal order
matching differs from market making at an exchange in that one may
liquidate in part one's inventory at the exchange, although with non-negligible
transaction costs \cite{galien2016}.}

\textcolor{black}{Trader grouping is often performed according to
their role in financial markets (e.g. individual investor, institutional
investor, etc). Average properties are then computed over whole sub-populations
\cite{grinblatt2000investment,jackson2004aggregate,GRINBLATT2009,barber2009just}. }

\textcolor{black}{Unsupervised clustering on the other hand rests
on the similarity of actions of traders determined }\textcolor{black}{\emph{ex
post}}\textcolor{black}{. The simplest approach is based on the computation
of the correlation matrix of trader inventory changes; then Principal
Component Analysis is used together with Random Matrix Theory (to account for the finite length of the available time series) to extract
eigenvalues outside the random spectrum \cite{Zovko2007,Lillo2008,Kertesz2012}.
It turns out that at a timescale of a single day, only one or two
eigenvalues stand out of the noise spectrum, the largest being associated
with previous price returns. This allows one to classify traders as
mean-reverting (the majority of them), trend-following, and non-classifiable.
The drawback of this approach is that linear correlations 
do not capture the full correlation structure of traders. In addition
it fixes the number of categories of traders.}

\textcolor{black}{An alternative approach is Statistically Validated
Networks (SVNs) \cite{Tumminello2011} which consists in computing
a similarity score between two agents according to how synchronous
and similar their activity and inactivity periods are. This method is generic
and works well provided that the number of possible states (active,
inactive, etc) is small: one determines a p-value of
synchronousness and establishes a link between two agents in a statistically
sound way. Once all the links between all pairs of agents are tested,
one obtains the full synchronousness network of agents. \cite{Michele2012}
find a surprising degree of synchronization within groups of Finnish
traders at a daily timescale. }

\textcolor{black}{Lead-lag relationships between traders are much
less studied. For lack of available trader-resolved data, the literature has focused
on lead-lag relationships between price returns \cite{kullmann2002time,toth2006increasing,huth2014high,curme2015emergence}.
Lead-lag relationships between traders are discussed in the context
of various time scales of contrarian behavior \cite{boudoukh1994tale,jegadeesh1995overreaction}.
We are not aware of any work on unsupervised inference of lead-lag
between traders.}

\section{{Data description and notations}}

\textcolor{black}{We work with datasets on foreign exchange (FX) transactions
from two independent sources: a large dealing bank (LB hereafter}\footnote{\textcolor{black}{LB's electronic market-marking desk provides liquidity
(i.e. quotes and volumes) on the currency rates to large clients such
as commercial companies, financial institutions, pension funds, hedge-funds.}}\textcolor{black}{), and a broker-dealer (Swissquote Bank SA, SQ hereafter}\footnote{\textcolor{black}{SQ acts as an on-line broker on thousands of financial
instruments, with a large market share in the global Foreign eXchange
activity in Switzerland. Its clients range from retail investors to
asset managers and institutions.}}\textcolor{black}{). We refer the reader to \cite{harris2002trading}
for more details about the FX market organization. \textcolor{black}{Both }datasets contain
information about all the trades of their clients over a given period:
anonymous client identification number, trade time with a millisecond
precision, traded currency pair, signed volume, and price (currency
rate). A summary of the datasets structure and contents is provided
in Table \ref{tab:Datasets-description.}.}

\begin{center}
\textcolor{black}{}
\begin{table}
\begin{centering}
\textcolor{black}{}%
\begin{tabular}{|c||c|c|c|c|}
\hline
\textcolor{black}{Dataset} & \textcolor{black}{Timespan} & \textcolor{black}{Instruments} & \textcolor{black}{Traders} & \textcolor{black}{Trades}\tabularnewline
\hline
\multirow{1}{*}{\textcolor{black}{SQ 2012}} & \multirow{1}{*}{\textcolor{black}{31 Jan. 2012 $\rightarrow$ 10 Aug. 2012}} & \textcolor{black}{68} & \textcolor{black}{$>10^{3}$} & \textcolor{black}{$>10^{6}$}\tabularnewline
\textcolor{black}{SQ 2014-6} & \textcolor{black}{01 July 2014 $\to$ 15 March 2016} & \textcolor{black}{206} & \textcolor{black}{$>10^{4}$} & \textcolor{black}{$>10^{7}$}\tabularnewline
\multirow{1}{*}{\textcolor{black}{LB}} & \multirow{1}{*}{\textcolor{black}{01 Jan 2013 $\rightarrow$ 17 Sep. 2014}} & \textcolor{black}{12} & \textcolor{black}{$>10^{4}$} & \textcolor{black}{$>10^{6}$}\tabularnewline
\hline
\end{tabular}
\par\end{centering}
\textcolor{black}{\caption{Basic statistics of the datasets.\label{tab:Datasets-description.}}
}
\end{table}
\par\end{center}

\textcolor{black}{This paper focuses on the most traded pairs only.
Accordingly, Table \ref{tab:Datasets-description.-1} gives a breakdown
of descriptive statistics of the studied pairs. For confidentiality
reasons, we cannot give more precise figures for SQ.}
\begin{center}
\textcolor{black}{}
\begin{table}
\begin{centering}
\textcolor{black}{}%
\begin{tabular}{|c||c|c|c|}
\hline
\textcolor{black}{Dataset} & \textcolor{black}{Pair} & \textcolor{black}{Traders} & \textcolor{black}{Trades}\tabularnewline
\hline
\multirow{1}{*}{\textcolor{black}{SQ 2012}} & \texttt{\textcolor{black}{EURUSD}} & \textcolor{black}{$>10^{3}$} & \textcolor{black}{$>5\times10^{5}$}\tabularnewline
\hline
\textcolor{black}{SQ 2014-6} & \texttt{\textcolor{black}{EURUSD}} & \textcolor{black}{$>10^{4}$} & \textcolor{black}{$>10^{6}$}\tabularnewline
 & \texttt{\textcolor{black}{EURGBP}} & \textcolor{black}{$>5\times10^{3}$} & \textcolor{black}{$>5\times10^{5}$}\tabularnewline
 & \texttt{\textcolor{black}{USDJPY}} & \textcolor{black}{$>10^{4}$} & \textcolor{black}{$>10^{5}$}\tabularnewline
\hline
\multirow{1}{*}{\textcolor{black}{LB}} & \texttt{\textcolor{black}{EURUSD}} & \textcolor{black}{$7300$} & \textcolor{black}{$5\times10^{5}$}\tabularnewline
\hline
\end{tabular}
\par\end{centering}
\textcolor{black}{\caption{Basic statistics of the datasets for the three studied pairs.\label{tab:Datasets-description.-1}}
}
\end{table}
\par\end{center}

\subsection{{Number of transactions per trader}}

\textcolor{black}{The number of transactions $n$ per trader per year
for a given asset in equity markets has been reported to have heavy
tails which may be approximated by a power-law $P(n)\sim n^{-\alpha}$
with tail exponent $\alpha\simeq2$ \cite{Michele2012}. This means
that some traders are orders of magnitude more active than others,
which implies that focusing on the most active traders may simplify
much the prediction of future order flows. We checked that the power-law
still holds on FX markets over about 2 decades of $n$; the exponent
$\alpha$ was estimated for each currency and each year with the method
introduced by \cite{clauset2009power} (see also \cite{powerlawR}).
For a given currency pair, we filtered out the years in which less
than 1000 traders were active, which left 53 estimates. We found $\alpha_{avg}\simeq1.99\pm0.07$
(95\% confidence interval) in the largest dataset (SQ 2014-6). }

\subsection{{Trade size}}

\textcolor{black}{The different nature of the respective clients of
LB and SQ influences the typical trade size. We present the results
in multiple of $1000$ of the base currency for SQ clients and in
multiple of $100'000$ for LB clients. In Fig.  \ref{fig:Distribution-Amount},
we plot the distribution of the transaction sizes for the }\texttt{\textcolor{black}{EURUSD}}\textcolor{black}{.
Similar results are obtained for other pairs. We observe peaks at
round size (10,20,50,...), with a stronger effect for SQ than for
LB, which is consistent with the fact that its clients are mostly
individual traders, therefore more prone to be affected by psychological
biases than institutional traders (see a discussion about this phenomenon
in the FX EBS market in \cite{lallouache2014tick}).}

\textcolor{black}{}
\begin{figure}
\begin{centering}
\textcolor{black}{\includegraphics[scale=0.3]{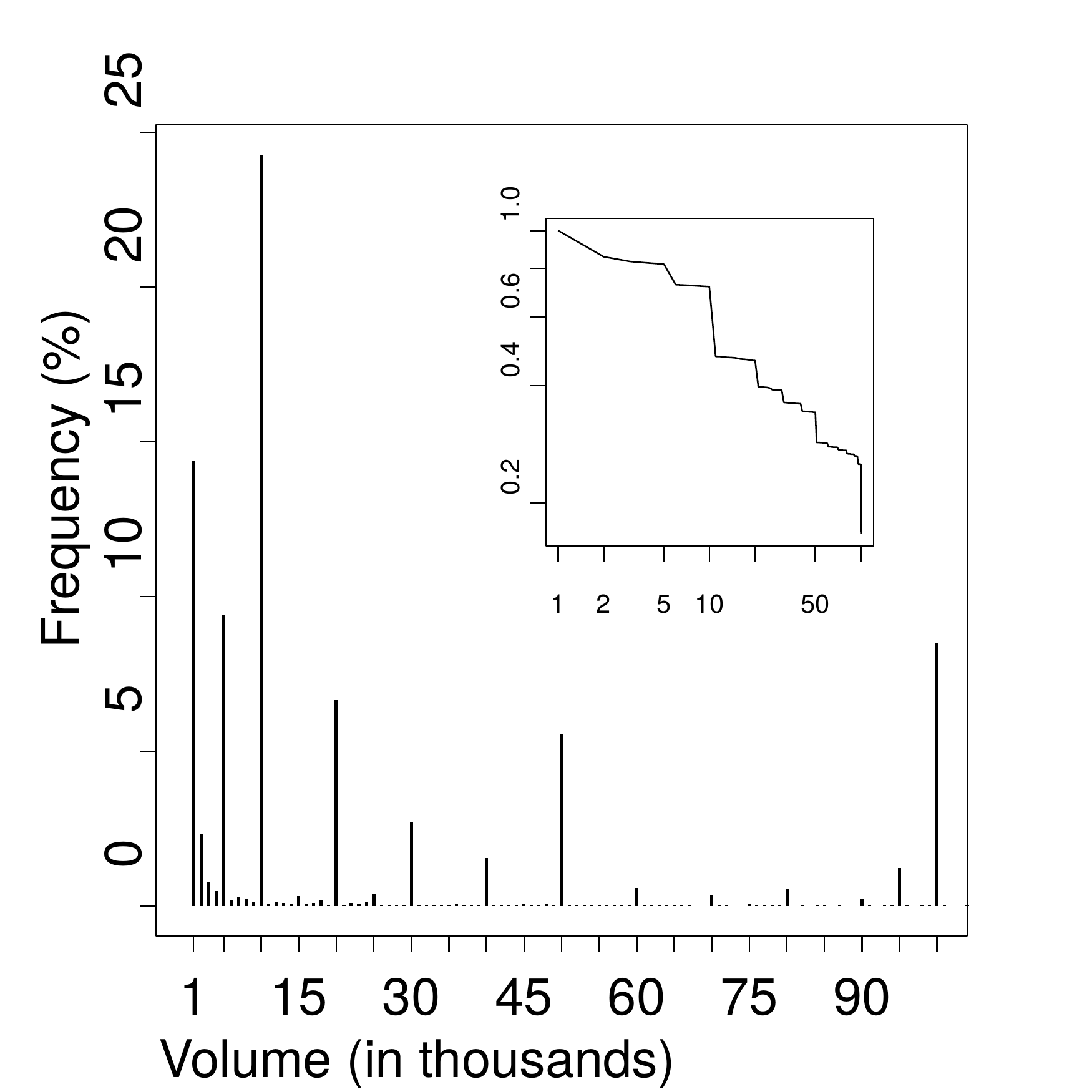}\includegraphics[scale=0.3]{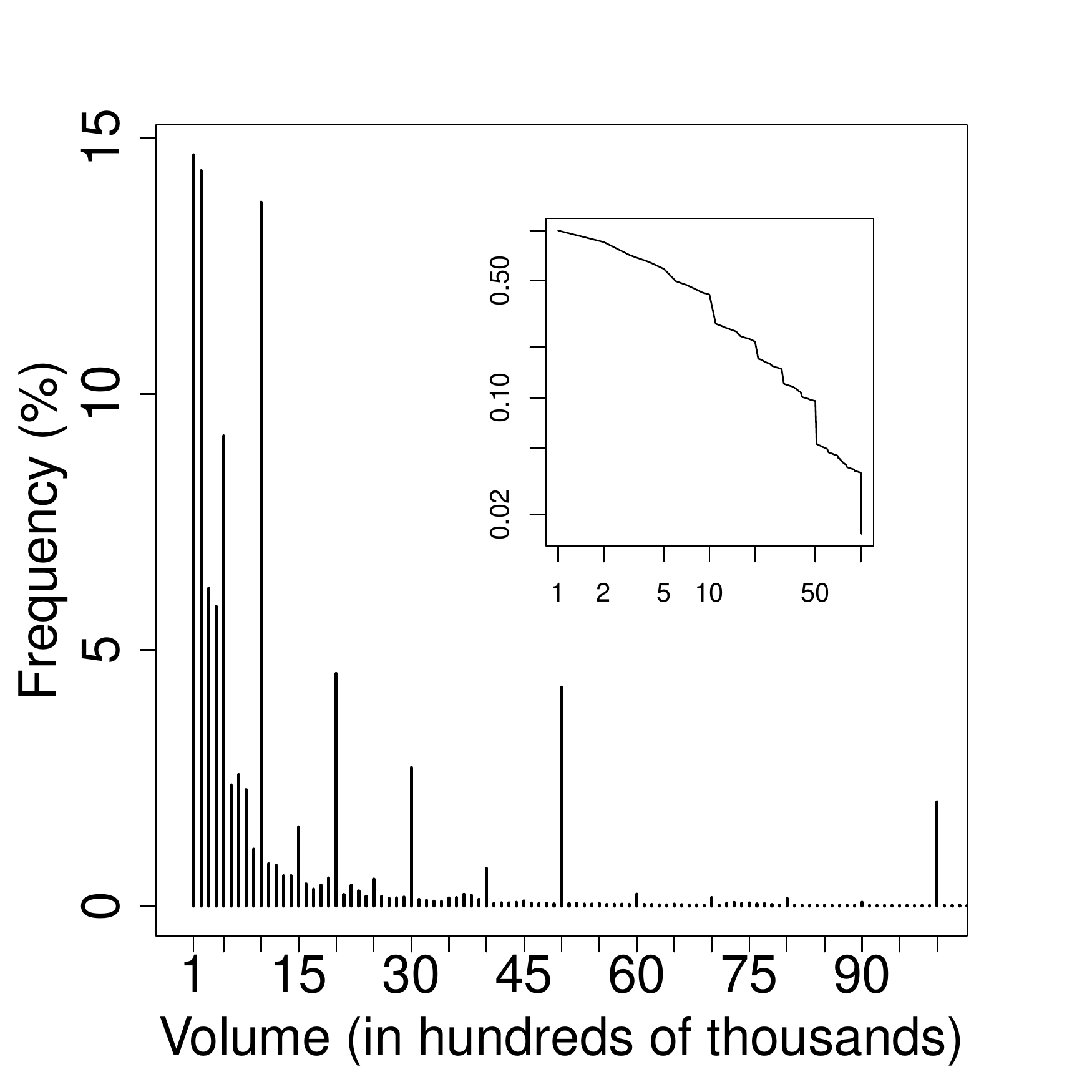}\caption{Distribution of \texttt{EURUSD} trade sizes. Left: SQ. Right: LB.
Inset: cumulative tail distribution, in log-log scale. \label{fig:Distribution-Amount}}
}
\par\end{centering}
\end{figure}

\section{Inference of lead-lag networks}

\textcolor{black}{Although lead-lag networks can be determined between the agents themselves, clustering the traders simplifies
the visualization of lead-lag networks and is very useful (and a standard procedure) before calibrating machine learning methods which often are confused if two or more predictors are very correlated. This is the case here: because many FX traders are algorithmic traders, their activity/inactivity
is very similar when they use the same algorithms.  Thus, grouping traders encodes the systematic use of the same set of strategies (which includes news sources) and does not reduce the predictive ability of our method.}

\subsection{Clustering traders by their synchronicity\label{subsec:bipartite}}

Since the lead-lag method we propose is an extension
of the SVNs, it is worthwhile explaining the method in some details.

\subsubsection{Method}

\textcolor{black}{Statistically Validated Networks (SVN) were introduced
in \cite{Tumminello2011} and applied to the clustering of Finnish
investors in \cite{Michele2012}. The technique aims at characterizing
the degree of synchronization between the actions of two traders,
thus, by extension, at identifying groups of traders who act in a
similar way. By definition, it is an unsupervised clustering method.}

\textcolor{black}{The first step is to cut time into slices of equal
length $\delta t$; we chose arbitrarily $\delta t=1$ hour. In practice,
the duration of each slice must be adapted to each situation. In the
case of the FX traders in our dataset, 1 hour is a reasonable choice
with respect to typical trading patterns of active traders. We define
time slice $t$ as the interval $[t,t+\delta t[$. For each time slice
$t$, we classify the state of all traders into buying (state +1),
selling (state -1), neutral (state 2), and inactive (0). Denoting
the total signed volume of trader $i$ during time slice $t$ by $V_{i}(t)$
and the sum of the absolute trading volume of the transactions of
trader $i$ during time slice $t$ by $G_{i}(t)$, one defines the
imbalance ratio $\rho_{i}(t)=V_{i}(t)/G_{i}(t)$.  The imbalance
ratio characterizes the trader as a net buyer ($\sigma_{i}(t)=1)$
if $\rho_{i}(t)>\rho_{0}$ ($\rho_{0}$ being a small threshold),
as a net seller ($\sigma_{i}(t)=-1$) if $\rho_{i}(t)<-\rho_{0}$,
as neutral ($\sigma_{i}(t)=2$) if $|\rho_{i}(t)|<\rho_{0}$, or as
inactive ($\sigma_{i}(t)=0$) if $V_{i}(t)=G_{i}(t)=0$. The choice
of $\rho_{0}\in[0.01,0.1]$ is not crucial; in the following we set
$\rho_{0}=0.01$. Because the following analysis focuses on the most
active traders, the inactive state will be dropped.}

\textcolor{black}{The synchronicity of a pair of traders is measured by
counting the co-occurrences in the time series of their states, and
attributing a p-value that reflects the statistical significance of
this synchronicity assuming pure randomness. To deal with the testing
of all pairs of traders for each of the 9 types of co-occurrences
of states $\{{\color{black}{-1,{\color{black}{\color{black}2}},1}\}\times\{}{-1,{\color{black}{\color{black}{\color{black}2}}},1}\}$,
a multiple hypothesis testing correction is needed. We choose to use
the False Discovery Rate \cite{Benjamini1995} with a rate set to
$p_{0}=0.05$. A network is built by validating links between pairs
of traders if the p-value of their synchronization is smaller than
the FDR-corrected threshold;}\textcolor{black}{ traders without any links are dropped (see \cite{Michele2012} for more details).} \textcolor{black}{Note that contrarily to one-shot statistical testing, for which such value of $p_0$ would be foolishly large and which does not control the false discovery rate (i.e. false positives), we are dealing here with a population of p-values in which the FDR is controlled. In other words, there is on average a $p_0$ fraction of  false links in the SVNs that we determine. Since we are mostly interested in groups, this value is not crucial.}

\textcolor{black}{The resulting network consists most of the time
in a large connected component (i.e. a large group of connected traders)
and other very small disconnected components. The large connected
component is further decomposed into communities (or modules). Many
methods have been designed to detect communities in complex networks
(see e.g. \cite{PhysRevE.80.056117} for a review). As in \cite{Michele2012},
we use the InfoMap method \cite{Rosvall2008}, which segments a connected
network according to a maximum entropy argument. While this method
is not suited for multi-links networks, it can deal with weighted
networks. Therefore an easy workaround consists in converting multi-links
into weighted links by assigning a weight equal to the number of validated
links between two traders. When applying community detection, we exclude
links between opposite action (buy-sell), as we are primarily interested
in finding groups of traders that act in the same direction at the same
time so as to be able to aggregate the volume of a given group and
compute a meaningful measure of its state. }

\subsubsection{\textcolor{black}{Trader synchronicity network descriptive statistics}}

\textcolor{black}{In the following, for each in-sample time window,
we keep the $500$ most active traders and filter out those with less
than 100 trades. We exclude weekends as trading activity then is markedly
different (and much smaller) than that of business days. In addition,
some traders do not use algorithmic trading, which restricts their
activity periods, or prefer to trade during the most active hours.
This is why we only keep trades from 9am to 4pm (London time). }

\textcolor{black}{Hourly time slices also allow the building of SVNs
over a few months, which opens the way to a large-scale investigation
both of the time evolution of SVNs and of prediction (see Sec.  \ref{sec:Flux-prediction}).
Figure \ref{fig:Examples-hour} shows representative examples of SVNs
computed with hourly time slices over a given time period. The number
of clusters is of particular interest: while the number of groups
of SQ traders stays roughly constant (and large), a peculiar phenomenon
occurs in the four months preceding Jan. 2014 in the LB dataset: Fig.
\ref{fig:Network_over} reports that the number of detected groups
reaches 1 then, with a similar decrease of the number of links between
traders. This implies that our method detected much less statistically
validated synchronization during this period. We were not able to
find a simple explanation for this phenomenon, but it is clear that
the presence of a single group prevents significant lead-lag. We thus
expect predictability to be minimal around January 2014 for LB data
(which is confirmed in section \ref{subsec:pred_results}). }

\textcolor{black}{}
\begin{figure}
\centering{}\textcolor{black}{\includegraphics[width=0.5\textwidth]{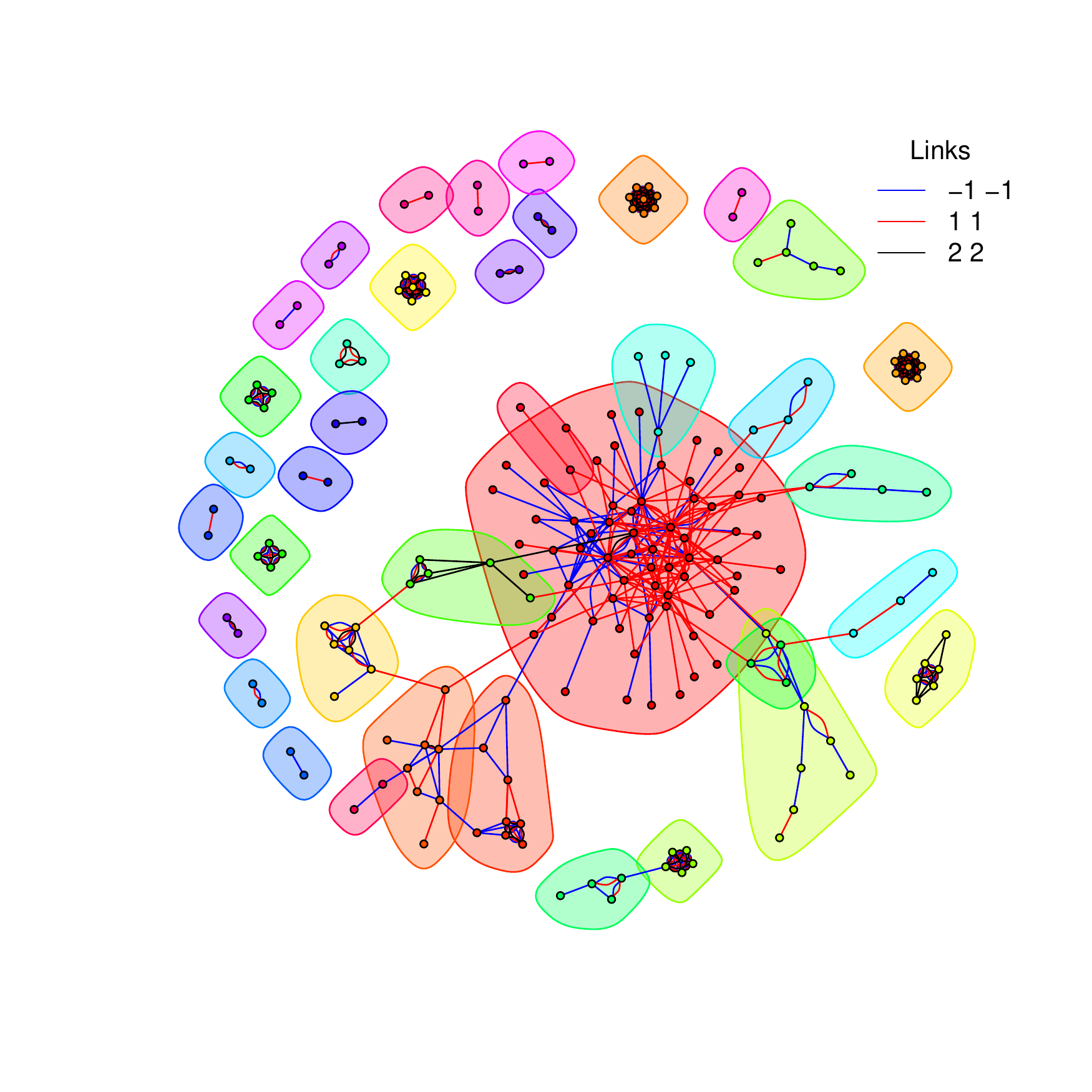}\includegraphics[width=0.5\textwidth]{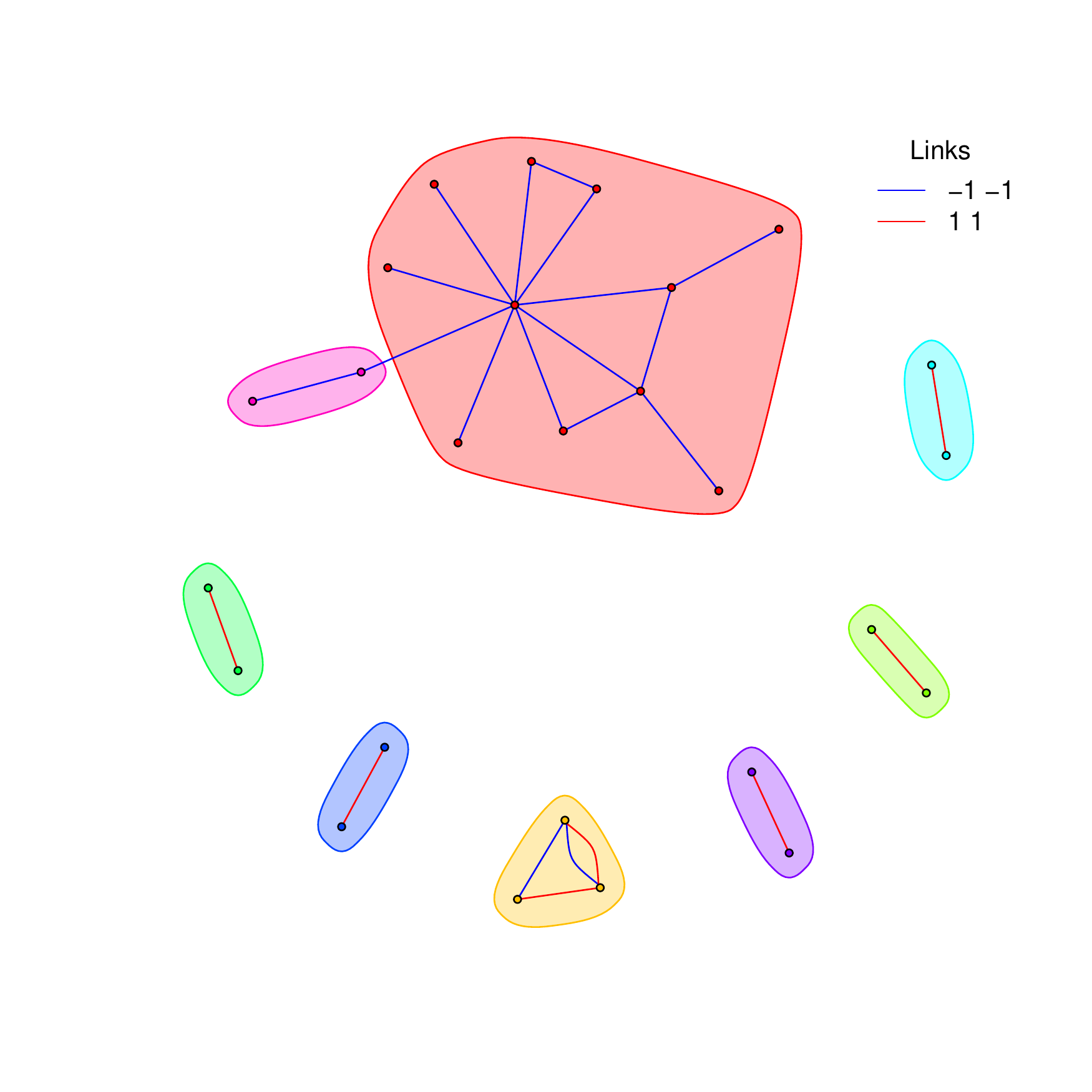}\caption{Projection of typical networks of traders, determined at the hourly
time scale (\texttt{EURUSD}). Every link between two traders is statistically
validated and labelled by the equal-time behavior of the pair: (1,1)
if both are net buyers, (-1,-1) if both are net sellers, (2,2) if
both are neutral. The InfoMap community detection method additionally
provides clusters, shown in colors. Left: SQ 2012. Right: LB.\label{fig:Examples-hour}.\textcolor{black}{{}
These are bi-dimensional projections of complex networks in which
horizontal and vertical directions have no special meaning.}}
}
\end{figure}

\textcolor{black}{}
\begin{figure}
\centering{}\textcolor{black}{\includegraphics[scale=0.4]{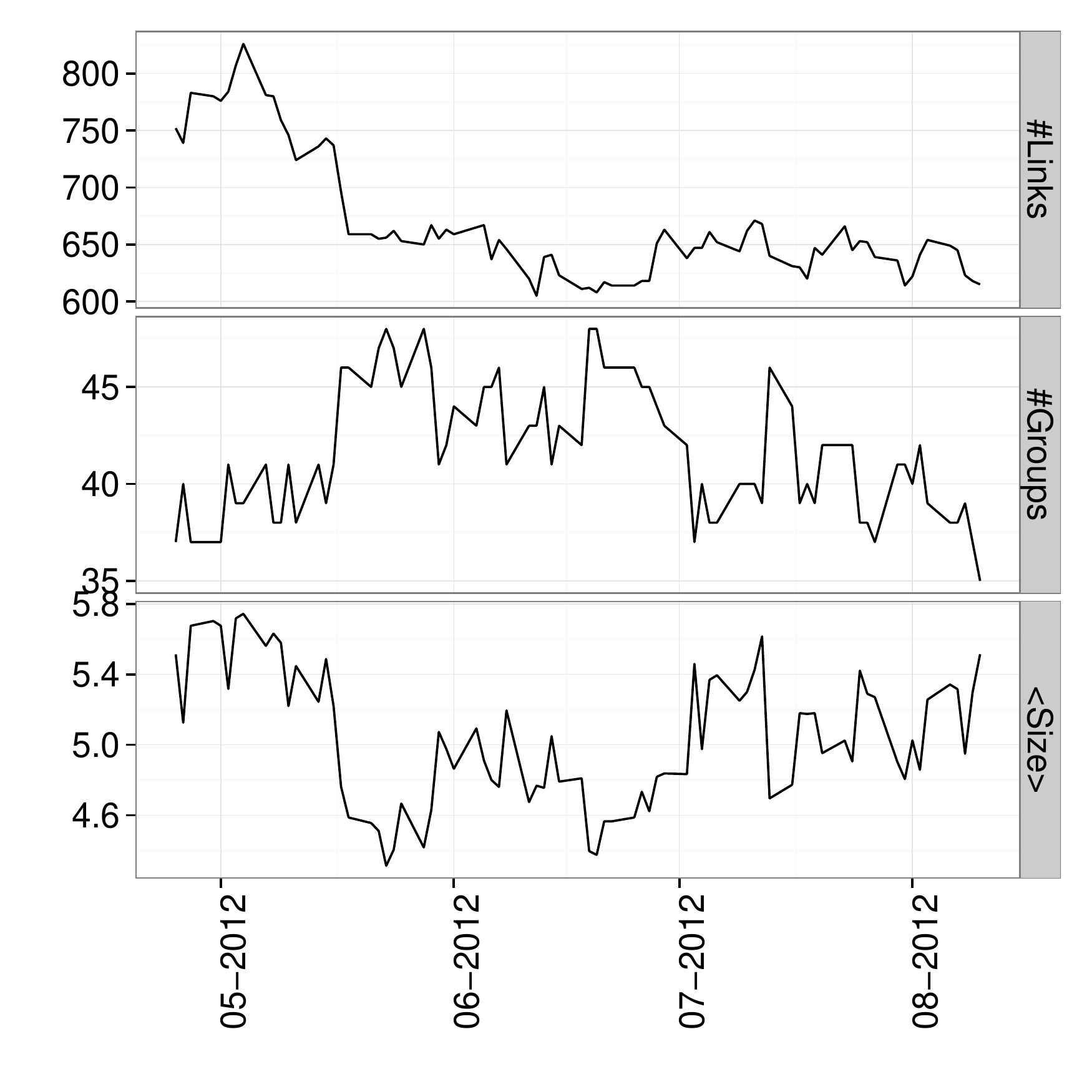}\includegraphics[scale=0.4]{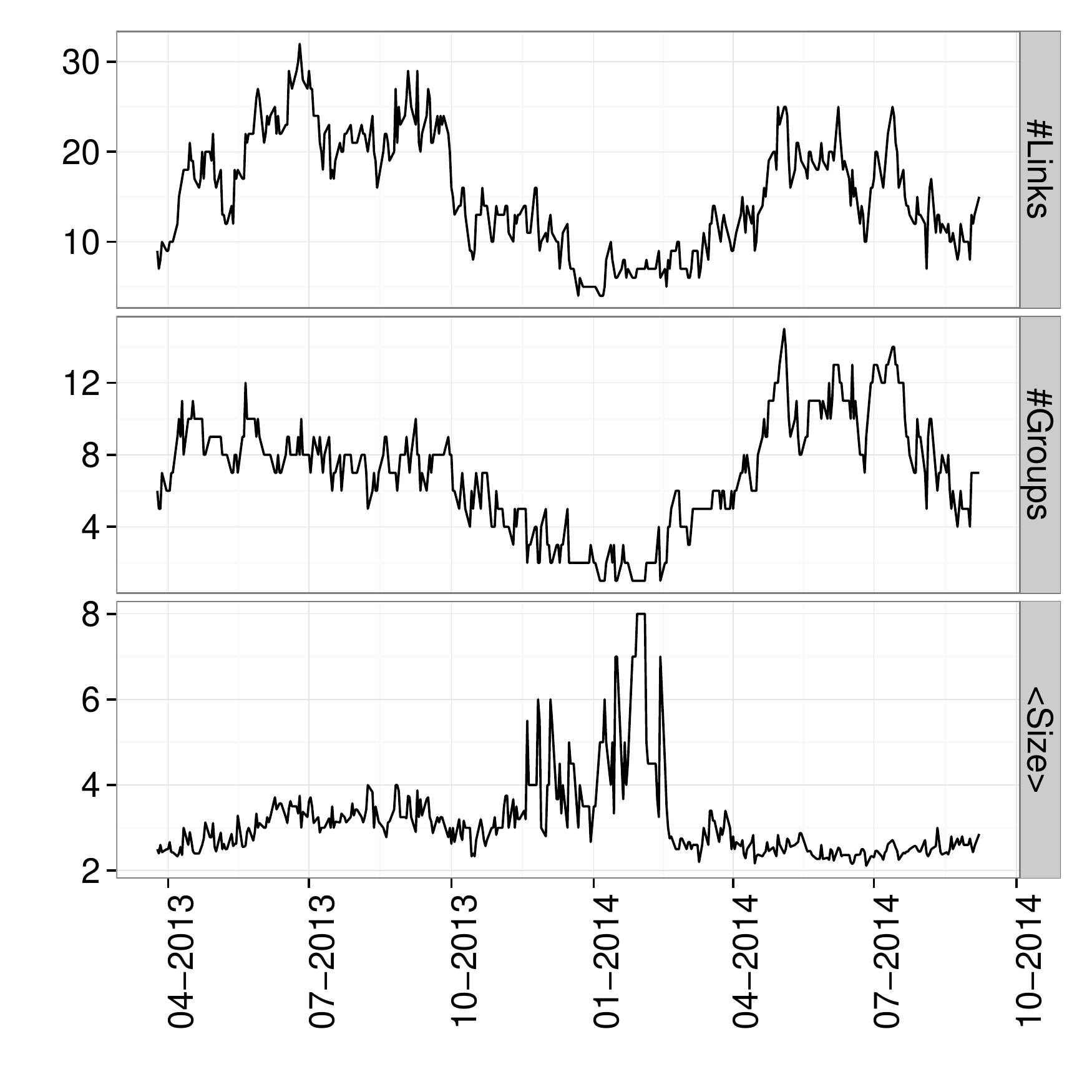}\caption{Network of traders: basic statistics over time \textcolor{black}{(}\texttt{\textcolor{black}{EURUSD}}\textcolor{black}{)}.
Left: SQ 2012. Right: LB.\label{fig:Network_over}}
}
\end{figure}

\subsubsection{\textcolor{black}{Stability of clustering as a function of time}}

\textcolor{black}{Sliding windows allow us to detect time-varying
clustering, which reflects changes of strategy use by the traders
as time goes on (as detect by a fixed 1-hour time slices). They also raise the question of the stability of
clustering, a necessary condition for the persistence of lead-lag
networks, thus for the possibility of predicting order flows. }

\textcolor{black}{Assessing clustering stability is made easier by
having consistent group labeling. The first step is to make sure that
the clustering tools guarantees the constant labeling of groups when
the grouping of traders is exactly the same one between two time slices.
In practice, a non-negligible proportion of traders does not belong
to the same group in the clustering performed at two consecutive times.
Thus, the first problem to solve is how to attribute coherent names
to clusters as time goes on. The simplest solution is to use a similarity
measure between of the grouping at time slices $t$ and $t-1$ and
to propagate the name of cluster $g_{t-1}$ to the most similar cluster
at time $t$. The similarity measure is based on the overlap of the
elements of two clusters and defined as }

\textcolor{black}{
\begin{equation}
OA(g_{t-1},g'_{t})=\left|g_{t-1}\cap g'_{t}\right|,\label{eq:OA_num}
\end{equation}
where $g$ and $g'$ are trader groups/clusters and where $|x|$ stands
for the number of elements of $x$. We shall use the normalized overlap
measure}

\textcolor{black}{
\begin{equation}
OP(g_{t-1},g'_{t})=\frac{OA(g_{t-1},g'_{t})}{\left|g_{t-1}\cup g'_{t}\right|}\label{eq:OA_frac}
\end{equation}
to account for the size of both clusters. }

\textcolor{black}{Consistent naming allows us to produce meaningful
visualizations. Figure \ref{fig:Cluster-dynamics-as} shows how the
traders switch between clusters as a function of time, using a so-called
``river chart'': at a given time, the traders belonging to the same
group are stacked together and form a continuous vertical dash, each
group being clearly separated from each other. One then adds the trajectory
of each trader from its group at time $t$ and its group at time
$t+1$. Having a consistent group labeling method ensures that if
there is strictly no change of group between two time steps, only
horizontal stripes appear. Thus, river charts allow us to visualize
at the same time group sizes and the evolution of group compositions,
i.e. distributional and dynamical properties. There clearly is a large
cluster whose size is relatively stable as a function of time. The
smallest clusters are much less stable: they merge and split again
as time goes on. It is noteworthy that clustering was performed every
week in this figure: the cluster structure is relatively stable even
at this sampling frequency.}

\begin{figure}[p]
\begin{centering}
\textcolor{black}{\includegraphics[scale=0.3]{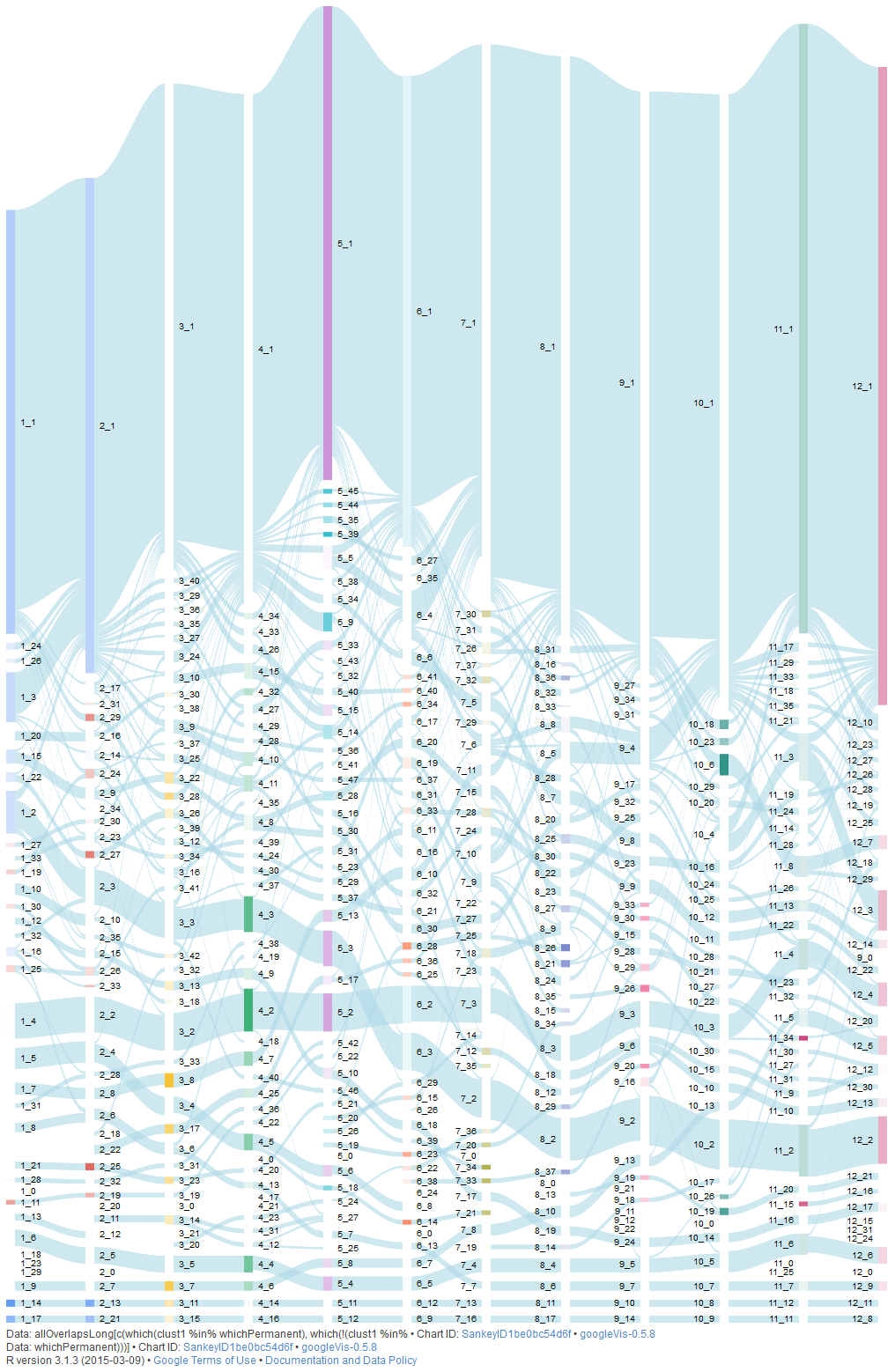}}
\par\end{centering}
\textcolor{black}{\caption{Agent clustering time evolution. Each elementary line width represents
the trajectory of an agent from one group to an other group. Hourly
time slices, 12 weeks in-sample, weekly clustering; SQ 2012, \texttt{\textcolor{black}{EURUSD}}.\label{fig:Cluster-dynamics-as} }
}
\end{figure}

\subsubsection{Cluster membership stability}

\textcolor{black}{We use the Adjusted Rand Index (ARI thereafter), a standard
global measure of clustering stability between two consecutive
clustering times \cite{rand1971objective,Fraley12mclustversion}.
An ARI of 1 denotes perfect clustering stability, while the expected
value of ARI for random clustering is 0. The stability of LB traders
is perfect (ARI=1) in about a third of the days, which underscores
a remarkable level of regularity of LB clients, also hinted at by
the large average ARI. Retail clients of SQ are more fickle (possibly
because a larger fraction of them do not use algorithmic trading), but their average
ARI is also remarkably high (see Table \ref{tab:Summary-statistics-of_ARI}).
In short, the ARI suggests a strong and encouraging level of clustering
stability in all data sets.}

\textcolor{black}{}
\begin{table}
\textcolor{black}{}%
\begin{tabular}{|c|c|c|c|c|}
\hline
\textcolor{black}{mean(std dev)} & \textcolor{black}{LB }\texttt{\textcolor{black}{EURUSD}} & \textcolor{black}{SQ }\texttt{\textcolor{black}{EURUSD}} & \textcolor{black}{SQ }\texttt{\textcolor{black}{EURGBP}} & \textcolor{black}{SQ }\texttt{\textcolor{black}{USDJPY}}\tabularnewline
\hline
\hline
\textcolor{black}{mean} & \textcolor{black}{0.87} & \textcolor{black}{0.83} & \textcolor{black}{0.91} & \textcolor{black}{0.84}\tabularnewline
\hline
\textcolor{black}{standard deviation} & \textcolor{black}{0.14} & \textcolor{black}{0.09} & \textcolor{black}{0.09} & \textcolor{black}{0.10}\tabularnewline
\hline
\textcolor{black}{fraction of perfect stability} & \textcolor{black}{0.34} & \textcolor{black}{0} & \textcolor{black}{0.01} & \textcolor{black}{0}\tabularnewline
\hline
\end{tabular}

\textcolor{black}{\caption{Summary statistics of the Adjusted Rand Index for some currency pairs
of both data sources. SQ means SQ2014-6.\label{tab:Summary-statistics-of_ARI}}
}
\end{table}

\subsection{{Statistically validated lead-lag networks}}

\textcolor{black}{Determining validated lead-lag relationships between
two time-series essentially consists in detecting synchronicity between
the first time series and the suitably lagged second time series.
In other words, one may apply the SVN machinery to the state of agent
$i$ and the lagged state of agent $j$ (including the case $i=j$).
In the context of investors, since some agents use the same systematic
strategies either to open or close a position, or both, and thus act
in a remarkably synchronous way, it makes sense to focus on the lead-lag
relationships between the groups of traders determined with SVNs and
a community detection method. }

\textcolor{black}{Once the groups of traders are determined via the
SVN method, the procedure works as follows:}
\begin{enumerate}
\item \textcolor{black}{volume imbalances are aggregated at the group level:
$V_{g}(t)=\sum_{i\in g}V_{i}(t)$ and the state of group $g$, denoted
by $\sigma_{g}(t)$, is determined in the same way as in the SVN method;}
\item \textcolor{black}{for each pair of group $(g,g')$ ($g=g'$ is allowed),
the p-value of the coincidence between $\sigma_{g}(t)$ and $\sigma_{g'}(t+1)$
is computed as in the SVN method;}
\item \textcolor{black}{the number of pairs of groups is $N_{groups}^{2}$
because we allow self-linking. Since the number pairs of states is
3x3, the significance level needs to account for $N_{tests}=9\times N_{groups}^{2}$
tests, thus FDR is used once more. Notice in particular that $g\rightarrow g$
links are not trivial and correspond to auto-correlated time series
of aggregated volume imbalance: these links appear as loops in the
directed network representation.}
\end{enumerate}
Grouping agents has the advantage of simplifying the description of
the system state, thus to reduce the dimensionality of the prediction
problem. In some cases, it may be useful to skip the grouping step
and determine lead-lag between agents.

\subsubsection{\textcolor{black}{Results}}

\textcolor{black}{The same parameters as in Fig.~\ref{fig:Examples-hour}
are used for link detection between groups. We display two representative
networks in Fig.~\ref{fig:SVLLN}. The most common type of link is
to oneself. More complex lead-lag relationships also exist: take group
22 in SQ data set; it typically buys }\texttt{\textcolor{black}{EURUSD}}\textcolor{black}{{}
within one hour after having sold }\texttt{\textcolor{black}{EURUSD}}\textcolor{black}{;
group 31 does the opposite. One notes that, \textcolor{black}{interestingly}, group
31 sells during the hour after group 22 has sold. This means that
both groups act in an opposite way provided that group 22 has sold
}\texttt{\textcolor{black}{EURUSD}}\textcolor{black}{{} in the previous
time slice. What is remarkable is this happens in a  systematic
way.}

\textcolor{black}{There are more lead-lag links between SQ groups than between LB groups, especially
self-links. Interestingly, more links that validate opposite directions
are present in the lead-lag case compared to the contemporaneous one.
The evolution of the number of links over time is shown in Fig. \ref{fig:LL_overtime}:
sudden drops of the number of lead-lag links are noticeable for all
data sets of SQ. Quite logically, there is no lead-lag relationships
around January 2014 for the LB dataset, which is to be related to
the detection of only one group (see Fig.~\ref{fig:Network_over}). }

\textcolor{black}{}
\begin{figure}
\begin{centering}
\textcolor{black}{\includegraphics[width=0.4\textwidth]{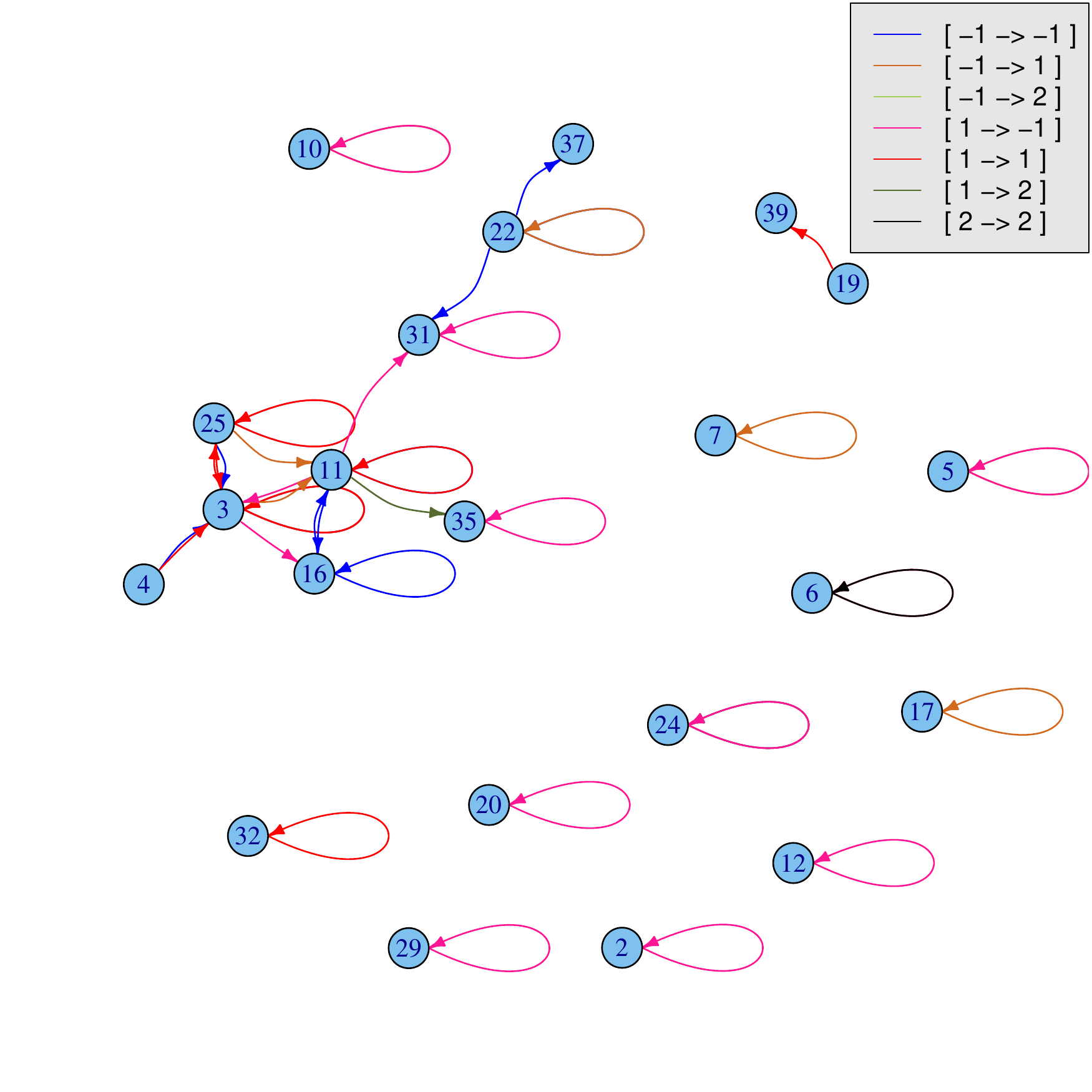}\includegraphics[width=0.4\textwidth]{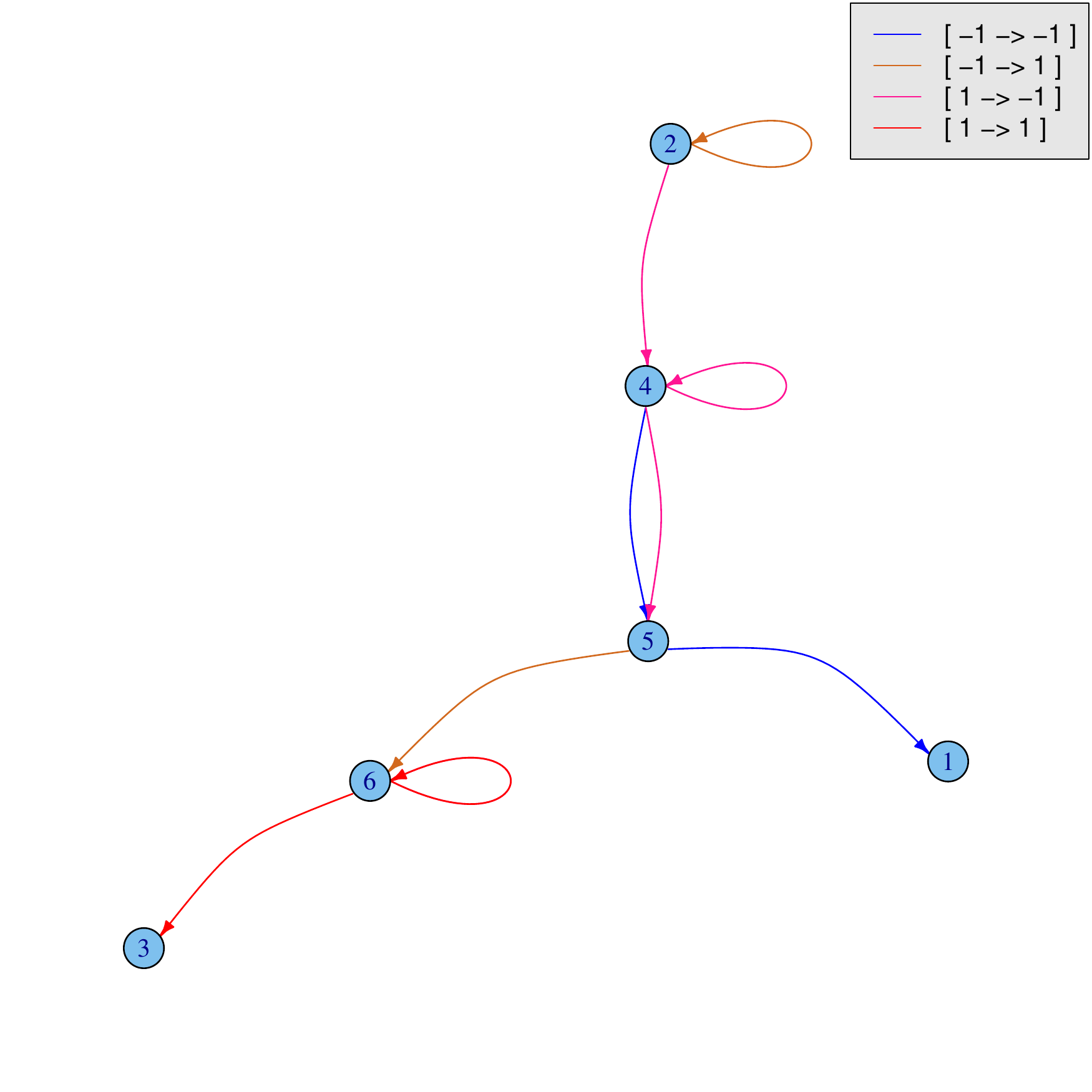}}
\par\end{centering}
\centering{}\textcolor{black}{\caption{Examples of lead-lag networks between groups at an hourly scale, for
a given given date and a calibration window of 50 days (\texttt{\textcolor{black}{EURUSD}},
left: SQ, right: LB). \label{fig:SVLLN} 2 labels the neutral state.}
}
\end{figure}
\textcolor{black}{}
\begin{figure}
\centering{}\textcolor{black}{\includegraphics[scale=0.45]{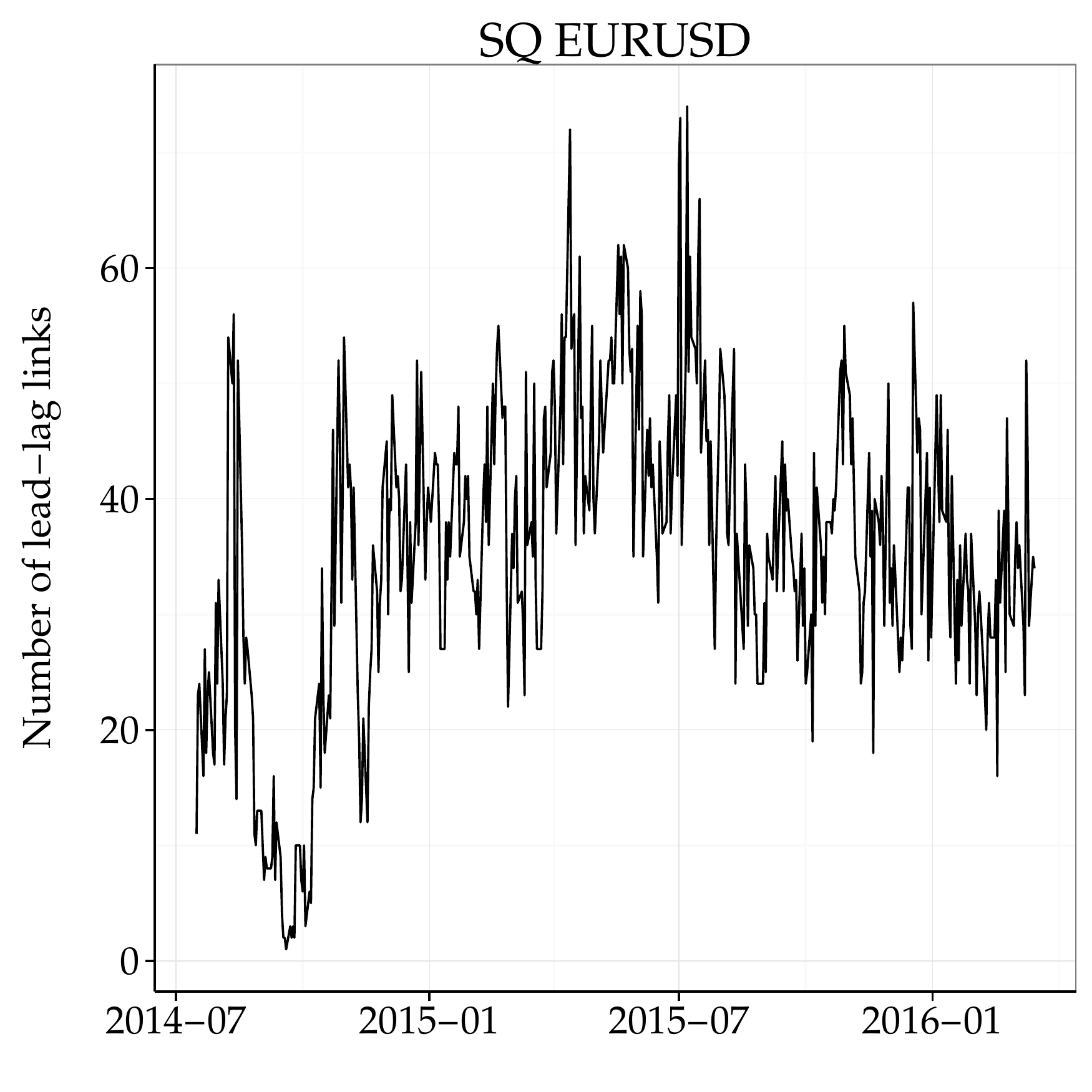}\includegraphics[scale=0.45]{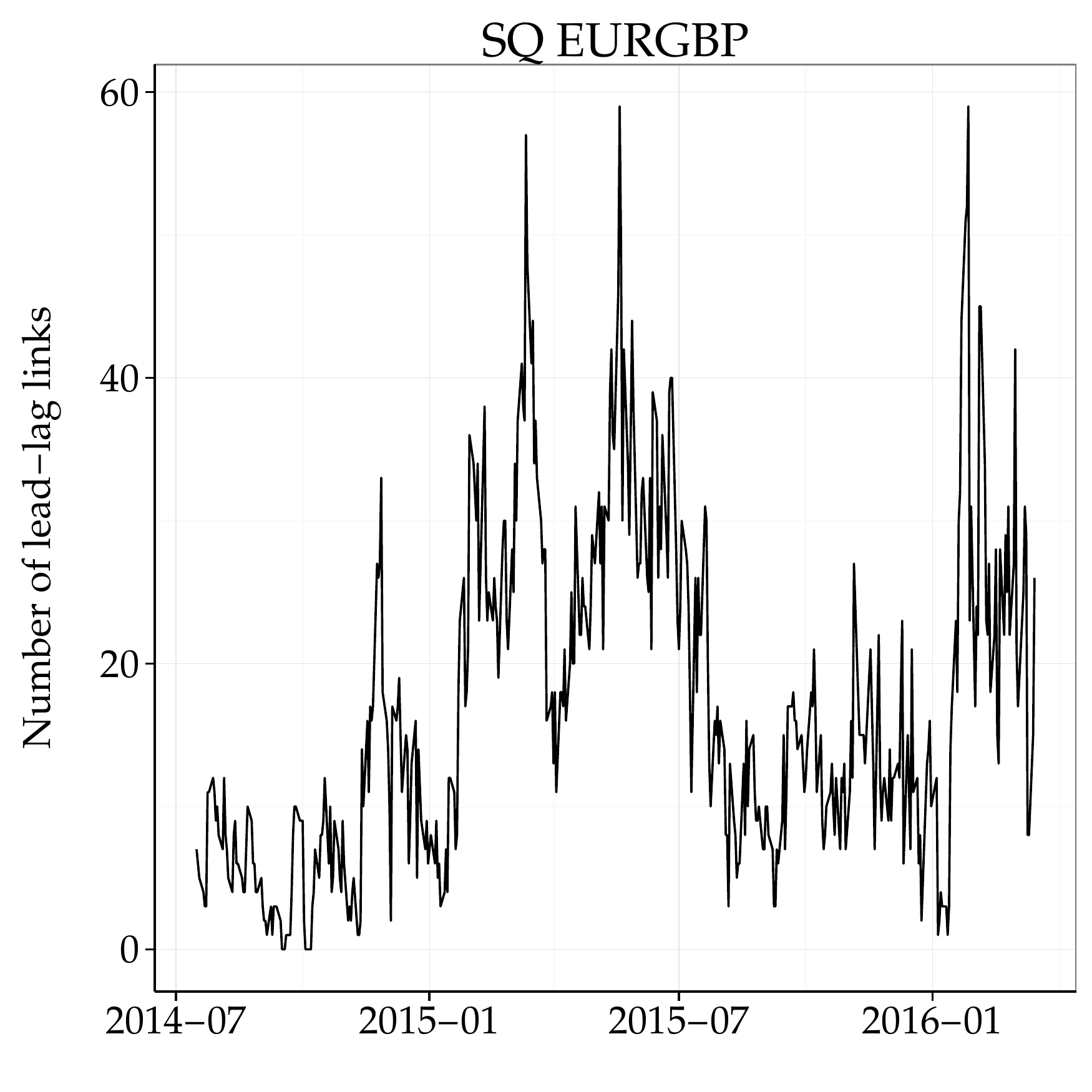}}\\
\textcolor{black}{\includegraphics[scale=0.45]{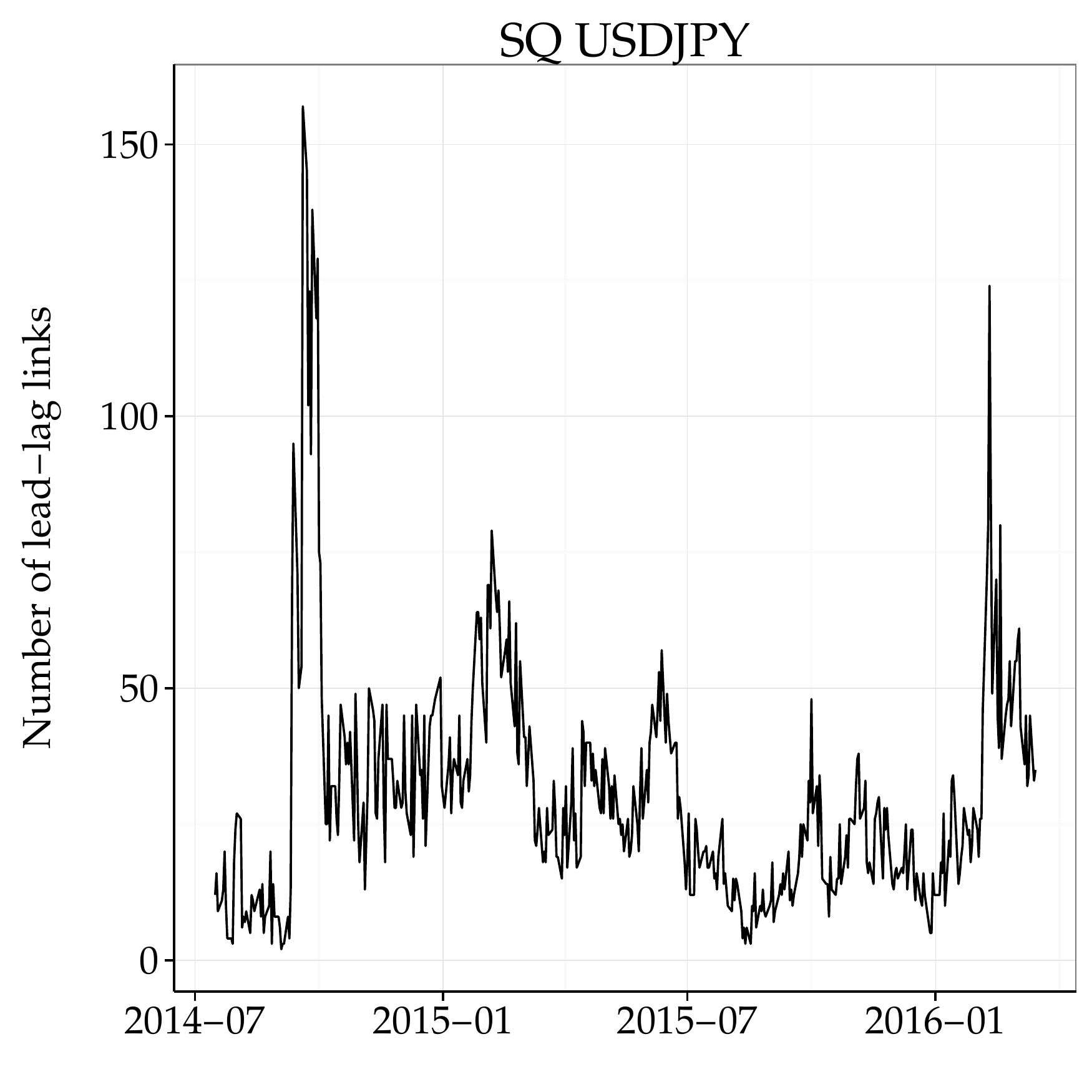}\includegraphics[scale=0.45]{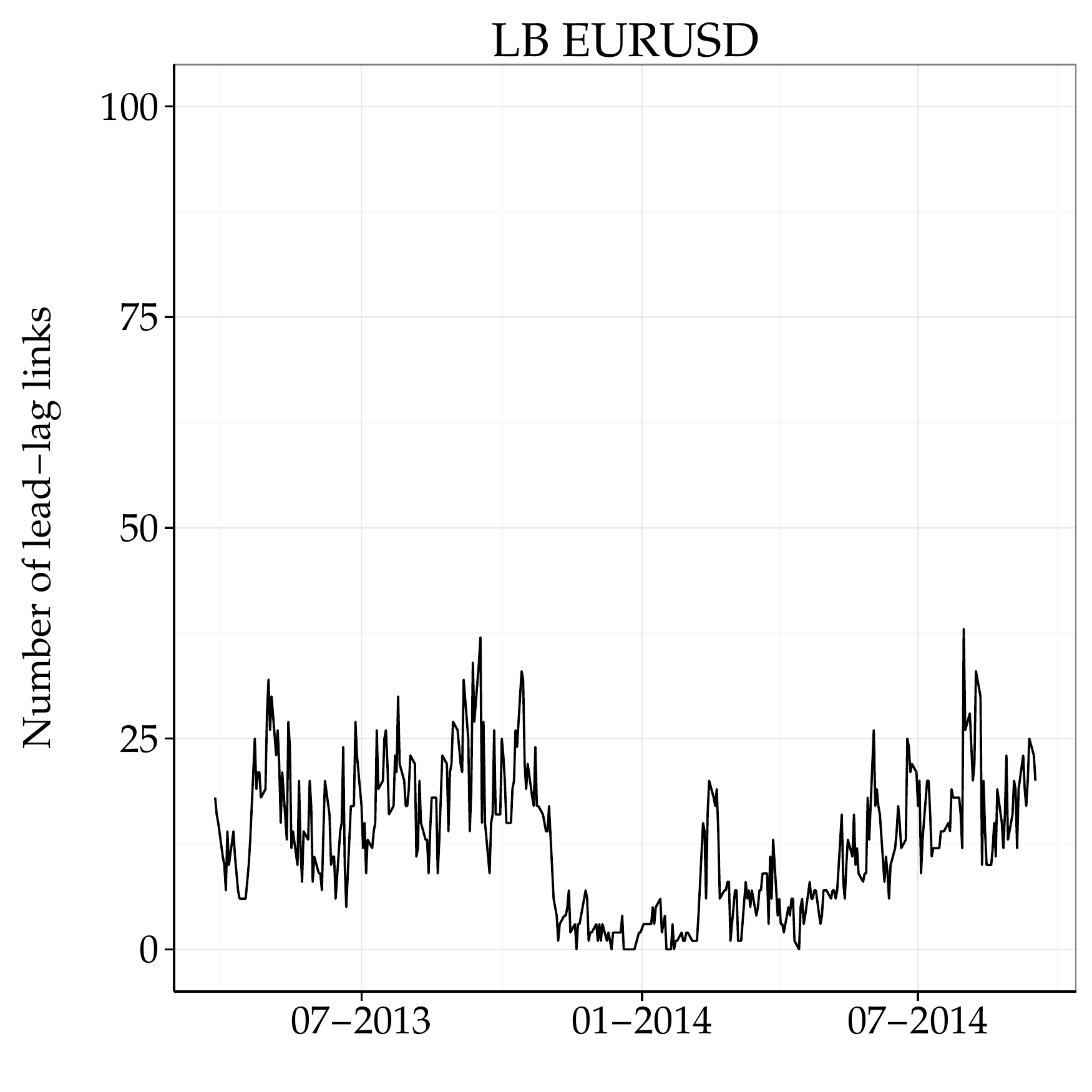}\caption{Lead-lag network of traders: number of links as a function of time.
Sliding windows of 45 days of calibration.\label{fig:LL_overtime}}
}
\end{figure}
\textcolor{black}{}

\textcolor{black}{The presence of links, valid under severe statistical
inspection, clearly demonstrate the existence of predictability in
the investors trading directions. However, predicting trading flows
not only requires equal-time clustering stability, but also lead-lag
stability. Measuring the latter at the level of traders is  more
informative than only using lead-lag between groups. Indeed, imagine
that at time $t$, group 1 includes Alice and Bob, and leads on group
2 which contains Carol and Dave. It may happen that at time $t+1$
group 1 splits into groups 1a and 1b, that the composition of group
2 is conserved, and that groups 1a and 1b still lead on group 2. The
point here is that the lead-lag between group 1 and 2 defines }\textcolor{black}{\emph{a
fortiori}}\textcolor{black}{{} lead-lag relationships between all the
traders of group 1 and those of group 2 at time $t$. The subsequent
splitting of group 1 into 1a and 1b does not change the lead-lag relationships
between traders. Thus, a suitable lead-lag stability measure is the
fraction of lead-lag links between traders that is conserved between
two successive clustering times, restricted to the traders that exist
at both times. Mathematically, let $\Lambda_{ij}(t\to t+1)$ denote
the adjacency matrix element of the lead-lag network at the trader
level between time $t$ and time $t+1$, i.e., $\Lambda_{ij}(t\to t+1)=1$
if trader $i$ leads on trader $j$ and 0 otherwise, then the stability
measure is defined as
\[
{\color{black}{\color{black}\beta(t\to t+1)=\frac{\sum_{ij}\Lambda_{ij}(t\to t+1)\Lambda_{ij}(t+1\to t+2)}{\sum_{i'j'}\Lambda_{i'j'}(t\to t+1)}}}
\]
Figure \ref{fig:Trader-trader-lead-lag-overlap} reports the time
evolution of $\beta$. It does fluctuate much, but never quite reaches
0 for long periods, except for LB in January 2014, which, on the whole,
leaves hope of successful predictions. However, one readily notices
that the number of both validated lead-lag network nodes and links
is much smaller for LB data. Whether a high value of $\beta$ is related
to a larger predictive power is investigated in Sec. \ref{sec:AUC_beta}.}

\textcolor{black}{}
\begin{figure}
\begin{centering}
\textcolor{black}{\includegraphics[width=0.4\textwidth]{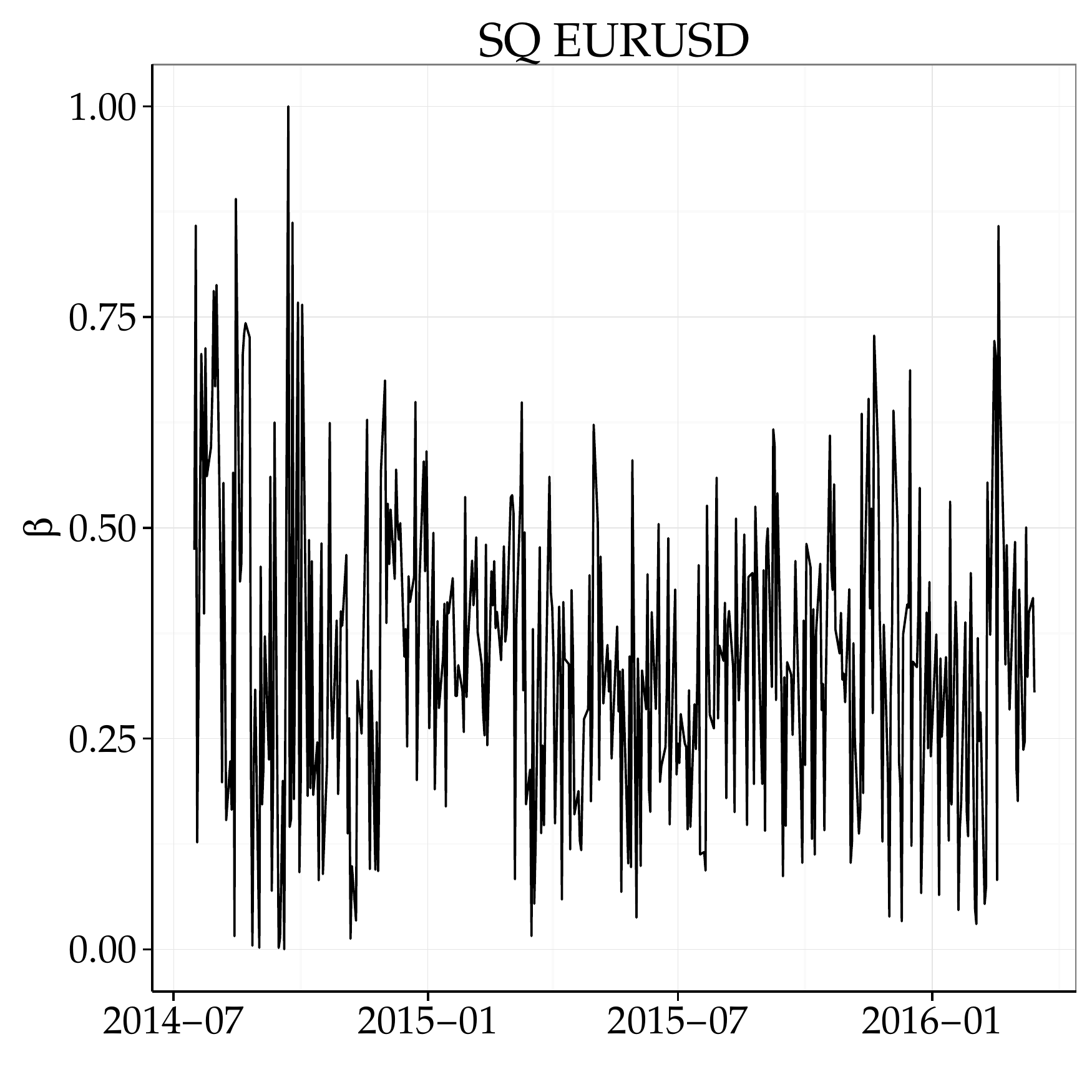}\includegraphics[width=0.4\textwidth]{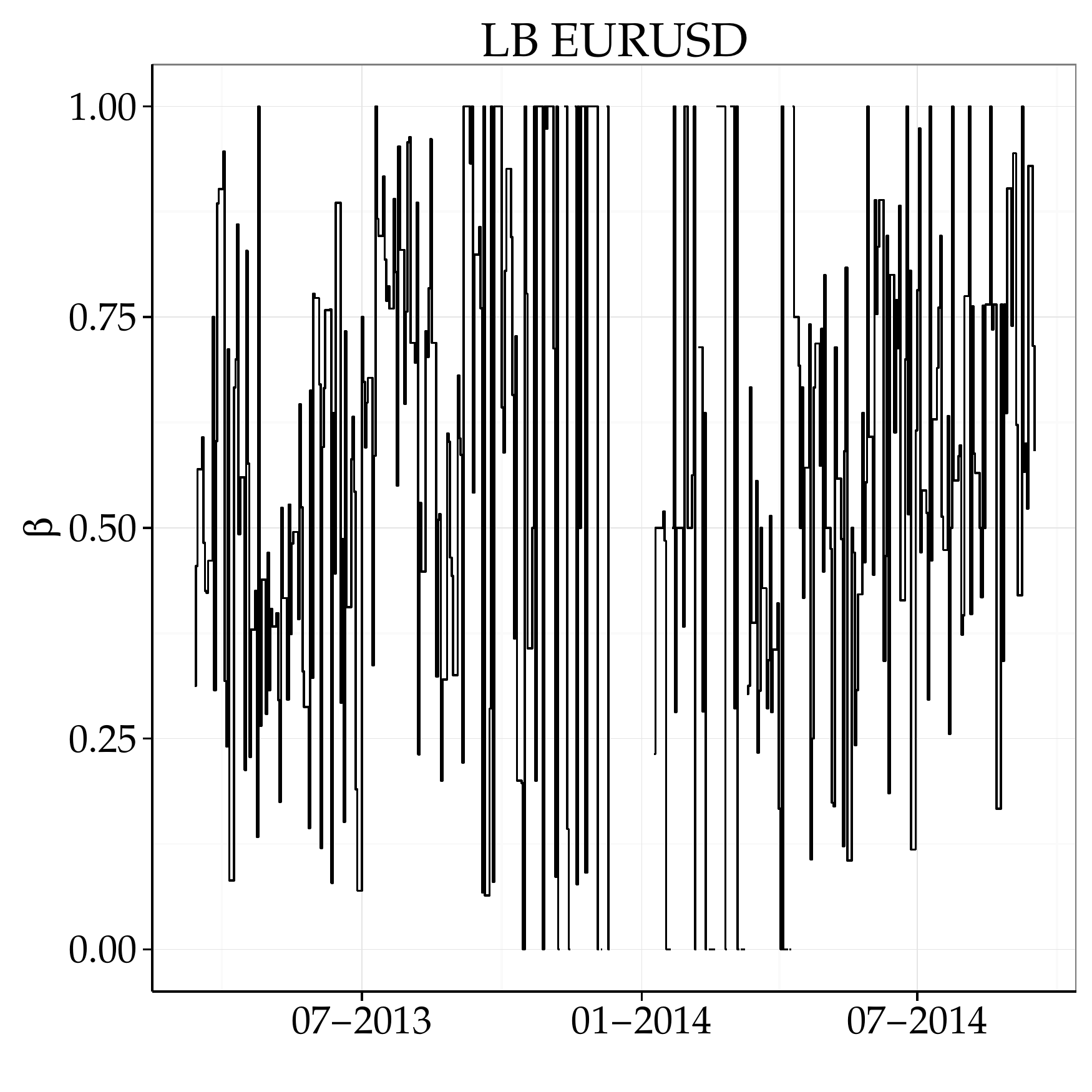}}
\par\end{centering}
\textcolor{black}{\caption{Trader-trader lead-lag overlap ratio $\beta$ as a function of time.
Same parameters as in Figure \ref{fig:LL_overtime}.\label{fig:Trader-trader-lead-lag-overlap}}
}

\end{figure}

\section{The predictability of inventory \label{sec:Flux-prediction}}

\textcolor{black}{We have so far shown the existence of a lead-lag
structure whose persistence implies that some quantities are predictable.
This does not however mean that e.g. order flows are simple autoregressive
models. When one fits ARIMA models and hours of the day as fixed factors,
the AIC criterion always suggests to use a ARIMA(0,0,0) model: at
this time scale, the only relevant factors are the hours of the day. }

\textcolor{black}{These networks also give crucial insights about
how to make predictions from trader-resolved data. While future actions
do depend on the current state of all groups, any prediction method
must also be fed with the lagged group states, so that the lead-lag
networks can be learned and exploited as well. }

\subsection{Order flow }

\textcolor{black}{For the sake of simplicity, we have restricted the
group states to their discrete values \{-1,2,+1\}. This strongly suggests
to predict the sign of the total order flow, instead of its exact
value. More precisely, we aim to predict $v(t+1)=\text{sign}(\sum_{i}V_{i}(t+1))$,
where $i$ sweeps over all the traders from a set $P$
of predictors consisting in the states of the groups, denoted by $\sigma_{g,t}$
 (determined in-sample by SVNs), their lagged values $\sigma_{g,t-1}$
and the time of the day. Including lagged group states is consistent
with the existence of group lead-lag of order one and is therefore necessary. }. This amounts  to classify ${v}(t+1)$
into $\{-1,0,+1\}$, from the knowledge of group states up to time
$t$. 

\textcolor{black}{The training phase is summarized by $P_{t_{0},t_{1}}\sim W_{t_0+1,t_{1}+1}$,
where $P_{t_{0},t_{1}}$ is a matrix of predictors and $W_{t_0+1,t_1+1}$
the vector of the quantity to be predicted; more precisely,}

\textcolor{black}{
\begin{equation}
P_{t_{0},t_{1}}=\left(\begin{array}{ccccc}
\sigma_{1,t_{0}} & \sigma_{2,t_{0}} & \sigma_{1,t_{0}-1} & \sigma_{2,t_{0}-1} & h(t_{0})\\
\vdots & \vdots & \vdots & \vdots & \vdots\\
\sigma_{1,t_{1}} & \sigma_{2,t_{1}} & \sigma_{1,t_{1}-1} & \sigma_{2,t_{1}-1} & h(t_{1})
\end{array}\right)\sim\left(\begin{array}{c}
{v}_{t_{0}+1}\\
\vdots\\
v_{t_{1}+1}
\end{array}\right)=W_{t_0+1,t_1+1},\label{eq:pred_probl}
\end{equation}
where the symbol $\sim$ implies that there is some kind of (possibly
highly non-linear) relationship between a line of $P_{t_{0},t_{1}}$
and the corresponding next global trading flow imbalance, as suggested
by the validated lead-lag networks plotted in Fig.~\ref{fig:SVLLN}.
Because $P_{t_{0},t_{1}}$ also contains the time of the day, subtle
hourly differences of these validated lead-lag networks may be detected
as well. Note that we do not feed the lead-lag networks to the machine learning method, but the latter exploits them in an implicit way. In addition, $P_{t_{0},t_{1}}$ may also include group states
lagged more than once. Many variations of Eq. \eqref{eq:pred_probl}
are relevant. First, instead of the group states, one can input the
actual volume (or log-volume), $v$ may also be the VWAP or future
price returns (see section \ref{sec:VWAP}}), etc. At all rates, we focus on the simplest possible setting here. }

\textcolor{black}{The discrete nature of $v_t$ suggests to infer the $\sim$ relationship
with logistic regression, which does not lead to satisfactory results
(see appendix A where they are reported). Off-the shelf machine learning
methods outperform logistic regression thanks to their more non-linear
nature; we thus will focus on such methods. Instead of trying and
comparing many machine learning methods and tuning their parameters
until seemingly finding predictability, we chose a single method known
for its robustness and performance: plain random forests (RF) \cite{Breiman2001,rfsrc},
which have many useful features in this context: first, they avoid
in-sample over-fitting, they are robust, non-linear and possess an
overall very good predictive power without tweaking any parameter
(at least on many ``standard'' data-sets \cite{JMLR:v15:delgado14a}).
A bonus is the availability of the relative importance of each predictor.
As a consequence, we will be able to check that group states, i.e.,
lead-lag, is on average more important than the time of the day. For
the sake of computation speed, calibration was performed every day,
not every hour. Thus, predictions for a given day rest on a calibration
that uses data up to the previous day.  } 

\textcolor{black}{We chose 10 calibrations window lengths, denoted
by $T_{in}$, ranging from 45 to 90 week days (9 to 18 weeks) with
common difference of 5 days (1 week). Although RFs output classification
probabilities, i.e., the probability that the sign of the next order
flow will be +1, say, we take  the most probable predicted state as the prediction of a given RF. In addition to computing the respective performance of each calibration length, we also performed a majority vote of the
predictions originating from each timescale. If $+1$ and $-1$ obtain
the same number of votes, the prediction is set to zero .} Finally, we used the package RandomForestSRC \cite{rfsrc} and did not tune any parameter of the RF calibration function.

\textcolor{black}{}

\textcolor{black}{}

\textcolor{black}{}

\subsubsection{\textcolor{black}{Results}}

\textcolor{black}{\label{subsec:pred_results}}

\textcolor{black}{}
\begin{figure}

\begin{centering}
\textcolor{black}{\includegraphics[width=0.4\textwidth]{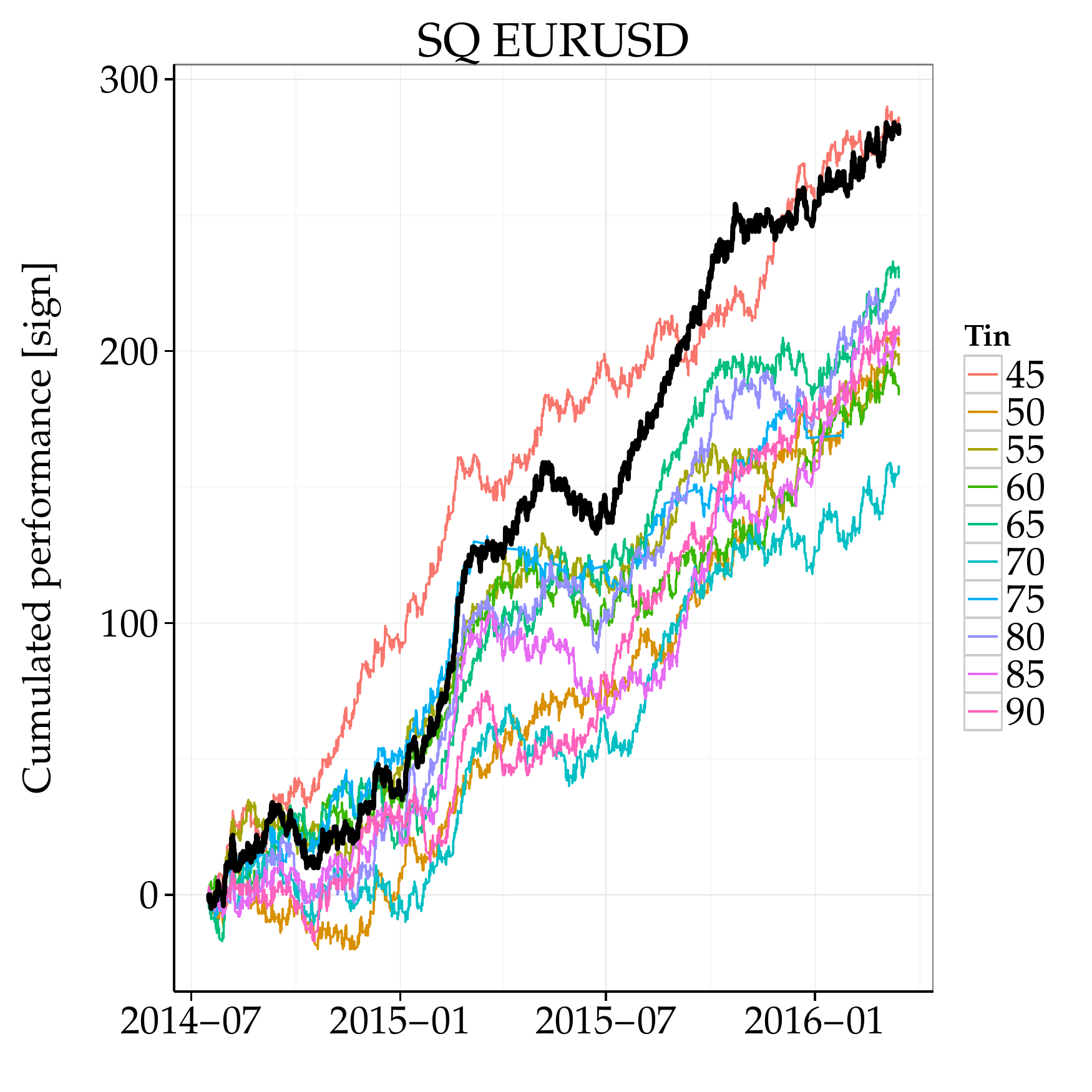}\includegraphics[width=0.4\textwidth]{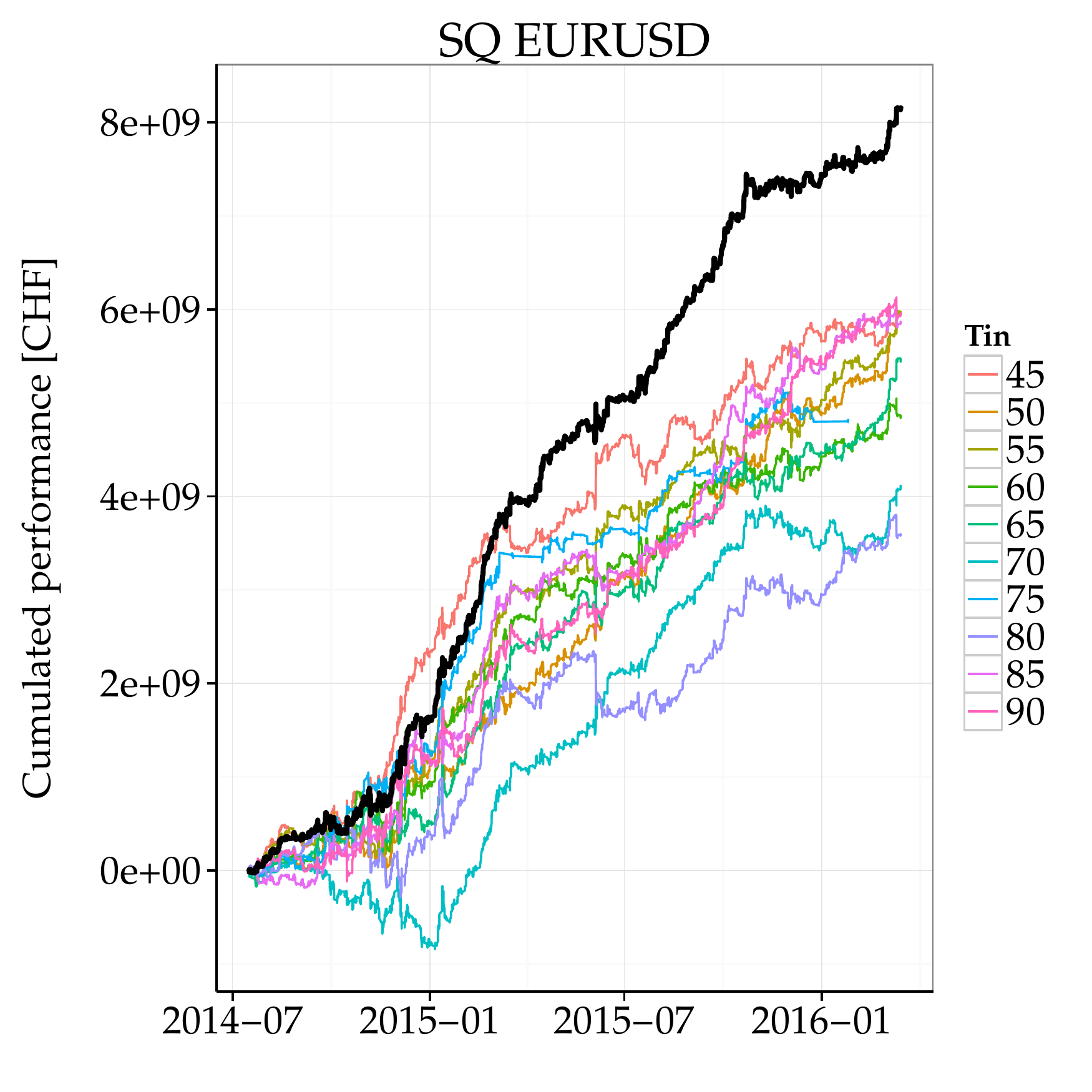}}\\
\par\end{centering}
\begin{centering}
\textcolor{black}{\includegraphics[width=0.4\textwidth]{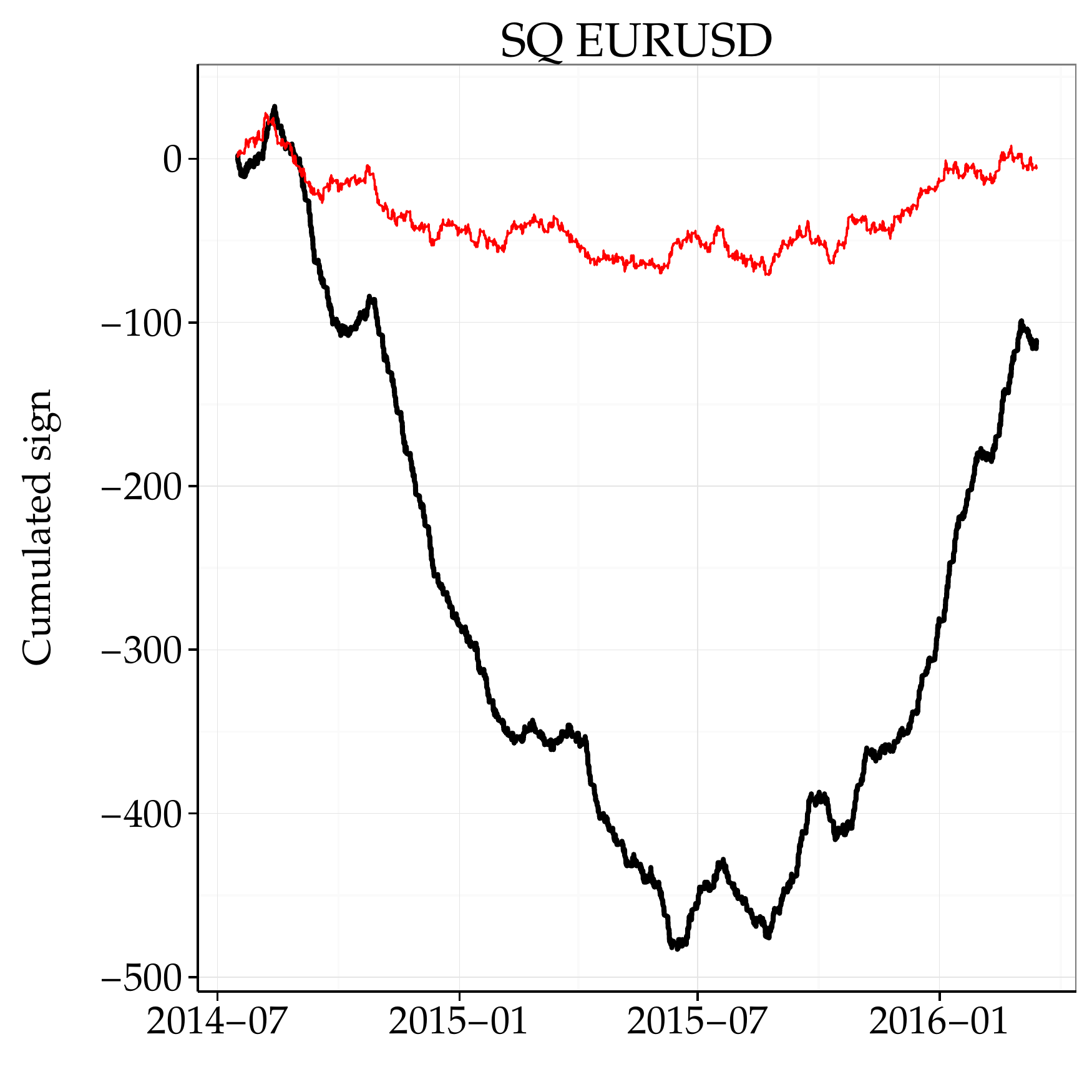}\includegraphics[width=0.4\textwidth]{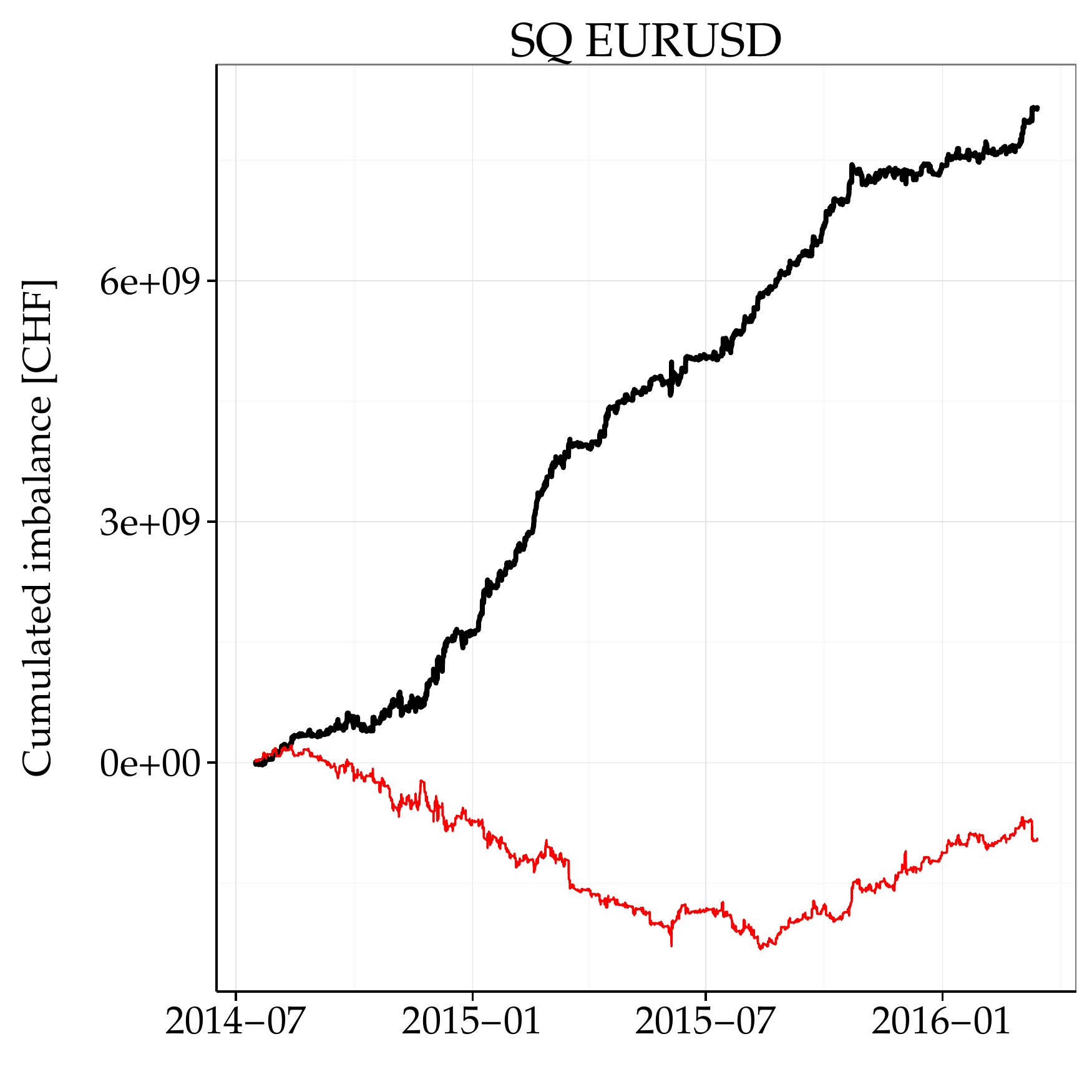}}
\par\end{centering}
\textcolor{black}{\caption{Out-of-sample performance: cumulated product of the predicted order
flow sign and actual sign (upper left), cumulated product of the predicted
order flow sign and actual order flow (upper right), cumulated product
of the predicted order flow sign and actual order flow, and the cumulated
actual flow (lower left), cumulated product of the predicted order
flow direction and actual order flow (black line) and cumulated actual
order flow (red line); \texttt{\textcolor{black}{EURUSD}}, Swissquote.
The thick black lines correspond to a majority vote between all calibration
window lengths.\label{fig:SQ_EURUSD_perf}}
}

\end{figure}
\textcolor{black}{}

\textcolor{black}{}
\begin{figure}
\begin{centering}
\textcolor{black}{\includegraphics[width=0.4\textwidth]{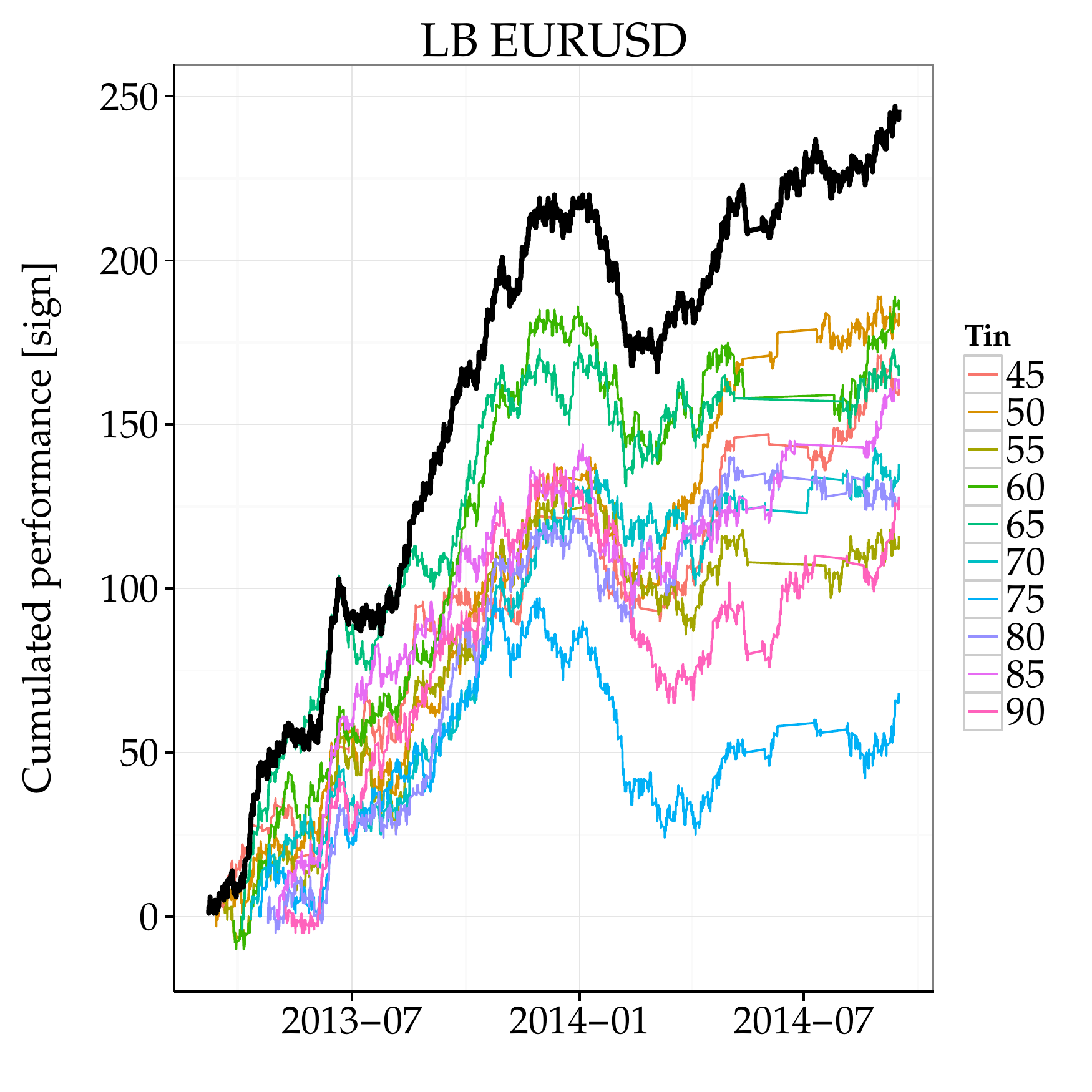}\includegraphics[width=0.4\textwidth]{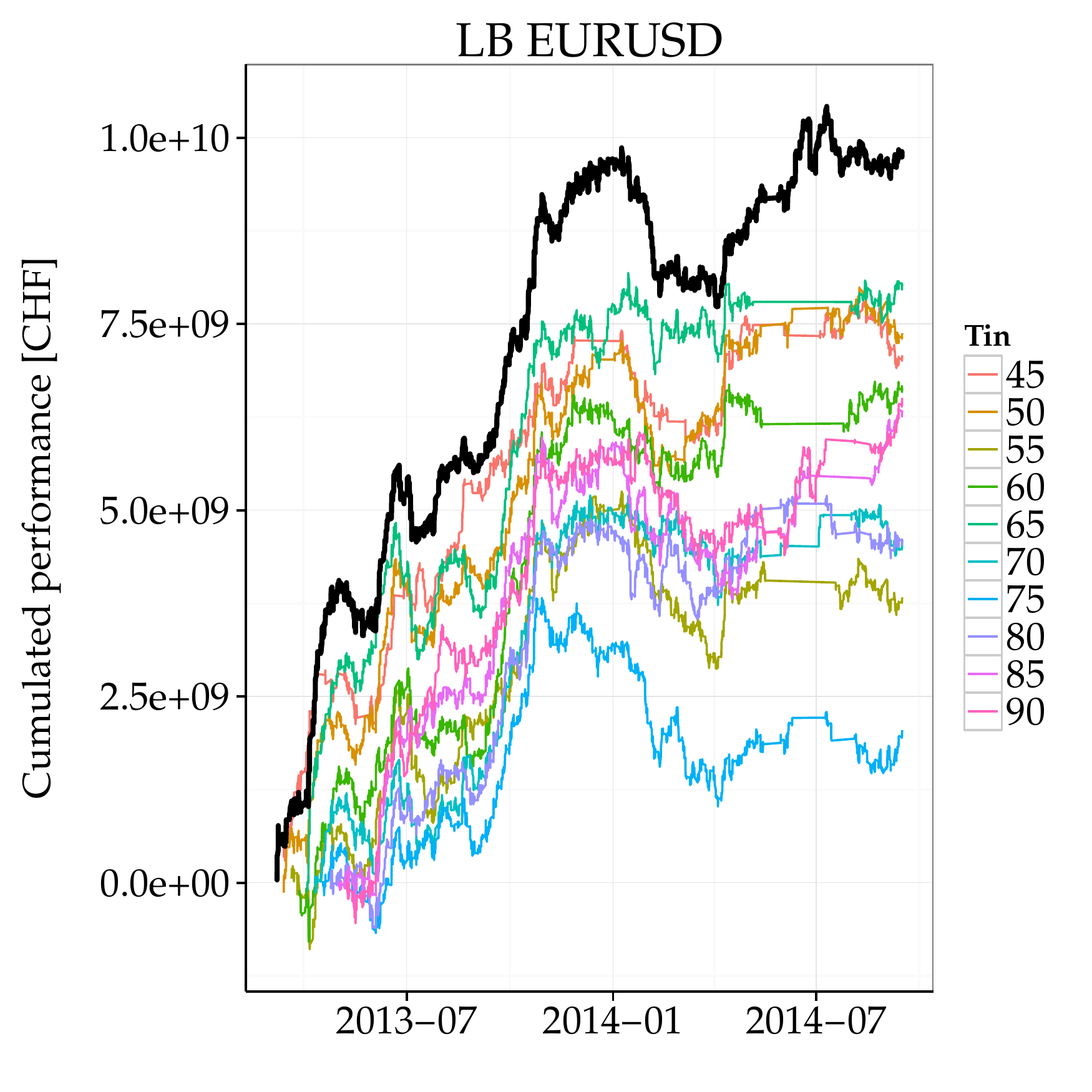}}\\
\par\end{centering}
\begin{centering}
\textcolor{black}{\includegraphics[width=0.4\textwidth]{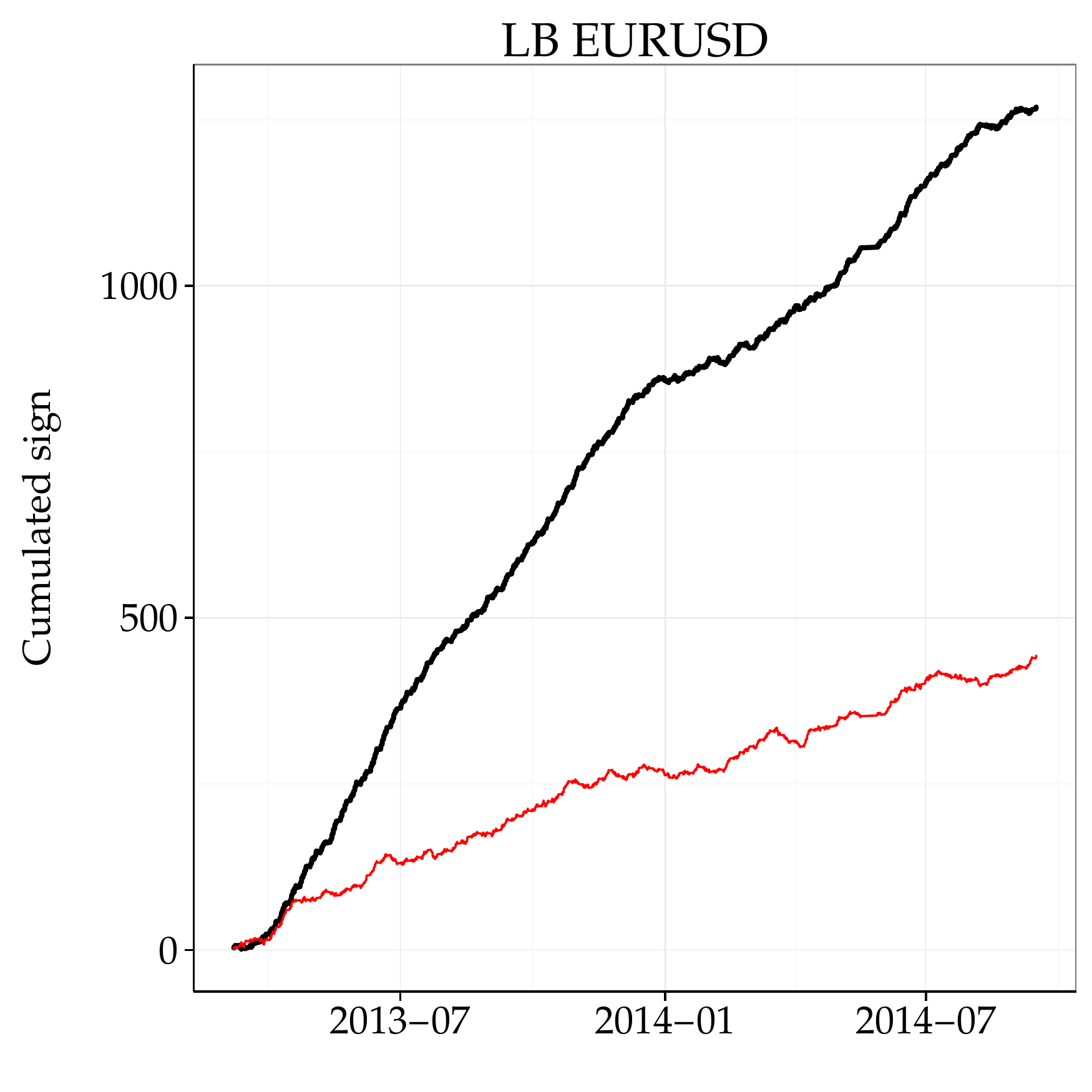}\includegraphics[width=0.4\textwidth]{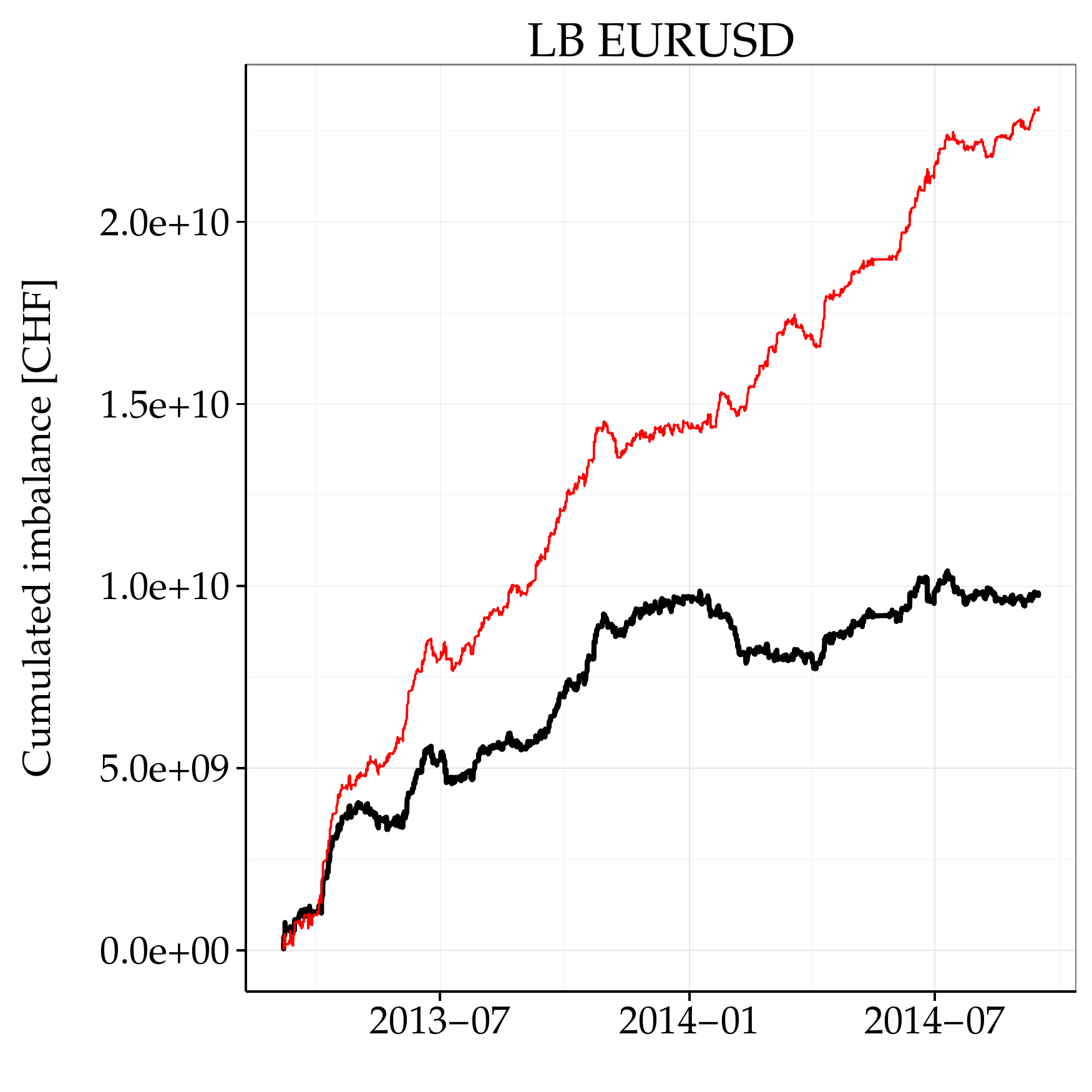}}
\par\end{centering}
\textcolor{black}{\caption{Out-of-sample performance: cumulated product of the predicted order
flow sign and actual sign (upper left), cumulated product of the predicted
order flow sign and actual order flow (upper right), cumulated product
of the predicted order flow sign and actual order flow, and the cumulated
actual flow (lower left), cumulated product of the predicted order
flow direction and actual order flow (black line) and cumulated actual
order flow (red line); \texttt{\textcolor{black}{EURUSD}}, LB. The
thick black lines correspond to a majority vote between all calibration
window lengths.\label{fig:LB_perf}}
}
\end{figure}

\textcolor{black}{We run the procedure on the three most traded pairs
of the SQ 2014-6 dataset and on }\texttt{\textcolor{black}{EURUSD}}\textcolor{black}{{}
LB dataset. Figures \ref{fig:SQ_EURUSD_perf} to \ref{fig:LB_perf}
report the out-of-sample performance of our method. Although we train
RFs on the signs of order flow, we also plot the cumulated performance
in currency units. In essence, if the prediction of the order flow
signs is successful during the most active periods, it also predicts
the actual order flow itself, on average. This the case of SQ clients,
but not for LB: the most precise sign predictions are during lunch
time, i.e. when the activity reaches a local minimum.}

\textcolor{black}{A possible explanation for this is the clear difference
between SQ and LB clients: whereas the cumulated net imbalance of
SQ clients is mostly mean-reverting, that of LB clients keeps on increasing.
This means that SQ clients are mostly speculators, while LB clients
use the FX market for other purposes than mere speculation, on average. }

\textcolor{black}{Let us now apply  statistical tests to the out-of-sample
performance in currency units: for both brokers and all pairs, the
out-of-sample performance is clearly significant. However, the real
question is the predictive power of our methodology regarding the
sign of the next order flow. Chou and Chu approach consists in testing
whether $A$ predicts $B$, where $A$ and $B$ are binomial variables,
taking into account auto-correlations, the null hypothesis being that
$A$ does not predict $B$ \cite{chou2010testing}. When applied
to SQ data, this test unambiguously shows that our method yields predictive
answers, and that it is not useful in the case of LB data. This may
mean that an hourly time scale may not be a wise choice for LB traders.}

\textcolor{black}{}
\begin{figure}
\centering{}\textcolor{black}{\centerline{\includegraphics[scale=0.3]{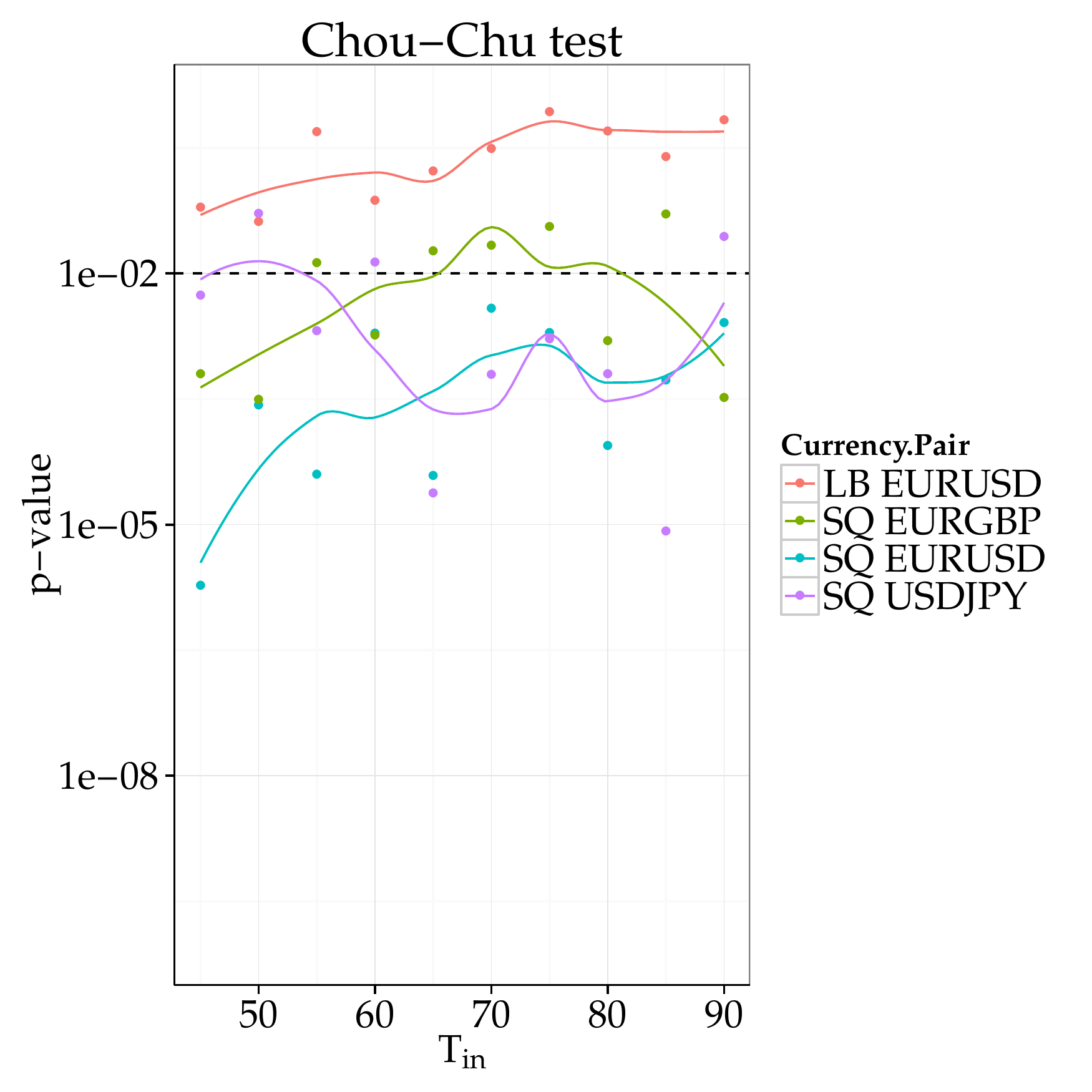}\includegraphics[scale=0.3]{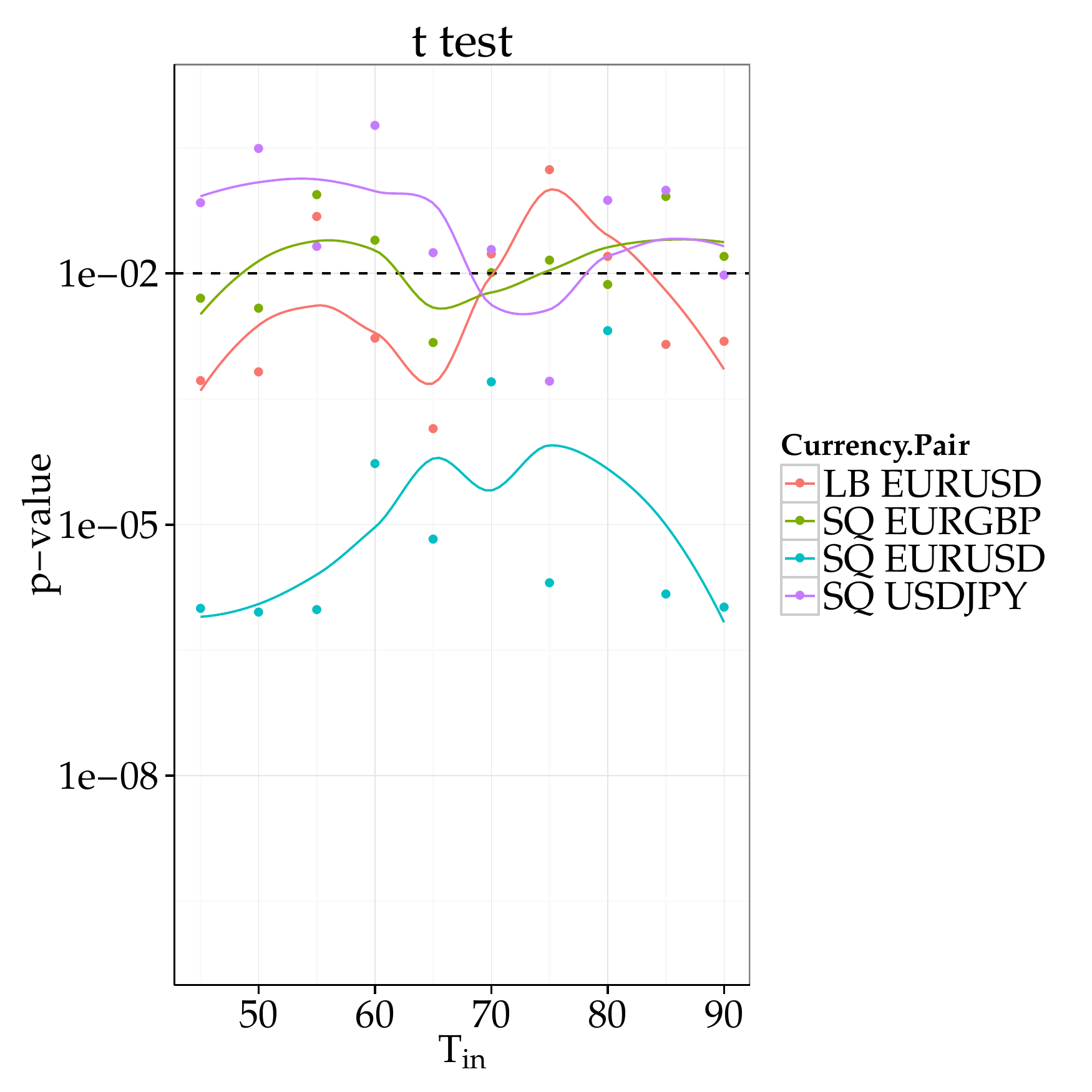}\includegraphics[scale=0.3]{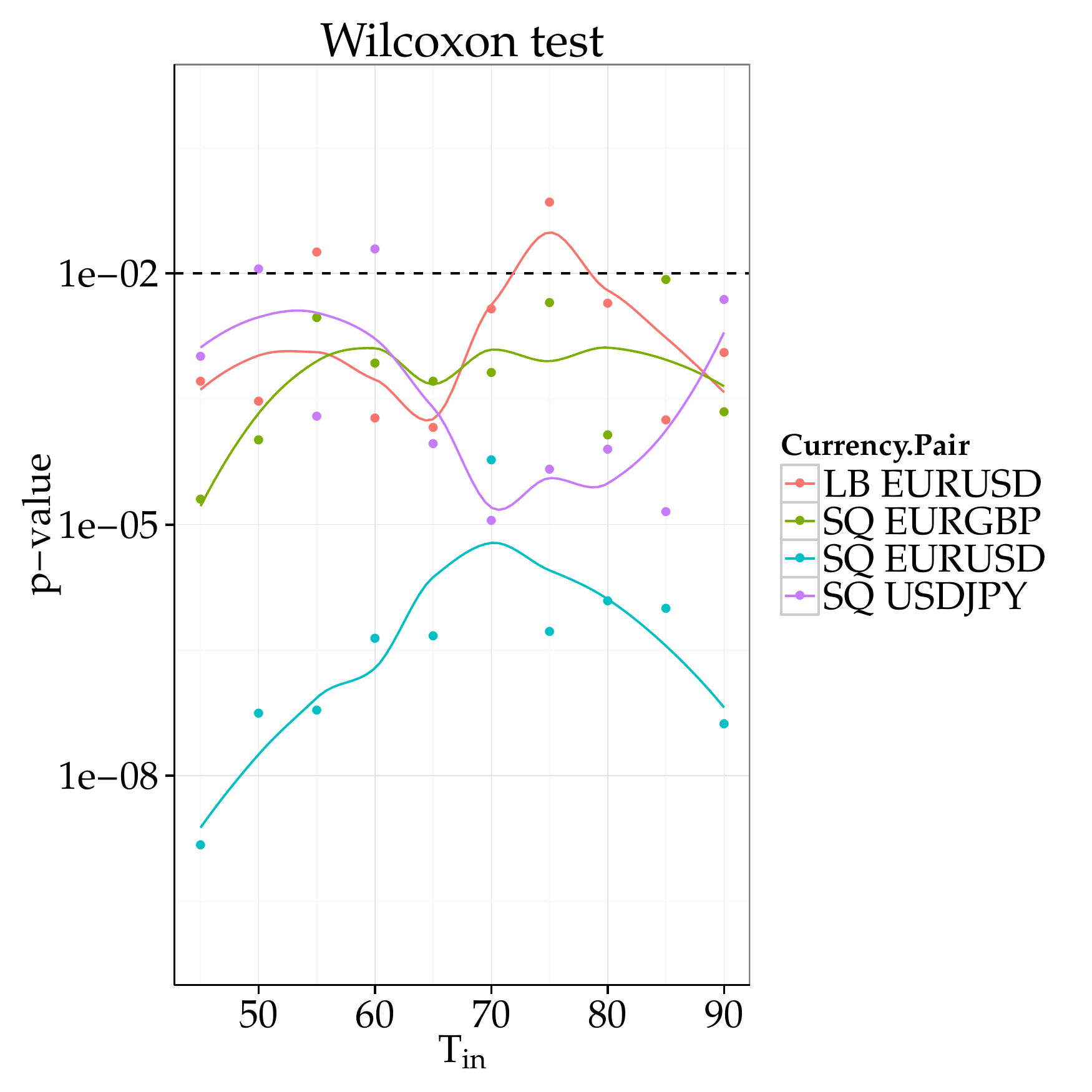}}\caption{Tests of the predictive power of the binary prediction (left plot)
and two tests of the out-of-sample flow prediction performance. Smoothed
curves and dashed lines at $y=0.01$ are for eye-guidance only. \label{fig:tstat-per-hour-2}}
}
\end{figure}

\textcolor{black}{}
\begin{figure}

\begin{centering}
\textcolor{black}{\includegraphics[scale=0.3]{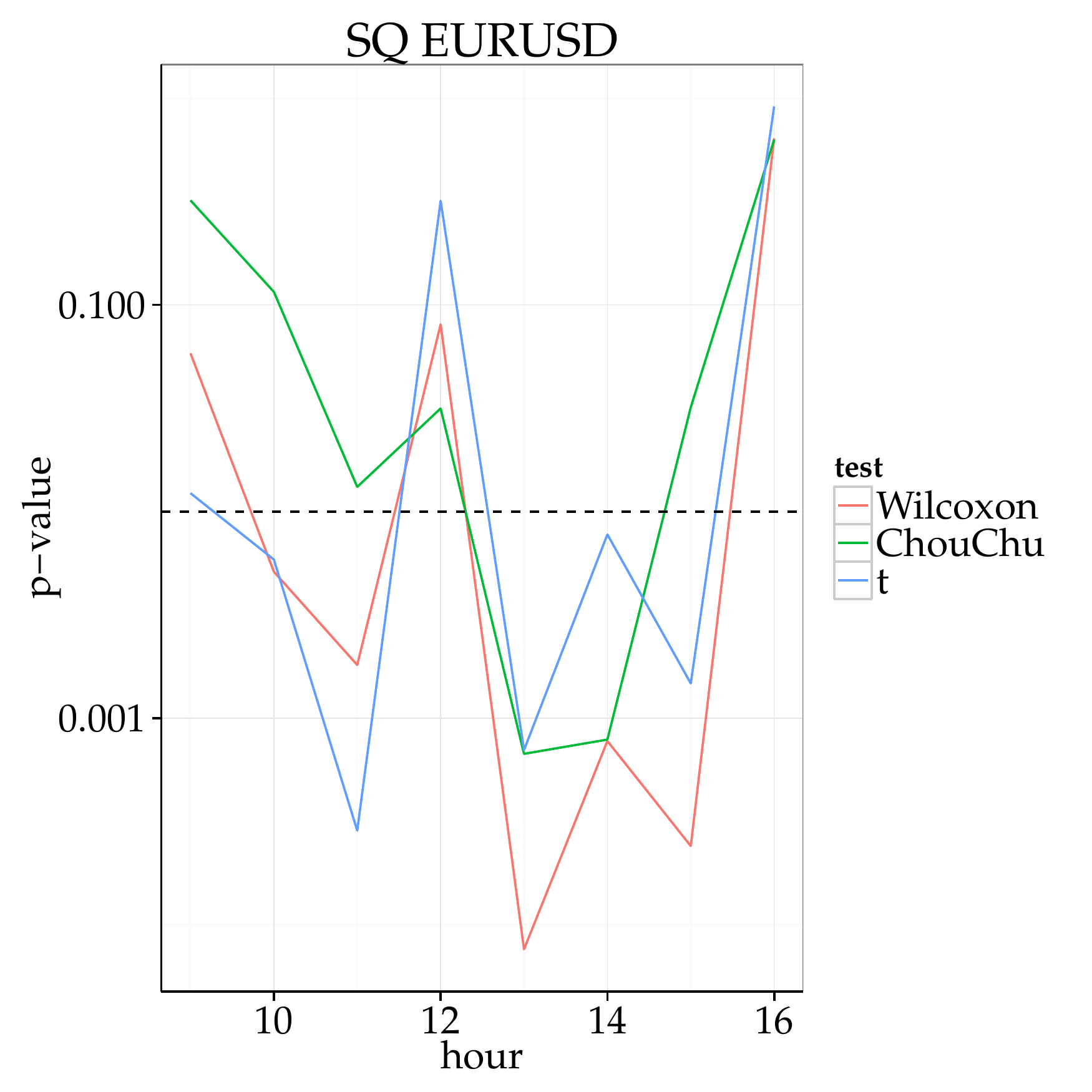}\includegraphics[scale=0.3]{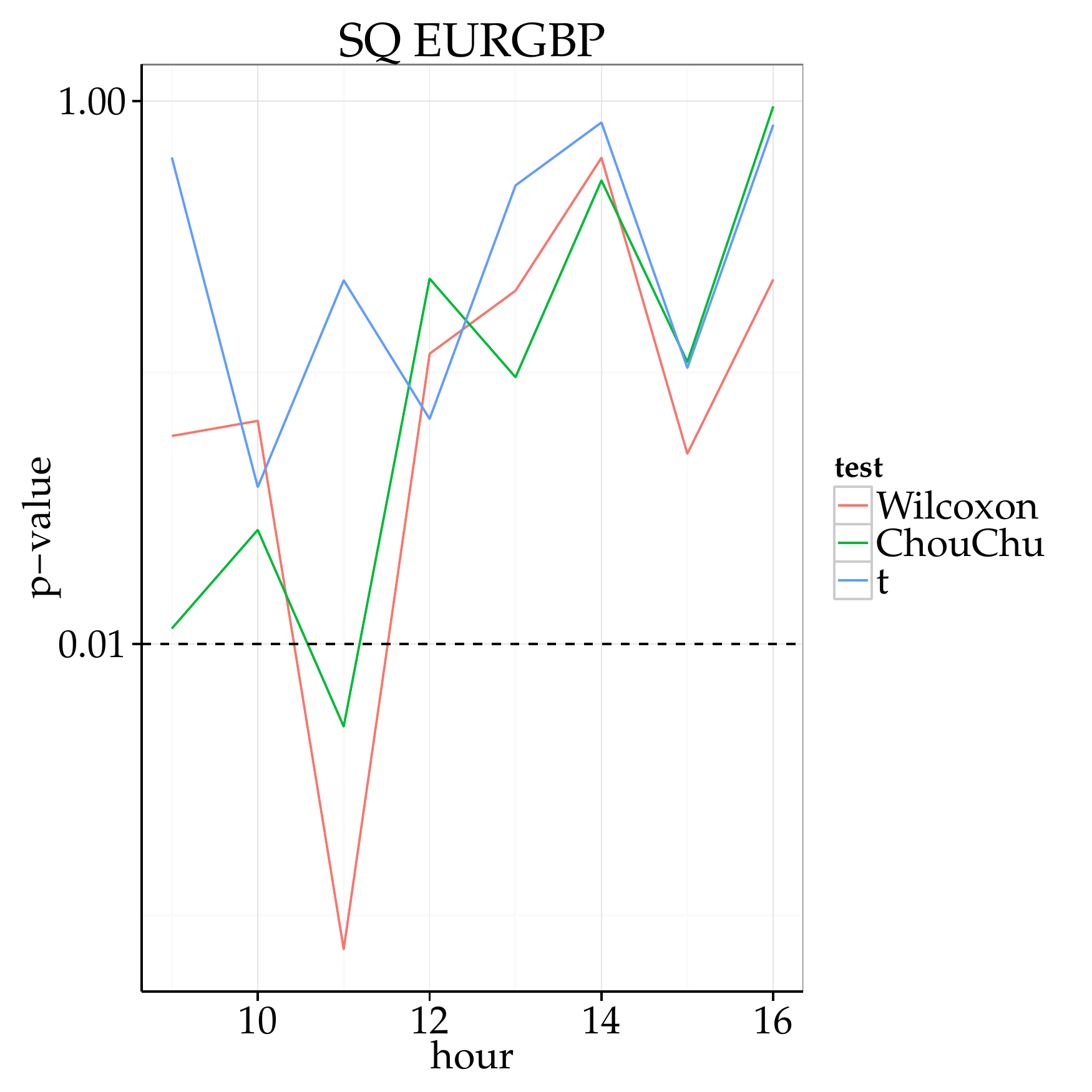}}\\
\textcolor{black}{\includegraphics[scale=0.3]{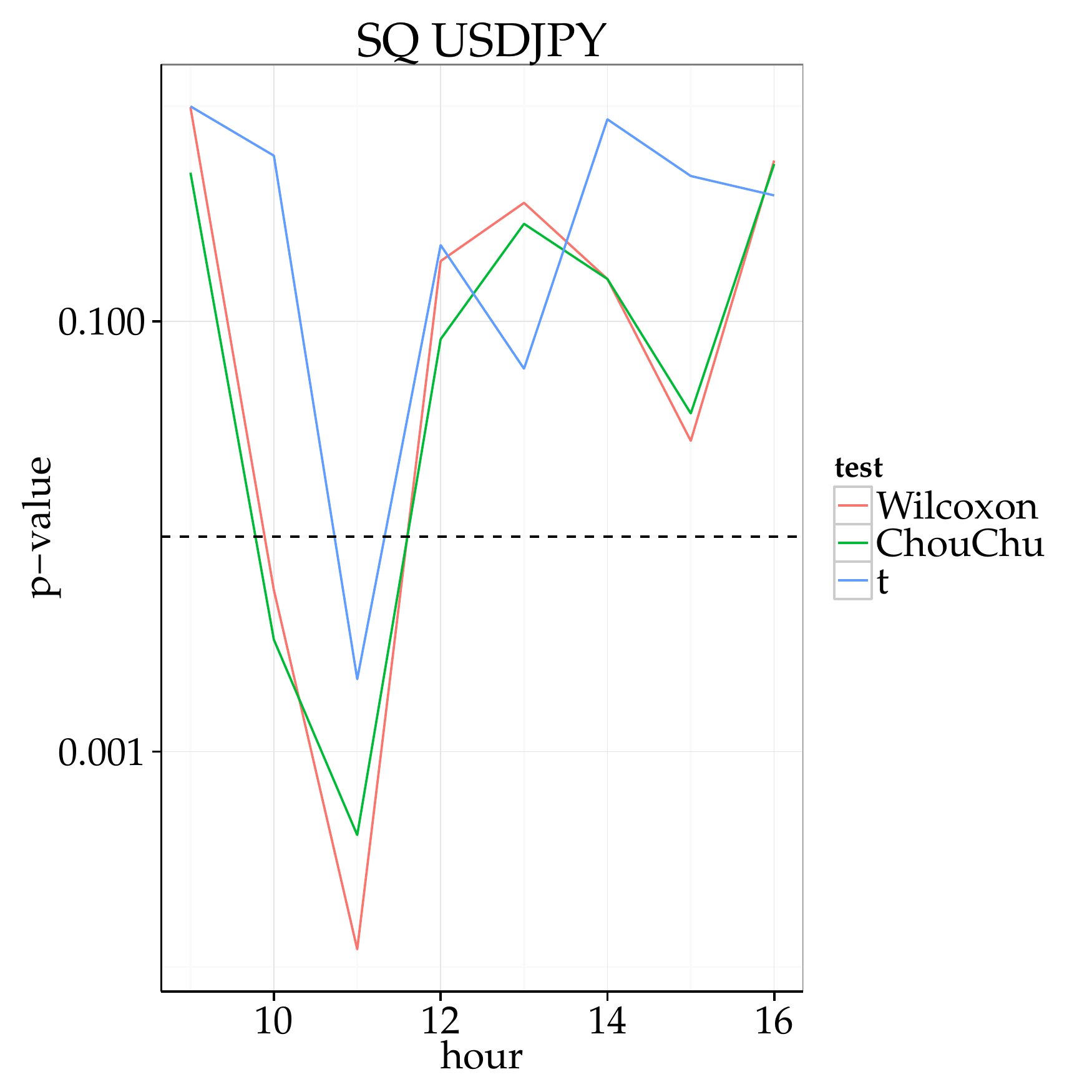}\includegraphics[scale=0.3]{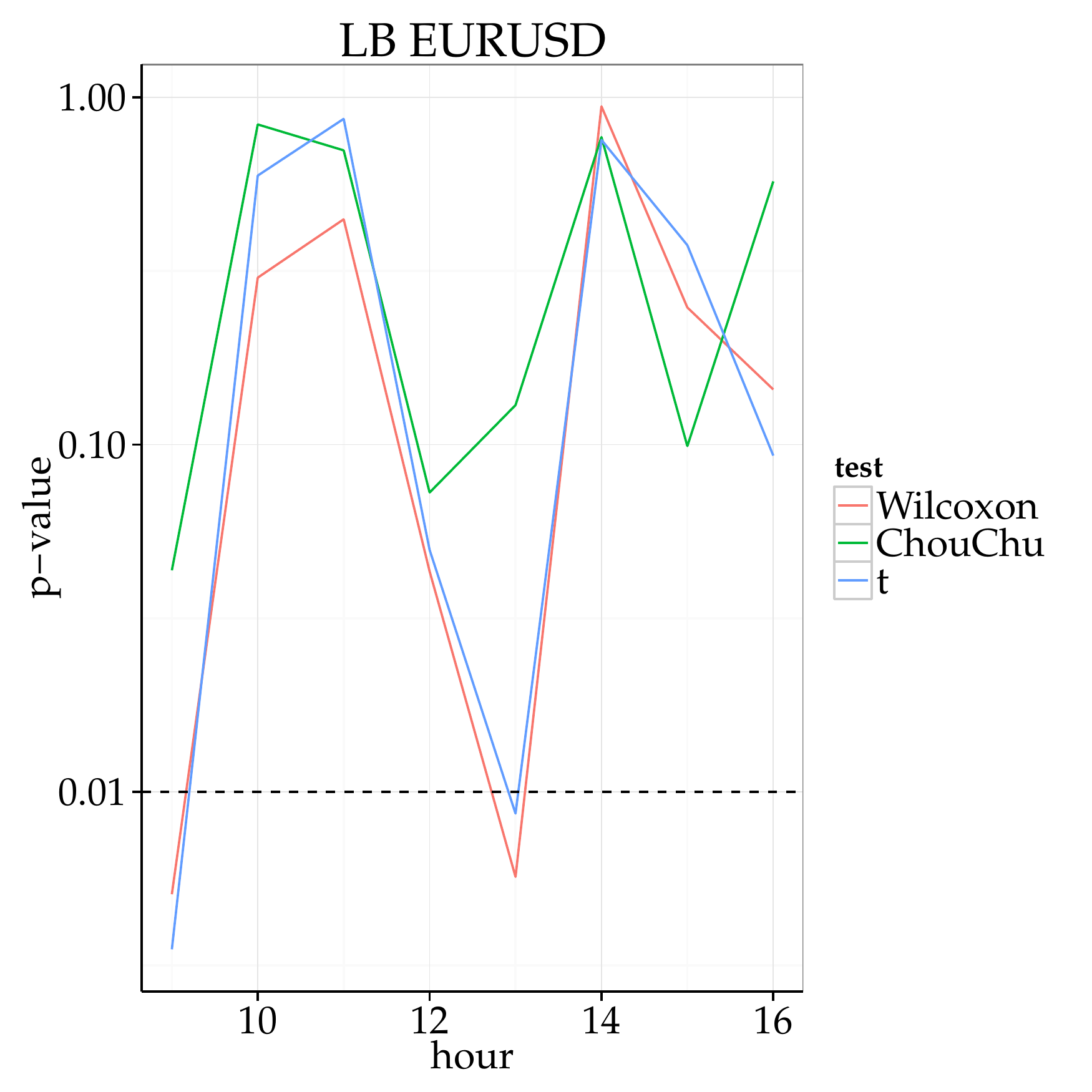}}
\par\end{centering}
\textcolor{black}{\caption{Test of the predictive power of the binary prediction (Chou-Chu) and
of the resulting cumulated predicted imbalance (Wilcoxon and t) as
a function of the time of the day.\label{fig:Test-of-the}}
}

\end{figure}

\textcolor{black}{A different way to look at these results is to analyse
the performance conditioned on the hour of the day. In Fig. \ref{fig:Test-of-the},
we report the p-values of out-of-sample performance conditional on
the hour of the day. We notice a tendency to perform well in periods
of high activity for SQ clients, not for LB clients.}

\textcolor{black}{Adding the returns to the feature set did not improve
the forecast, probably because the returns effect are already embedded
in groups actions, in line with the factor analysis of \cite{Lillo2008}.
Remarkably, multiple runs of the procedure revealed that there is
no need to update the model every day: an update every $5$ days approximately
gave the same results, which is consistent with the large persistence,
on average, of lead-lag networks. }

\subsubsection{\textcolor{black}{Importance of predictors}}

\textcolor{black}{In order to check that the order flow imbalance
depends on the group states we check }\textcolor{black}{\emph{a contrario
}}\textcolor{black}{that the hour of the day is not the most important
predictor most of the time. That the non-linear relationship between
the predictors and future order imbalance signs sometimes depends
on the hour of the day is not surprising, as indeed some groups may
be more active at certain times of the day. However, given the existence
of lead-lag networks, we do not expect the hour of the day to top
the list of variable importance very often. }

\textcolor{black}{There are several ways to measure variable importance
in Random Forests; we have used the standard Breiman-Cutler measure,
which (roughly speaking) consists in shuffling the elements of a given
column of the predictors matrix and determining how the in-sample
prediction error changes (see \cite{Breiman2001} for more details).
We shall focus on the rank of the hour of the day relative to the
one of all the other columns. Since we use an unsupervised method
to cluster the traders, the number of groups varies as a function
of time. As a consequence, we define the rank ratio of column $h$
as the ratio between the rank of the importance of $h$, denoted by
$\mbox{rank}(h)$, 1 being the most important, and the number of columns
of $P$, denoted by $K_{t_{1}}$. Because a rank of 1 would correspond
to $1/K_{t_{1}}$, a time varying quantity, and might therefore make
it unnecessarily difficult to compare two rank ratios, we define the
adjusted rank ratio of $h$ as
\[
r_{h}=\frac{\mbox{rank}(h)-1}{K_{t_{1}}-1}.
\]
}

\begin{figure}
\begin{centering}
\textcolor{black}{\includegraphics[width=0.35\textwidth]{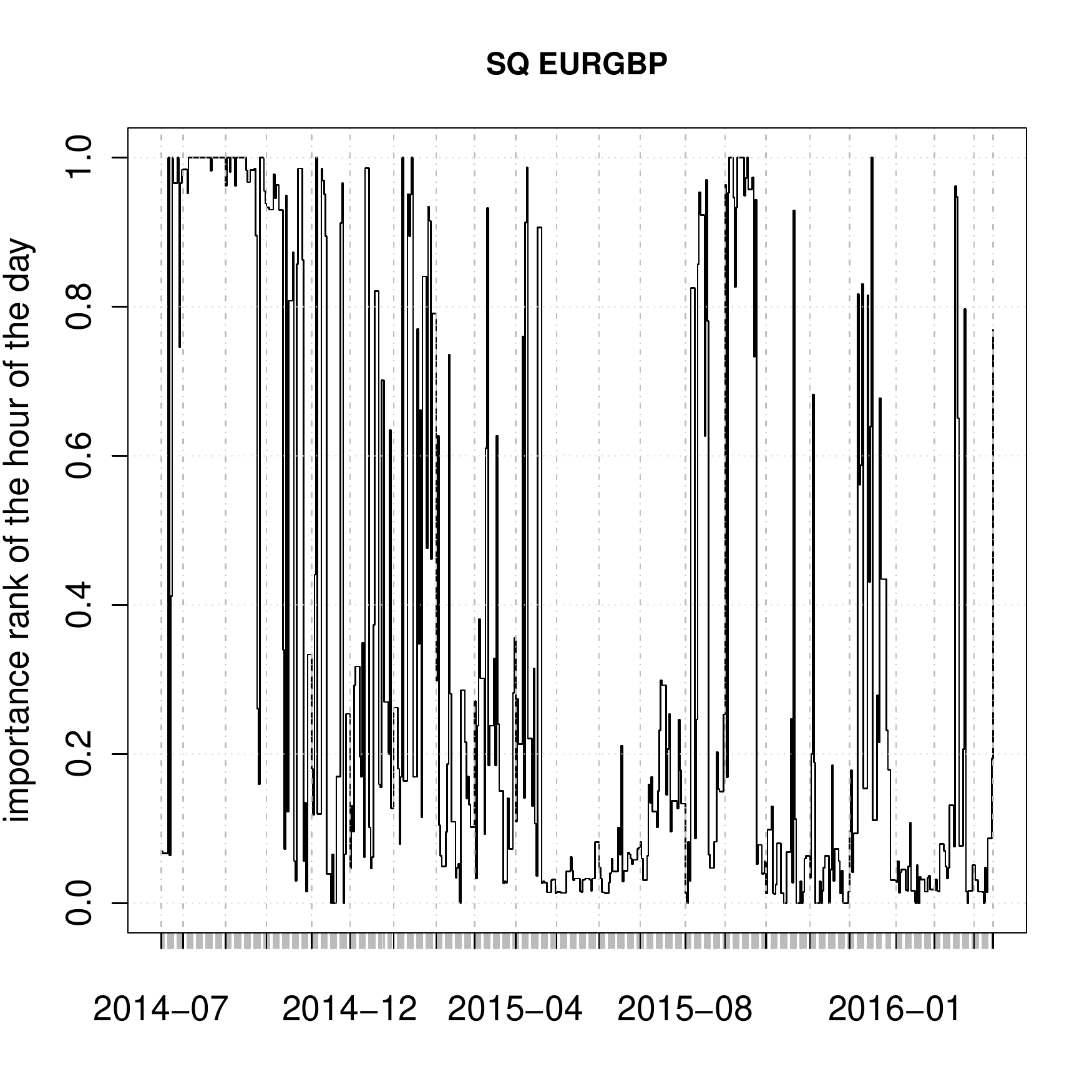}\includegraphics[width=0.38\textwidth]{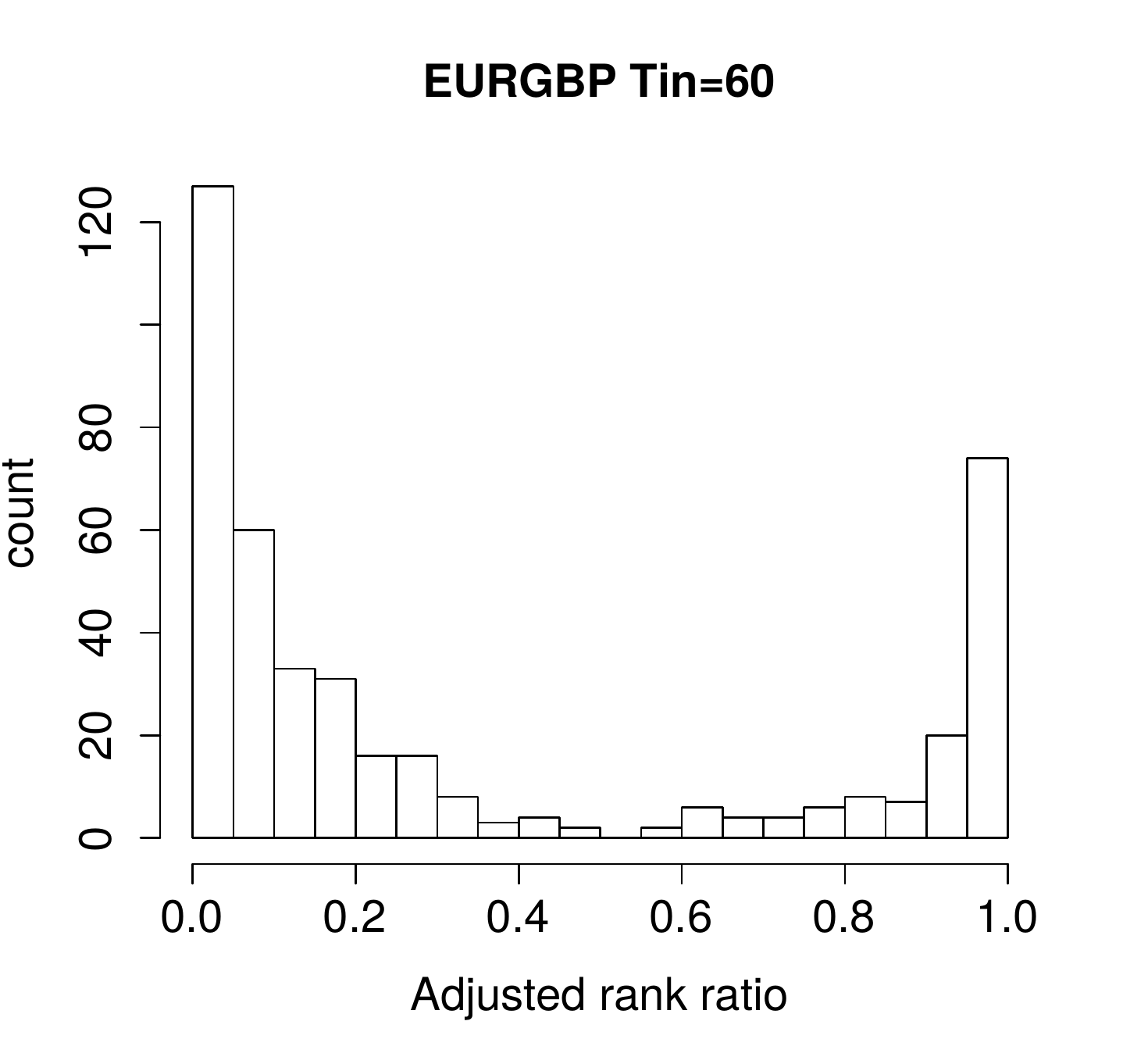}\includegraphics[width=0.35\textwidth]{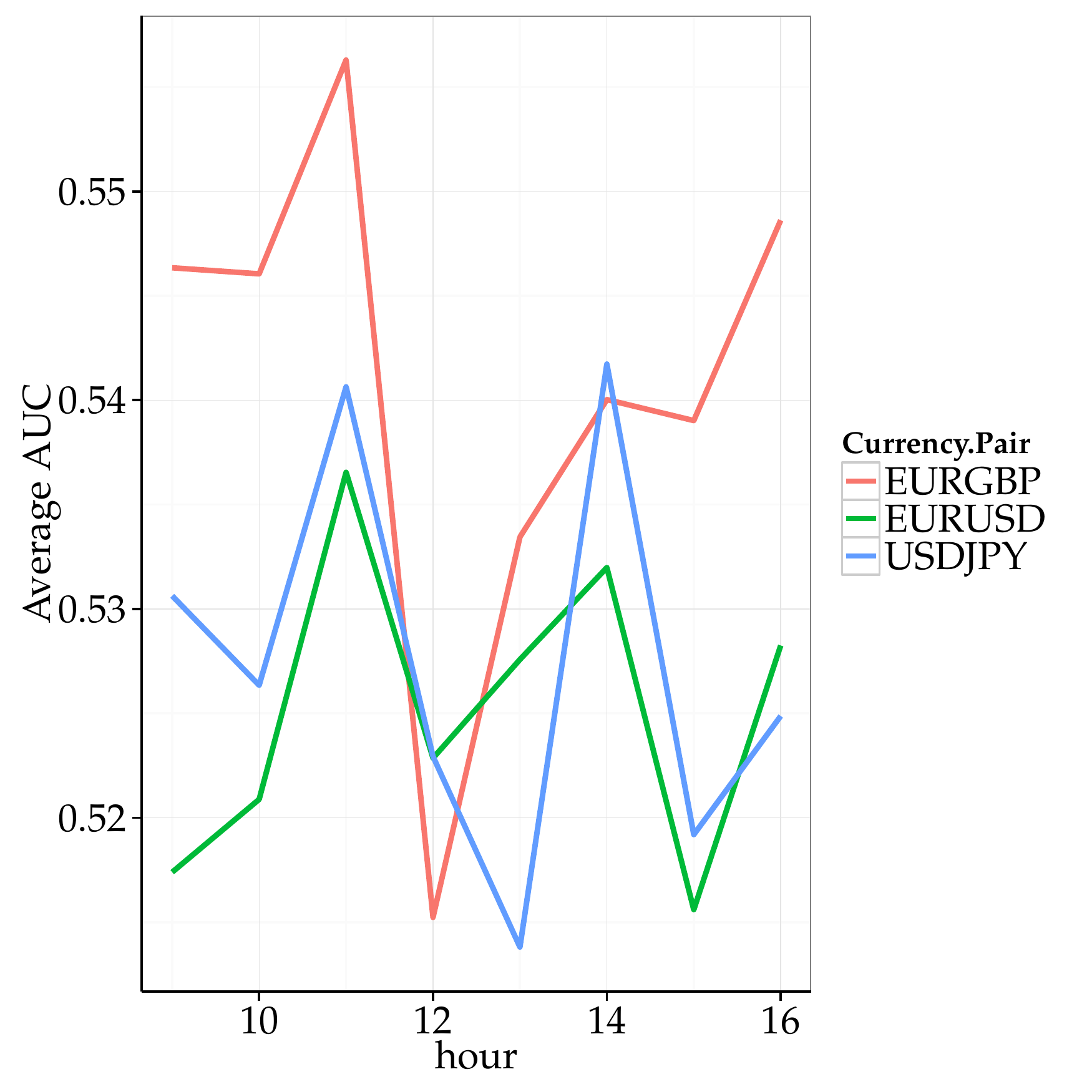}}
\par\end{centering}
\textcolor{black}{\caption{Adjusted rank ratio of importance of the hour of the days, $r_{h}$
, for \texttt{\textcolor{black}{EURGBP}} as a function of time (left
plot) and its histogram (middle plot). The right plot reports the
dependence on the Area Under Curve (AUC) of $r_{h}$, averaged over
all $T_{in}$, as a function the hour of the day. Dataset: SQ2014-6
\label{fig:avgAUC_vs_hour_fracRank}}
}
\end{figure}

\textcolor{black}{With this convention, the hour column is the most
important predictor if $r_{h}=0$ and the least important one if $r_{h}=1$.
The left plot of Fig~\ref{fig:avgAUC_vs_hour_fracRank} plots $r_{h}$
as a function of time for }\texttt{\textcolor{black}{EURGBP}}\textcolor{black}{.
It turns out that the distribution of the relative importance of $h$
is rather bimodal (see the histogram in the middle plot of the same
figure) and persistent. The bimodal nature of $r_{h}$ is less pronounced
for other $T_{in}$ and currency pairs. At all rates, one may wonder
if prediction is more successful in periods of low $r_{h}$ or high
$r_{h}$, or equivalent, how predictive of success $r_{h}$ is. As
we deal with a classification problem, the tool of choice is Receiver
Operating Characteristic (ROC) curves and their associated Area Under
Curve (AUC), which, in a nutshell, quantifies how different the distribution
of $r_{h}$ is when the prediction is correct and when it fails. By
definition, an AUC of 0.5 corresponds to the absence of predictive
power of $r_{h},$ while an AUC of 1 implies that $r_{h}$ fully determines
the success of predictions. The right plot of the same figure makes
it clear that there is some predictability associated to $r_{h}$,
and that it is once again larger during the most active hours. Since
there is only one value of $r_{h}$ for each day, and since one makes
one prediction per hour, we have computed average AUC conditional
on the hour of the day.}

\textcolor{black}{}
\textcolor{black}{ }

\begin{figure}

\centering{}\textcolor{black}{\includegraphics[width=0.5\textwidth]{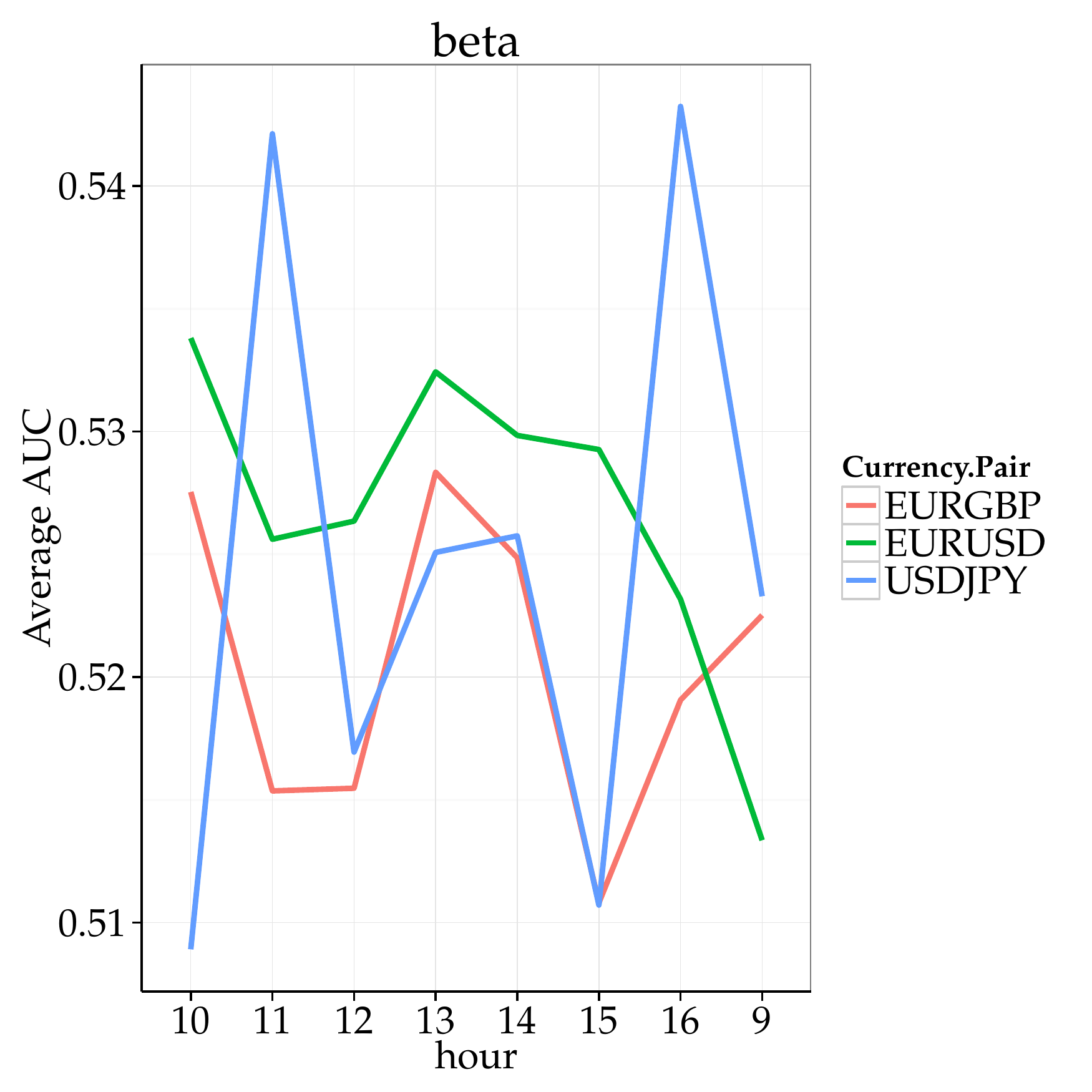}\caption{Area Under Curve (AUC) vs the hour of the day for the trader-trader
lead-lag persistence measure $\beta$; SQ 2014-6 dataset.\label{fig:Area-Under-Curve}}
}
\end{figure}

\textcolor{black}{Finally, we compute the hourly average AUC of the
trader-trader lead-lag persistence $\beta.$ Figure \ref{fig:Area-Under-Curve}
reports the AUC for the three SQ currency pairs as a function of the
number of weeks of the calibration window, restricted to the same
out-of-sample period. Once again, the predictive power of such measure
is weak.}

\textcolor{black}{\label{sec:AUC_beta}}

\textcolor{black}{}

\subsection{VWAP prediction}\label{sec:VWAP}

Managing one's imbalance requires more than predicting the sign of
the order flow. Indeed,  succeeding in predicting the average direction
of the trades during the next period is also necessary. Instead of
predicting separately the sign of the next price return, we focus
on the VWAP of the trades of the broker's clients during each time
slice as it combines both volume and price. If a broker can predict
the evolution of his VWAP in the next time slice, then it will be
able to manage much more efficiently its inventory.

\begin{figure}
\begin{centering}
\textcolor{black}{\includegraphics[width=0.4\textwidth]{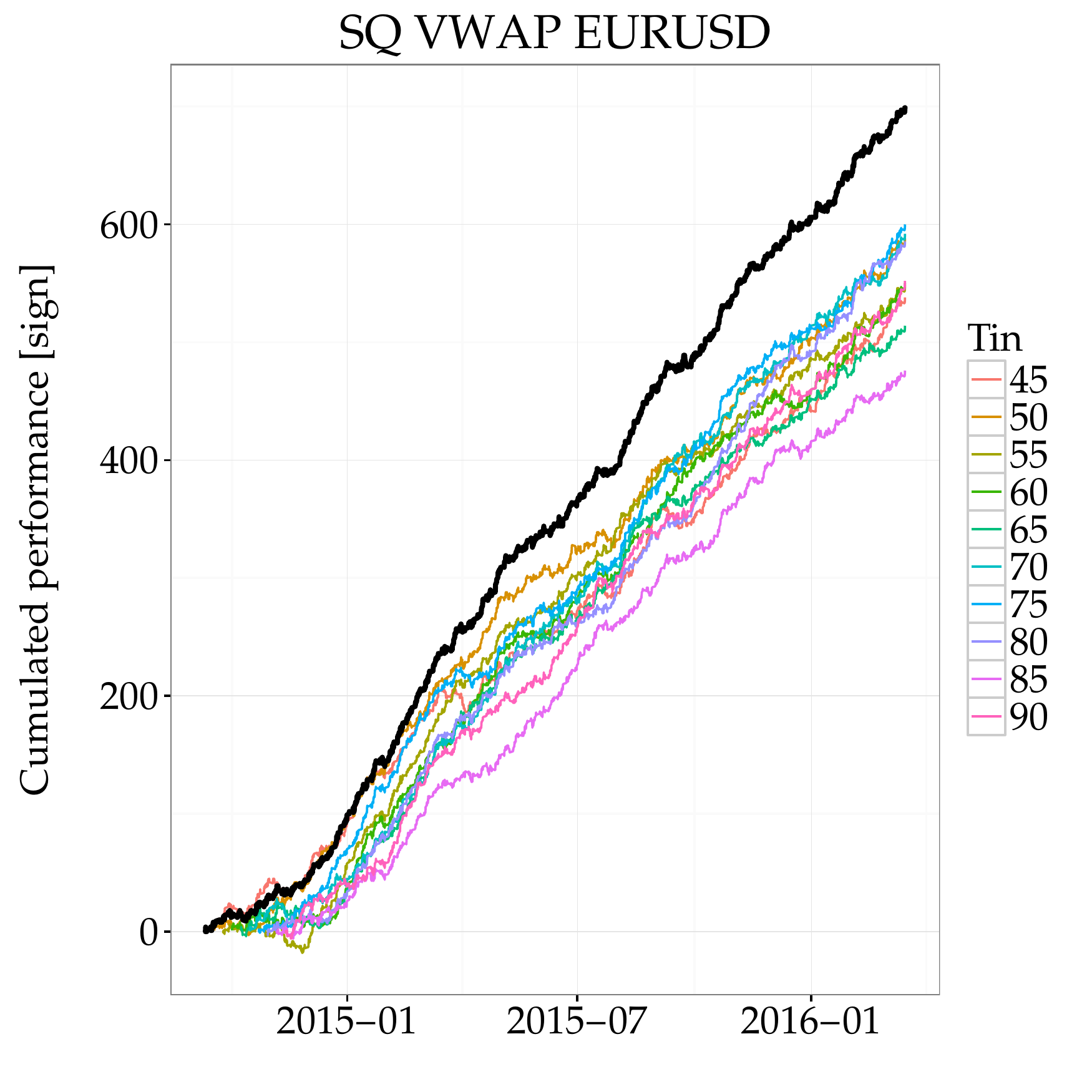}\includegraphics[width=0.4\textwidth]{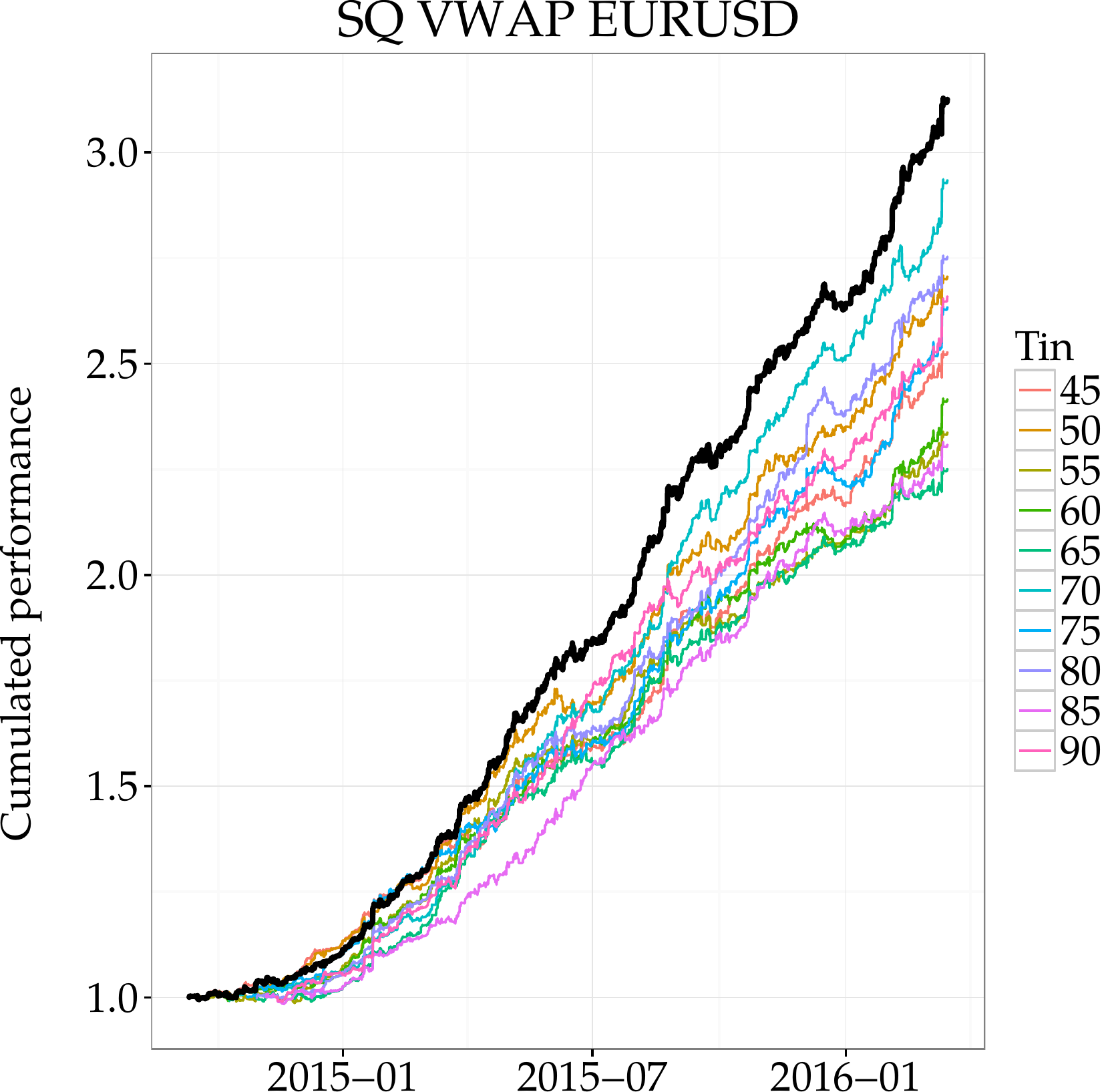}}\\
\par\end{centering}
\textcolor{black}{\caption{Out-of-sample performance: cumulated product of the predicted VWAP
sign change and actual sign (upper left), cumulated product of the
predicted VWAP sign change and actual sign change (upper right); SQ
\texttt{\textcolor{black}{EURUSD}}. The thick black lines correspond
to a majority vote between all calibration window lengths.\label{fig:SQ_EURUSD_perf-VWAP}}
}
\end{figure}

\begin{figure}

\textcolor{black}{\includegraphics[width=0.5\textwidth]{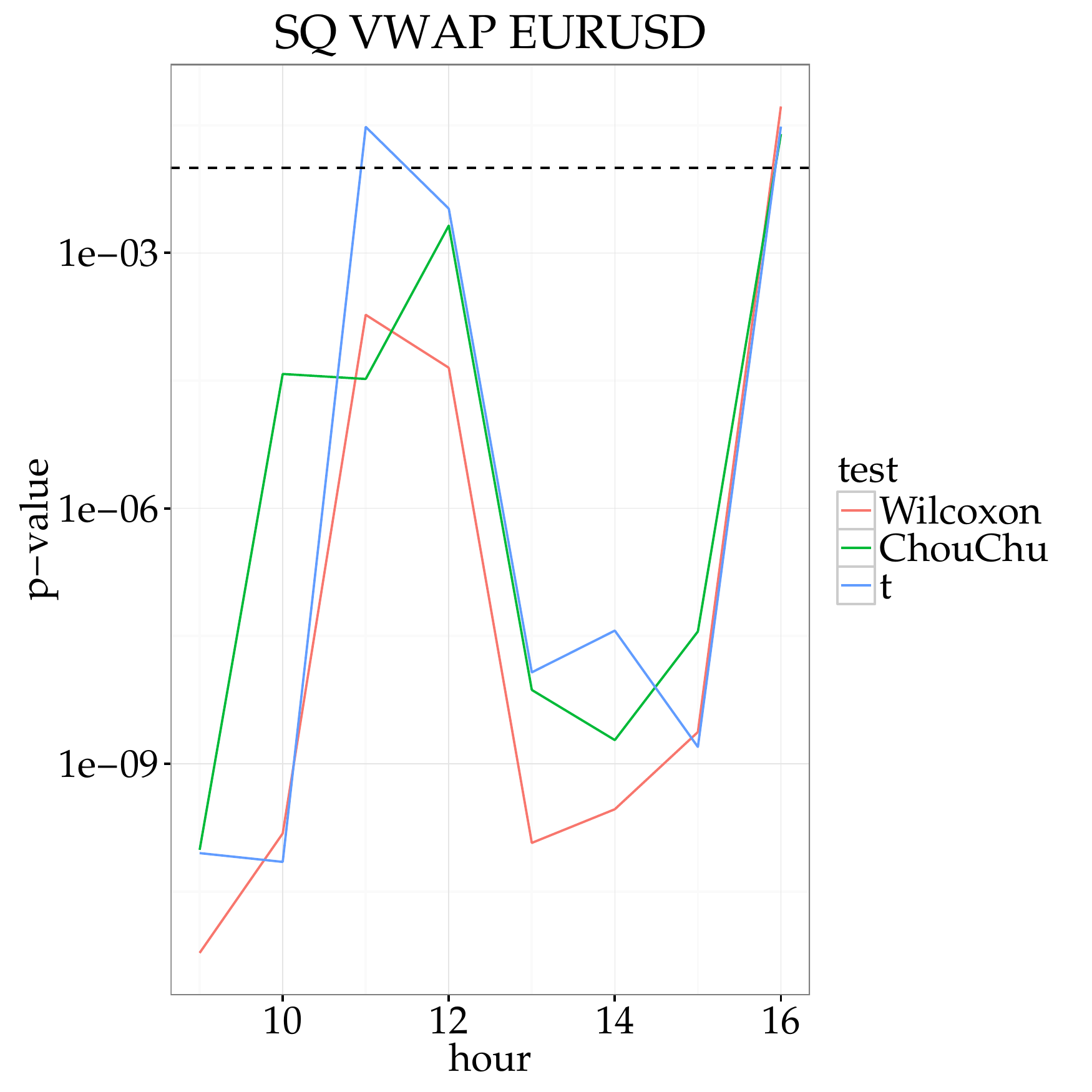}\includegraphics[width=0.5\textwidth]{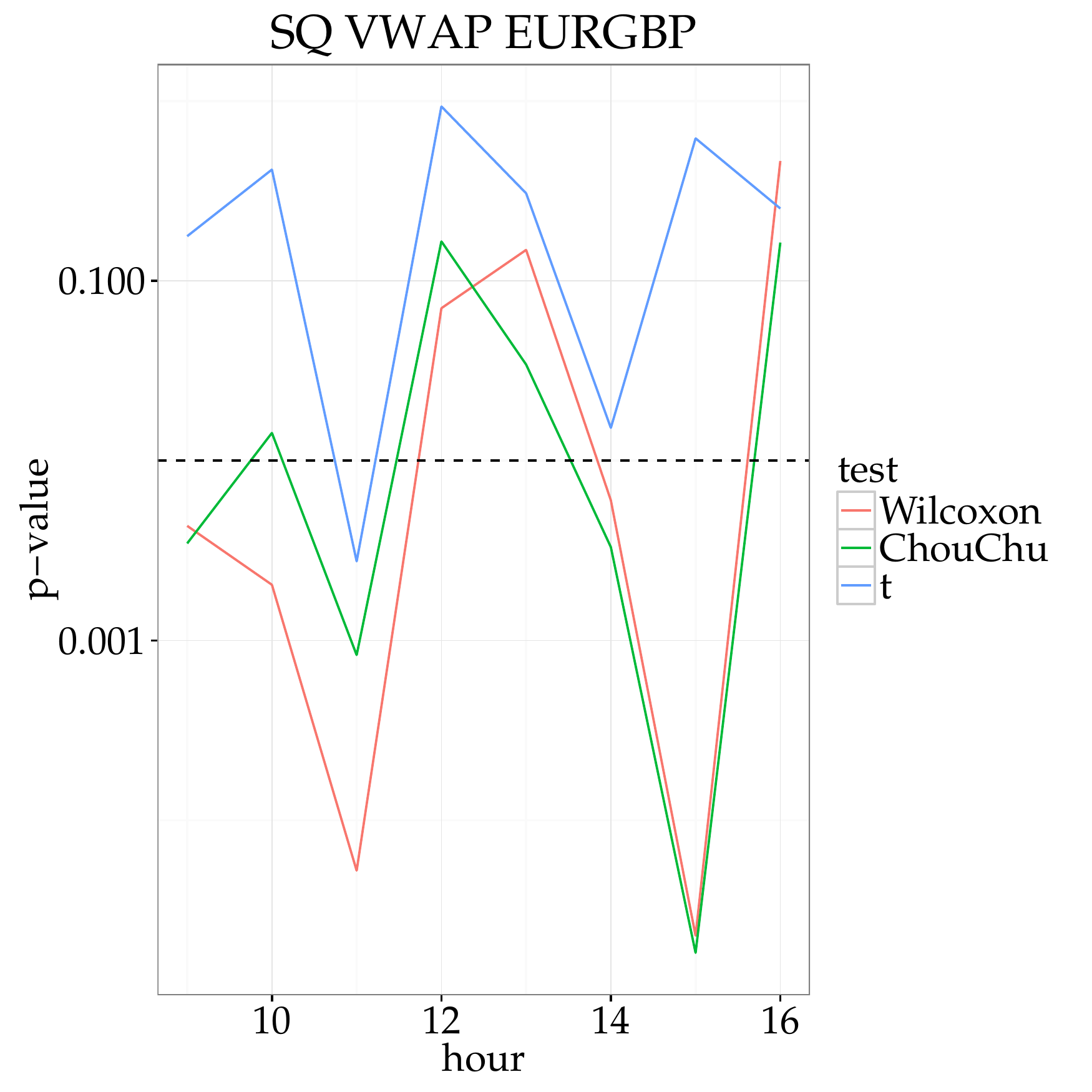}}

\textcolor{black}{\includegraphics[width=0.5\textwidth]{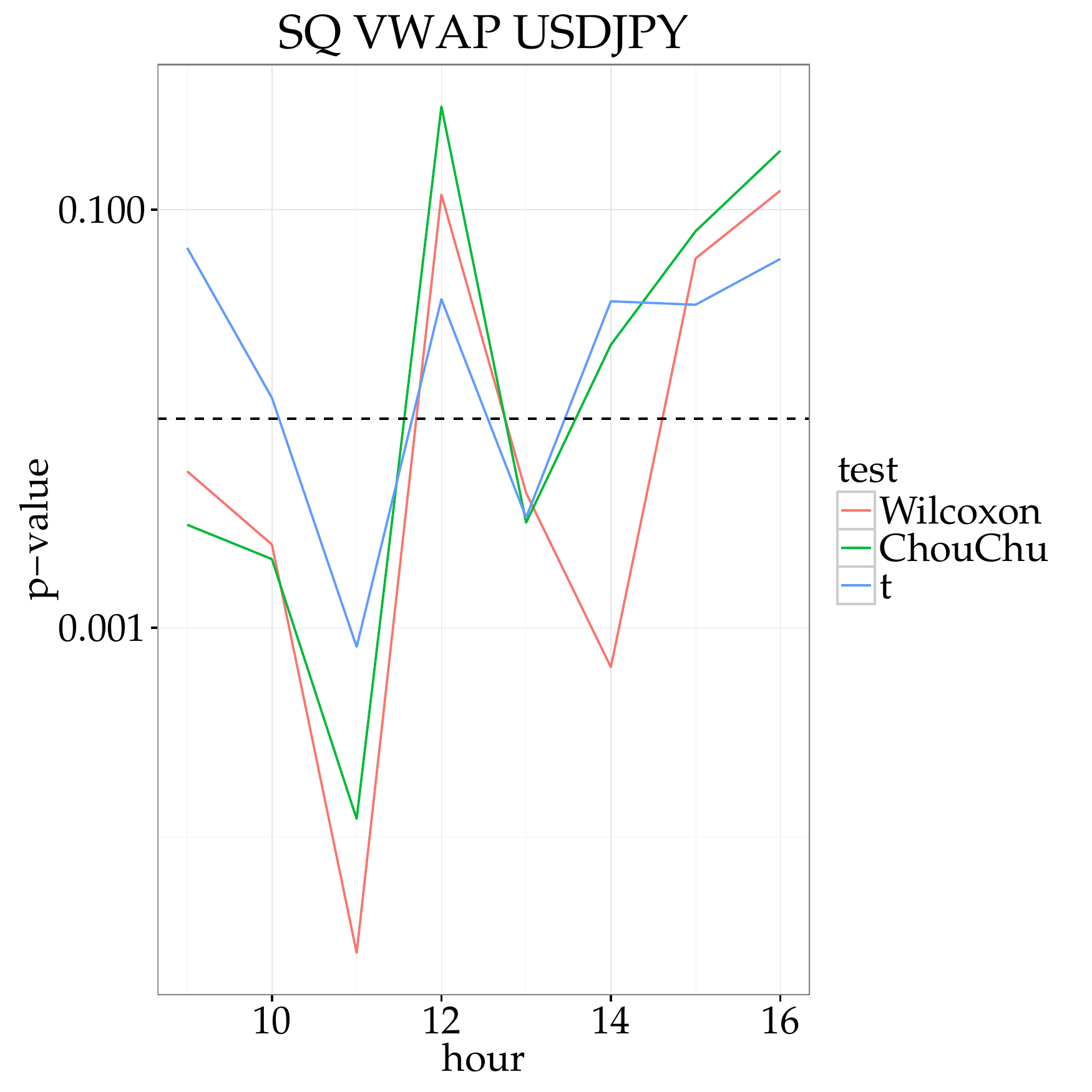}}\caption{VWAP change prediction: test of the predictive power of the binary
prediction (Chou-Chu) and of the resulting cumulated predicted performance
(Wilcoxon and t) as a function of the time of the day; SQ.\label{fig:Test-of-the-VWAP}}

\end{figure}

The setup is very similar as that for order flow prediction, except
that the vector $v_{t_{1}+1}$ to predict now contains the signs of
the changes of the broker's VWAP. In other words, we still face a
classification problem, but for the VWAP signed difference this time.
We keep the same parameters as before. Figure \ref{fig:SQ_EURUSD_perf-VWAP}
displays the results for \texttt{\textcolor{black}{EURUSD}} and SQ:
predictability is clearly significant and even better than for the
sign of the order flow. Figure \ref{fig:Test-of-the-VWAP} shows that
the predictability of the sign of VWAP change is particularly significant
at the most active hours for all currency pairs for SQ. LB customers
on the hand show once again exactly the opposite behavior: prediction
ability is the worst during the most active hours: VWAP change is
not predictable if the order flow signs are not.

\textcolor{black}{}

\textcolor{black}{}
\begin{figure}
\begin{centering}
\textcolor{black}{\includegraphics[width=0.4\textwidth]{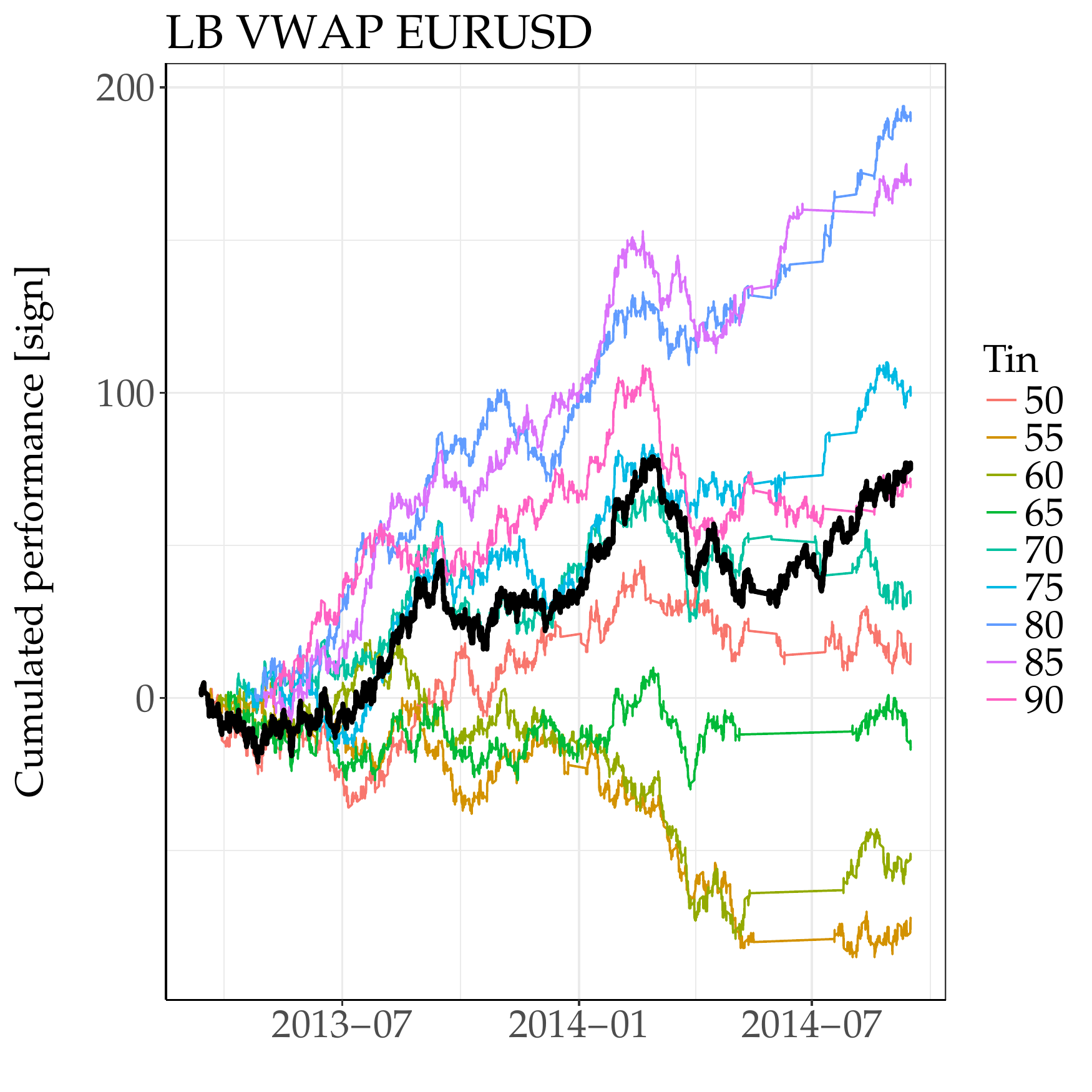}\includegraphics[width=0.4\textwidth]{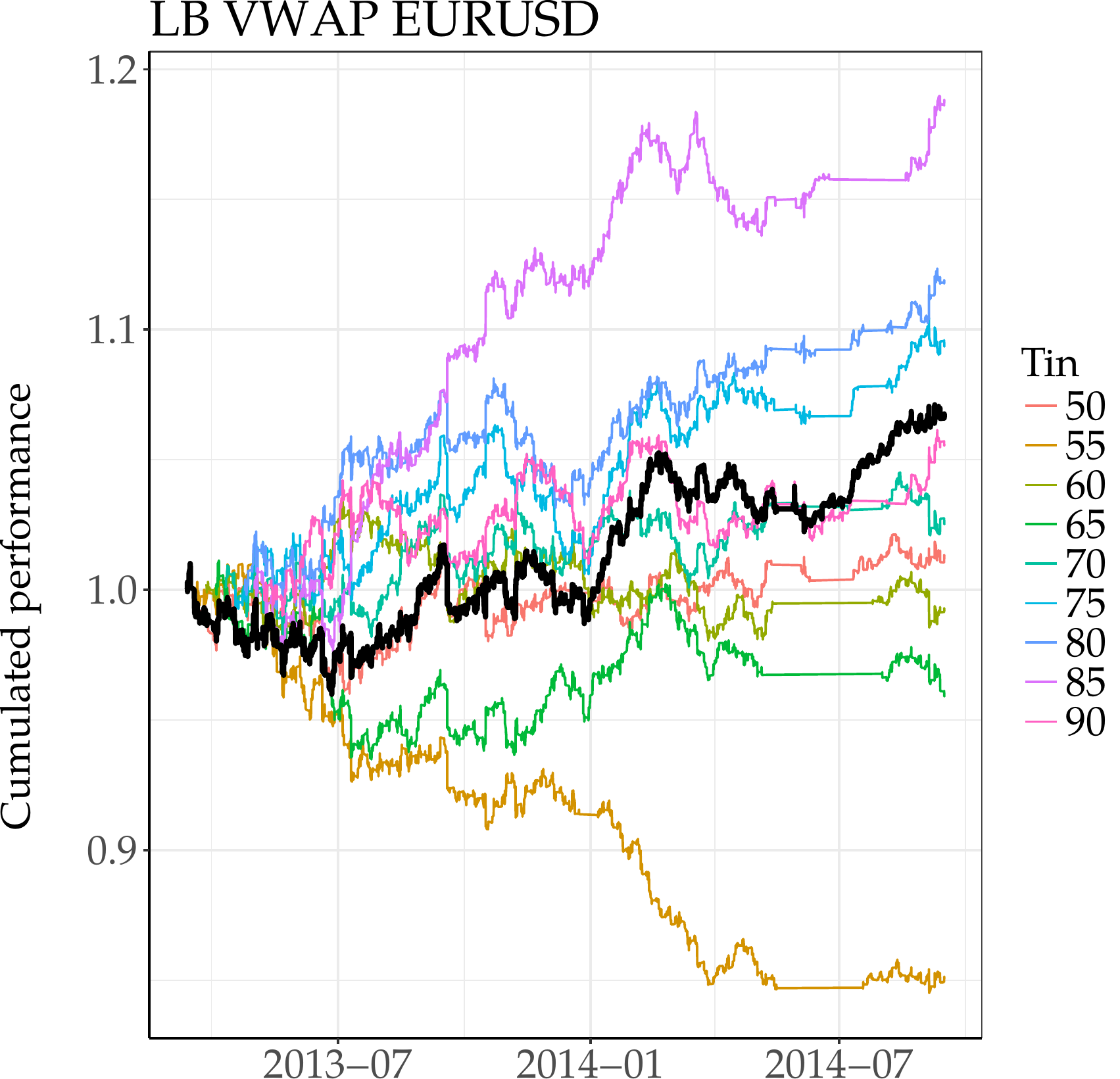}}\\
\par\end{centering}
\textcolor{black}{\caption{Out-of-sample performance: cumulated product of the predicted VWAP
sign change and actual sign (upper left), cumulated product of the
predicted VWAP sign change and actual sign change (upper right); LB
\texttt{\textcolor{black}{EURUSD}}. The thick black lines correspond
to a majority vote between all calibration window lengths.\label{fig:LB_EURUSD_perf-VWAP}}
}
\end{figure}

\begin{figure}

\centerline{\includegraphics[width=0.5\textwidth]{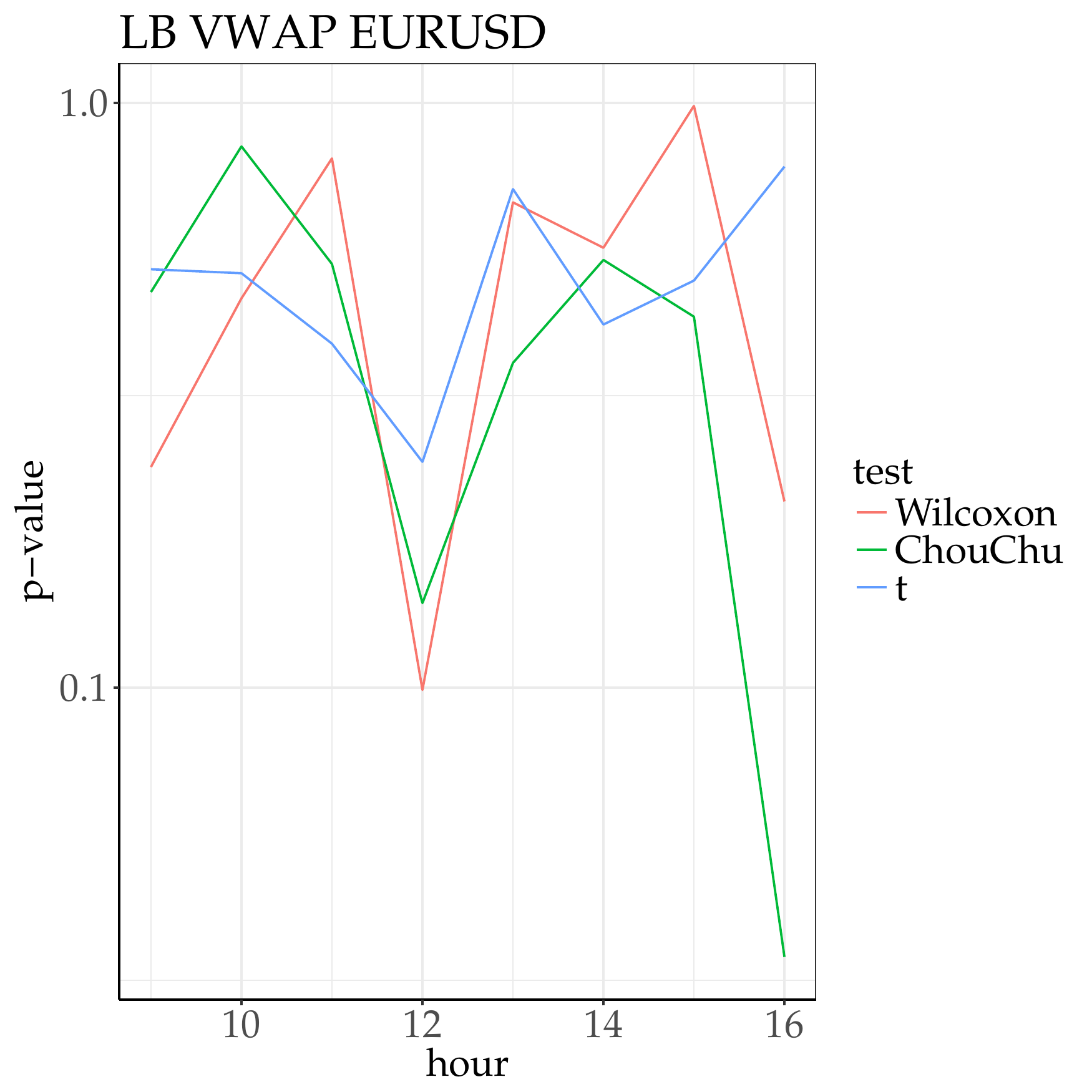}}

\caption{VWAP change prediction: test of the predictive power of the binary
prediction (Chou-Chu) and of the resulting cumulated predicted performance
(Wilcoxon and t) as a function of the time of the day; LB, EURUSD.\label{fig:Test-of-the-VWAP-LB}}
\end{figure}

\section{Conclusion and perspectives}

\textcolor{black}{Our aim was first to introduce an unsupervised lead-lag
network inference method and show the existence of persistent agent
lead-lag networks in financial data. Quite remarkably, these hidden
causal networks open the way to useful predictions at a one-hour
time horizon in one of the most unpredictable complex systems. Admittedly,
we used more detailed (and private) information than most market participants can obtain:
the point is that this kind of information makes the origin of predictability
explicit. Reversely, one then understands why predictability is not
significant when lead-lag networks are too sparse, either because
the trader activity similarity is too small or because the chosen
timescale does not correspond to the natural activity rate of the
agents. }

\textcolor{black}{Being able to predict the evolution of one's inventory
is very useful in practice for improving internal order matching and
inventory management, in particular with respect to risk constraints.
This implies that theoretical results about inventory management must
be generalized in order to include predictability both of order flow
and VWAP (e.g. \cite{galien2016}).}

\textcolor{black}{Avoiding overfitting was one of the focus of the
applied part of this paper. However, substantial improvements to the
learning methods can easily be achieved. First, more relevant information
can be added to predictors, such as the group states lagged more than
once. This will be helpful if some traders have typical holding periods
larger than the duration of time slices. In addition, we have not
exploited the full potential of random forests, which do not output
binary predictions, but the fraction of trees that predict a given
state. Finally, simple learning schemes such as follow-the-leader
may improve much the aggregation of the predictions from each time-window calibration lengths, hence the overall prediction performance. These kind of methods simple are likely to exploit better the available information without adding sources of overfitting.}

\textcolor{black}{We have also avoided method overfitting by using
a single machine learning method without trying to vary any of its
parameters. Improvements to random forests such as oblique Random
Forests \cite{Menze2011}, or boosted trees \cite{friedman2000additive},
may yield better performance. This would also open the way to ensemble
learning. Finally, while we arbitrarily chose one-hour time slices,
one should investigate lead-lag networks at various time scales at
the same time.}

\textcolor{black}{On a more philosophical note, our work is a first
step towards the understanding of what triggers the activity of an
investor, one of the current mysteries in Finance. We cannot explain
yet why a group of traders acts at time $t$. However, their activity
depends on the past activity of some other groups or themselves, which
is expected since most trading activity is self-referential in financial
markets. The fact that VWAP changes are predictable points out that
traders react at least in part to past the direction of past price
changes. This recursive (thus indirect) answer to a fundamental question
also deserves further investigation, in particular with a multiple
timescales approach.}

\textcolor{black}{Finally, coming back to complex systems, this work
shows how an unpredictable complex system becomes predictable once
agent-resolved data is simplified and prepared so as to make lead-lag
networks learnable, which promises to improve ones' tools to manage
risk dynamically. Future work will estimate how predictable agent
behavior is in other contexts such as communication networks and on-line
shopping.}

\bibliographystyle{ws-acs}
\bibliography{implicit_networks}

\begin{thebibliography}{10}
\providecommand{\urlprefix}{}
\expandafter\ifx\csname urlstyle\endcsname\relax
  \providecommand{\doi}[1]{doi:\discretionary{}{}{}#1}\else
  \providecommand{\doi}{doi:\discretionary{}{}{}\begingroup
  \urlstyle{rm}\Url}\fi

\bibitem{avellaneda2008high}
Avellaneda, M. and Stoikov, S., High-frequency trading in a limit order book,
  \emph{Quantitative Finance} \textbf{8} (2008) 217--224.

\bibitem{barber2009just}
Barber, B., Lee, Y., Liu, Y., and Odean, T., Just how much do individual
  investors lose by trading?, \emph{Review of Financial Studies} \textbf{22}
  (2009) 609--632.

\bibitem{Benjamini1995}
Benjamini, Y. and Hochberg, Y., {Controlling the False Discovery Rate: A
  Practical and Powerful Approach to Multiple Testing}, \emph{Journal of the
  Royal Statistical Society. Series B (Methodological)} \textbf{57} (1995)
  289--300.

\bibitem{bouchaud2017have}
Bouchaud, J.-P. and Challet, D., Why have asset price properties changed so
  little in 200 years, in \emph{Econophysics and Sociophysics: Recent Progress
  and Future Directions} (Springer, 2017), pp. 3--17.

\bibitem{BouchaudFarmerLillo}
Bouchaud, J.-P., Farmer, J.~D., and Lillo, F., How markets slowly digest
  changes in supply and demand, in \emph{Handbook of Financial Markets:
  Dynamics and Evolution}, eds. Hens, T. and Schenk-Hopp\`e, K. (Elsevier,
  2009), pp. 57--160.

\bibitem{boudoukh1994tale}
Boudoukh, J., Richardson, M.~P., and Whitelaw, R., A tale of three schools:
  Insights on autocorrelations of short-horizon stock returns, \emph{Review of
  financial studies} \textbf{7} (1994) 539--573.

\bibitem{Breiman2001}
Breiman, L., {Random Forests}, \emph{Machine Learning} \textbf{45} (2001)
  5--32.

\bibitem{chou2010testing}
Chou, C. and Chu, C.-S.~J., Testing independence of two autocorrelated binary
  time series, \emph{Statistics \& probability letters} \textbf{80} (2010)
  69--75.

\bibitem{clauset2009power}
Clauset, A., Shalizi, C.~R., and Newman, M.~E., Power-law distributions in
  empirical data, \emph{SIAM review} \textbf{51} (2009) 661--703.

\bibitem{curme2015emergence}
Curme, C., Tumminello, M., Mantegna, R.~N., Stanley, H.~E., and Kenett, D.~Y.,
  Emergence of statistically validated financial intraday lead-lag
  relationships, \emph{Quantitative Finance}  (2015) 1--12.

\bibitem{JMLR:v15:delgado14a}
Fern\'{a}ndez-Delgado, M., Cernadas, E., Barro, S., and Amorim, D., {Do we Need
  Hundreds of Classifiers to Solve Real World Classification Problems?},
  \emph{Journal of Machine Learning Research} \textbf{15} (2014) 3133--3181.

\bibitem{Filimonov2012}
Filimonov, V. and Sornette, D., {Quantifying reflexivity in financial markets:
  Toward a prediction of flash crashes}, \emph{Physical Review E} \textbf{85}
  (2012) 056108.

\bibitem{Fraley12mclustversion}
Fraley, C., Raftery, A.~E., Murphy, T.~B., and Scrucca, L., mclust version 4
  for {R}: Normal mixture modeling for model-based clustering, classification,
  and density estimation (2012).

\bibitem{friedman2000additive}
Friedman, J., Hastie, T., Tibshirani, R., \emph{et~al.}, Additive logistic
  regression: a statistical view of boosting (with discussion and a rejoinder
  by the authors), \emph{The annals of statistics} \textbf{28} (2000) 337--407.

\bibitem{galien2016}
Gallien, F., Kassibrakis, S., Malamud, S., and Passerini, F., Managing
  inventory with proportional transaction costs, \emph{Available at SSRN
  2788593}  (2016).

\bibitem{powerlawR}
Gillespie, C.~S., Fitting heavy tailed distributions: The {poweRlaw} package,
  \emph{Journal of Statistical Software} \textbf{64} (2015) 1--16.

\bibitem{grinblatt2000investment}
Grinblatt, M. and Keloharju, M., The investment behavior and performance of
  various investor types: a study of {F}inland's unique data set, \emph{Journal
  of Financial Economics} \textbf{55} (2000) 43--67.

\bibitem{GRINBLATT2009}
Grinblatt, M. and Keloharju, M., {Sensation Seeking, Overconfidence, and
  Trading Activity}, \emph{The Journal of Finance} \textbf{64} (2009) 549--578.

\bibitem{hardiman2013critical}
Hardiman, S.~J., Bercot, N., and Bouchaud, J.-P., Critical reflexivity in
  financial markets: a hawkes process analysis, \emph{The European Physical
  Journal B} \textbf{86} (2013) 442.

\bibitem{harris2002trading}
Harris, L., \emph{Trading and exchanges: Market microstructure for
  practitioners} (Oxford University Press, 2002).

\bibitem{huth2014high}
Huth, N. and Abergel, F., High frequency lead/lag relationships--empirical
  facts, \emph{Journal of Empirical Finance} \textbf{26} (2014) 41--58.

\bibitem{rfsrc}
Ishwaran, H. and Kogalur, U., \emph{Random Forests for Survival, Regression and
  Classification (RF-SRC)} (2016),
  \urlprefix\url{https://cran.r-project.org/package=randomForestSRC}, r package
  version 2.4.1.

\bibitem{jackson2004aggregate}
Jackson, A., The aggregate behaviour of individual investors  (2004).

\bibitem{jegadeesh1995overreaction}
Jegadeesh, N. and Titman, S., Overreaction, delayed reaction, and contrarian
  profits, \emph{Review of Financial Studies} \textbf{8} (1995) 973--993.

\bibitem{kaniel2008individual}
Kaniel, R., Saar, G., and Titman, S., Individual investor trading and stock
  returns, \emph{The Journal of Finance} \textbf{63} (2008) 273--310.

\bibitem{kelley2013wise}
Kelley, E.~K. and Tetlock, P.~C., How wise are crowds? insights from retail
  orders and stock returns, \emph{The Journal of Finance} \textbf{68} (2013)
  1229--1265.

\bibitem{kullmann2002time}
Kullmann, L., Kert{\'e}sz, J., and Kaski, K., Time-dependent cross-correlations
  between different stock returns: A directed network of influence,
  \emph{Physical Review E} \textbf{66} (2002) 026125.

\bibitem{lallouache2014tick}
Lallouache, M. and Abergel, F., Tick size reduction and price clustering in a
  {FX} order book, \emph{Physica A: Statistical Mechanics and its Applications}
  \textbf{416} (2014) 488--498.

\bibitem{johnsonlargechanges}
Lamper, D., Howison, S.~D., and Johnson, N.~F., Predictability of large future
  changes in a competitive evolving population, \emph{Physical Review Letters}
  \textbf{88} (2001) 017902.

\bibitem{PhysRevE.80.056117}
Lancichinetti, A. and Fortunato, S., {Community detection algorithms: A
  comparative analysis}, \emph{Phys. Rev. E} \textbf{80} (2009) 56117.

\bibitem{farmersign}
Lillo, F. and Farmer, J.~D., The long memory of efficient markets,
  \emph{Non-lin. Dyn. and Econometric} \textbf{8} (2004).

\bibitem{Lillo2008}
Lillo, F., Moro, E., Vaglica, G., and Mantegna, R.~N., {Specialization and
  herding behavior of trading firms in a financial market}, \emph{New Journal
  of Physics} \textbf{10} (2008) 43019.

\bibitem{Menze2011}
Menze, B., Kelm, B., Splitthoff, D., Koethe, U., and Hamprecht, F., {On Oblique
  Random Forests}, in \emph{Machine Learning and Knowledge Discovery in
  Databases SE - 29}, eds. Gunopulos, D., Hofmann, T., Malerba, D., and
  Vazirgiannis, M., \emph{Lecture Notes in Computer Science}, Vol. 6912
  (Springer Berlin Heidelberg, 2011), ISBN 978-3-642-23782-9, pp. 453--469.

\bibitem{rand1971objective}
Rand, W.~M., Objective criteria for the evaluation of clustering methods,
  \emph{Journal of the American Statistical association} \textbf{66} (1971)
  846--850.

\bibitem{Rosvall2008}
Rosvall, M. and Bergstrom, C.~T., {Maps of random walks on complex networks
  reveal community structure.}, \emph{Proceedings of the National Academy of
  Sciences of the United States of America} \textbf{105} (2008) 1118--23.

\bibitem{SanMiguel2012}
San~Miguel, M., Johnson, J.~H., Kertesz, J., Kaski, K., D{\'\i}az-Guilera, A.,
  MacKay, R.~S., Loreto, V., {\'E}rdi, P., and Helbing, D., Challenges in
  complex systems science, \emph{The European Physical Journal Special Topics}
  \textbf{214} (2012) 245--271.

\bibitem{toth2006increasing}
T{\'o}th, B. and Kert{\'e}sz, J., Increasing market efficiency: Evolution of
  cross-correlations of stock returns, \emph{Physica A: Statistical Mechanics
  and its Applications} \textbf{360} (2006) 505--515.

\bibitem{Michele2012}
Tumminello, M., Lillo, F., Piilo, J., and Mantegna, R.~N., {Identification of
  clusters of investors from their real trading activity in a financial
  market}, \emph{New Journal of Physics} \textbf{14} (2012).

\bibitem{Tumminello2011}
Tumminello, M., Miccich\`{e}, S., Lillo, F., Piilo, J., and Mantegna, R.~N.,
  {Statistically Validated Networks in Bipartite Complex Systems}, \emph{PLoS
  ONE} \textbf{6} (2011) e17994.

\bibitem{Kertesz2012}
Zhou, W.-X., Mu, G.-H., and Kertesz, J., {Random matrix approach to the
  dynamics of stock inventory variations}, \emph{New Journal of Physics}
  \textbf{14} (2012) 93025.

\bibitem{Zovko2007}
Zovko, I. and Farmer, J.~D., {Correlations and clustering in the trading of
  members of the London Stock Exchange}, in \emph{AIP Conference Proceedings},
  Vol. 965 (AIP, 2007), ISSN 0094243X, pp. 287--299, \doi{10.1063/1.2828747},
  \urlprefix\url{http://scitation.aip.org/content/aip/proceeding/aipcp/10.1063/1.2828747}.

\end{thebibliography}

\appendix

\section{\textcolor{black}{Prediction with logistic regression}}

\textcolor{black}{For the sake of meaningful comparisons, we gave
to logistic regression the same inputs as to random forests (but split
the hour of the day into separate factors), used the same calibration
window lengths, and fitted a logistic model every day. The prediction
of the model is then rounded to either -1 or +1. The performance summary
statistics corresponding to the majority vote between all the calibration
window lengths both for logistic regression and random forests fare
are reported in Table \ref{tab:P-values-of-various}. It is quite
clear that logistic regression is out-performed by random forests. }

\begin{figure}
\begin{centering}
\includegraphics[width=0.4\textwidth]{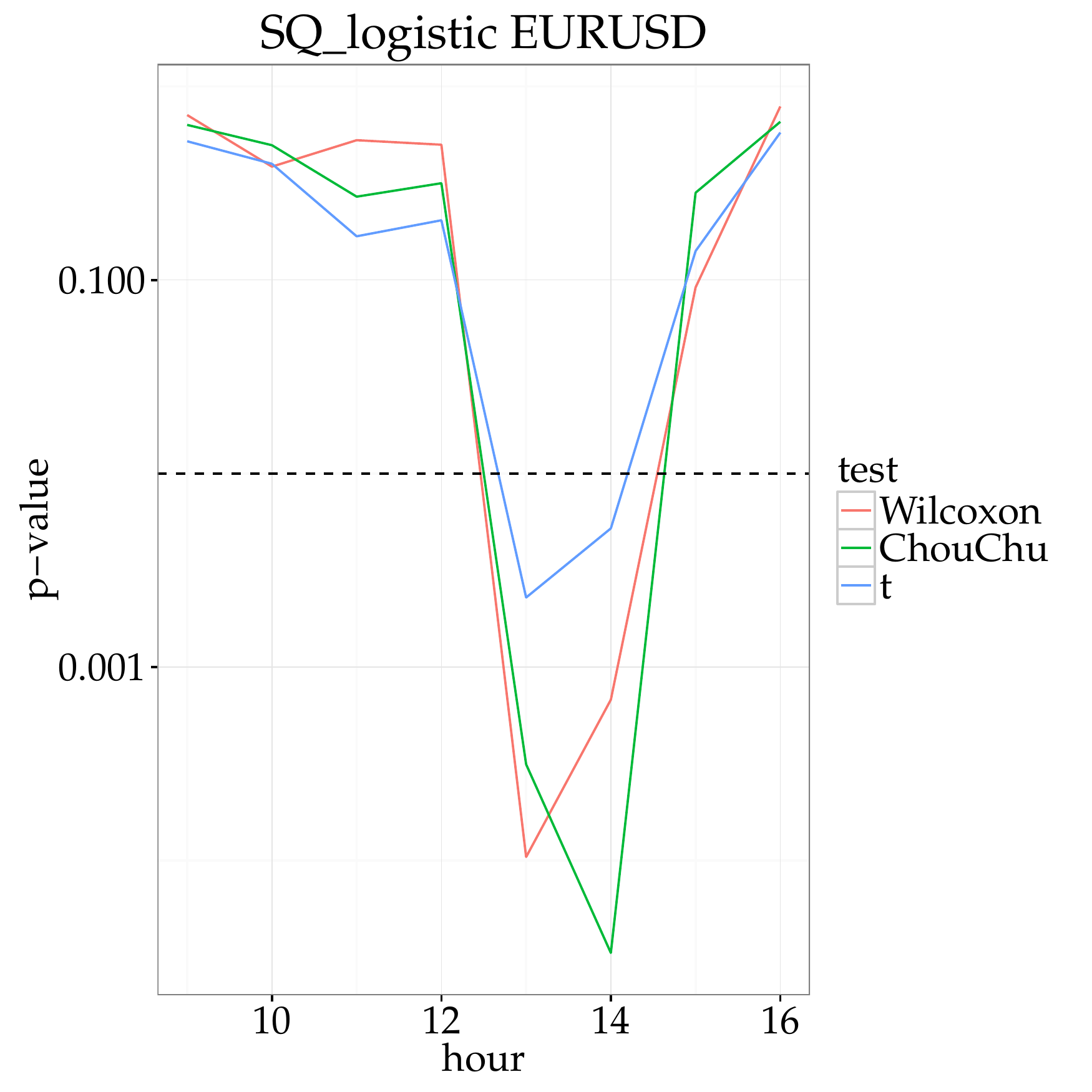}\includegraphics[width=0.4\textwidth]{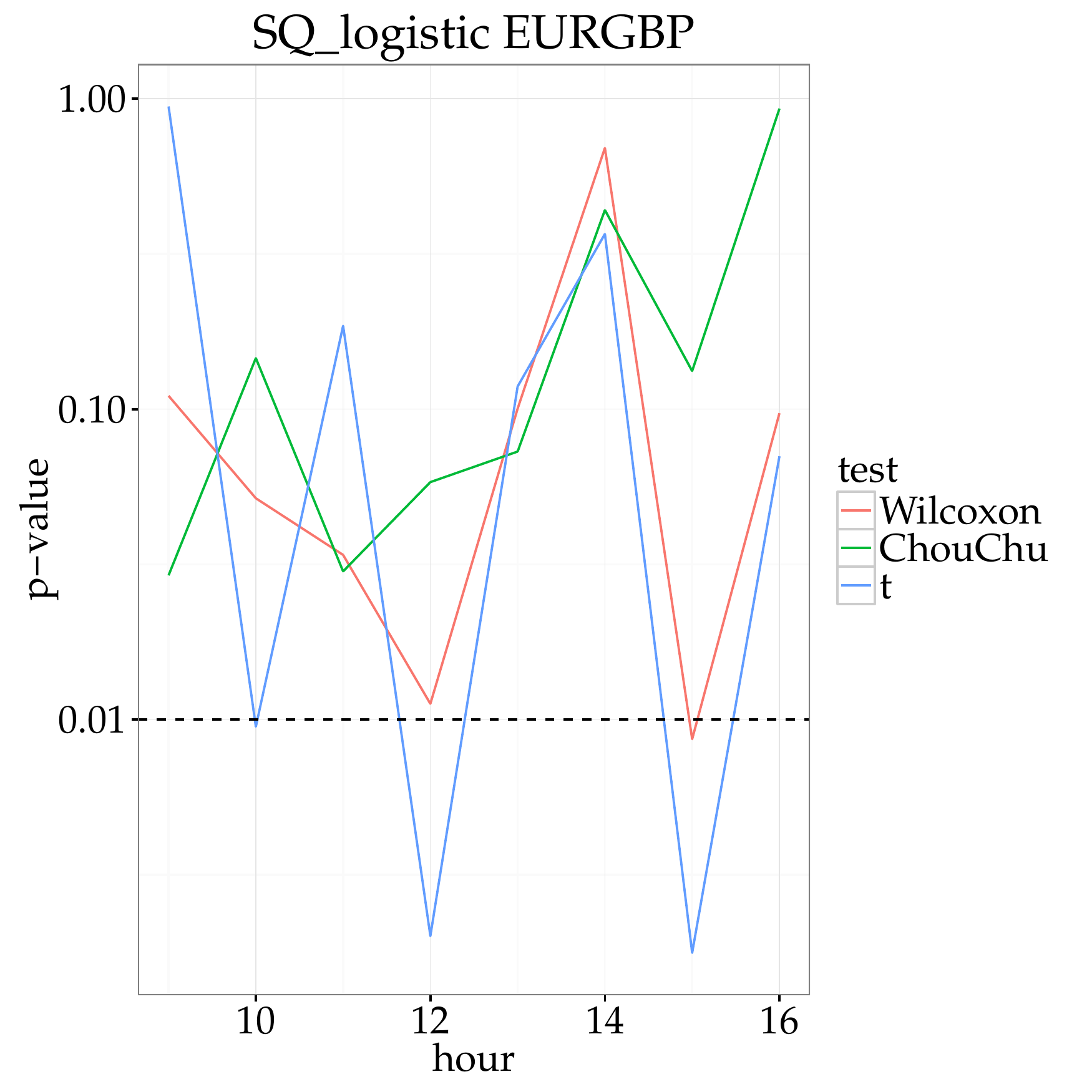}\\
\includegraphics[width=0.4\textwidth]{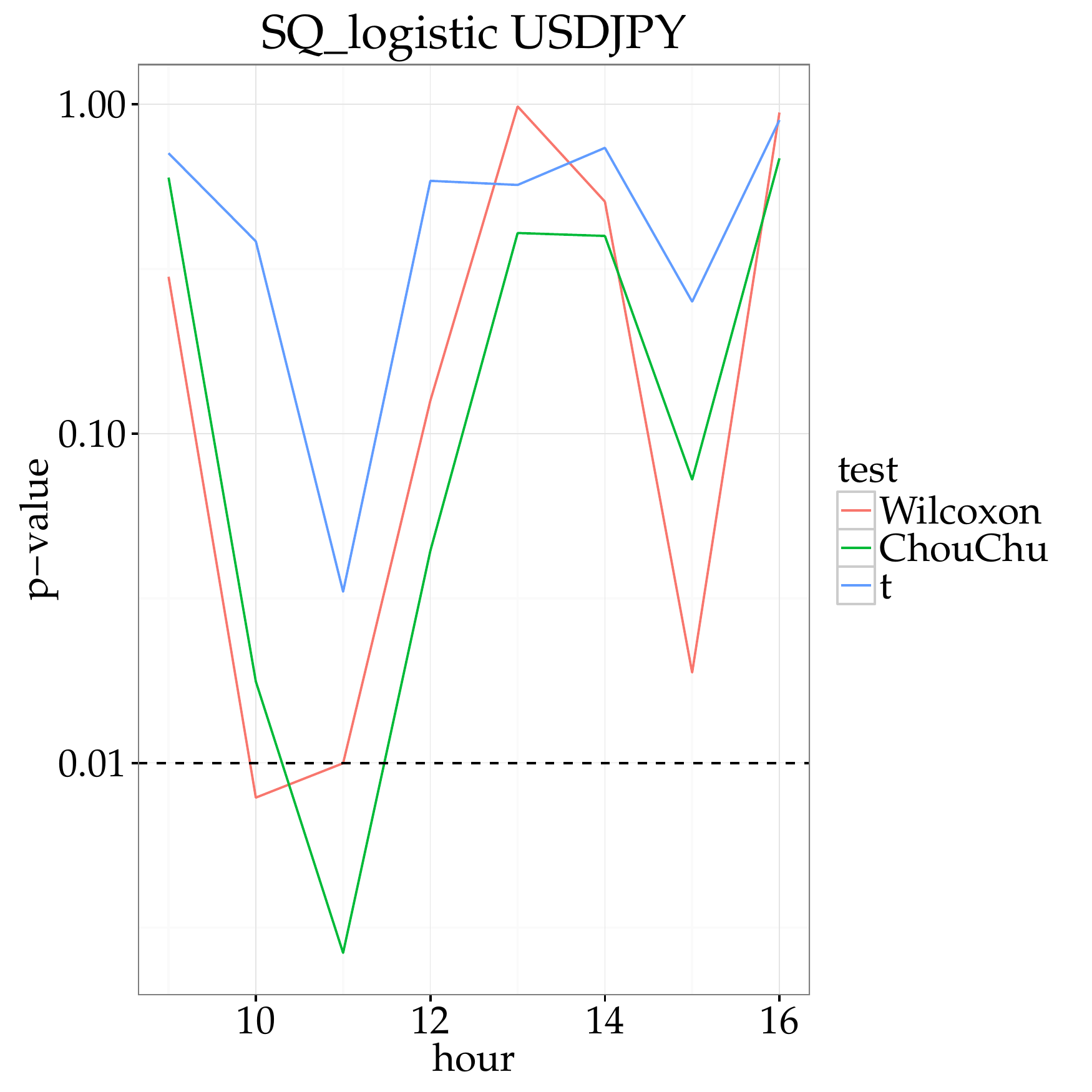}
\par\end{centering}
\caption{\textcolor{black}{Hourly statistics of the prediction performance
of a logistic regression; SQ2014-6.\label{fig:SQ_EURUSD_perf-logistic}}}
\end{figure}

\textcolor{black}{}
\begin{table}
\textcolor{black}{}%
\begin{tabular}{|c|c|c|}
\hline
\texttt{\textcolor{black}{EURCHF}} & \textcolor{black}{RF} & \textcolor{black}{logistic}\tabularnewline
\hline
\hline
\textcolor{black}{Chou Chu} & \textcolor{black}{3.5e-06} & \textcolor{black}{4.8e-05}\tabularnewline
\hline
\textcolor{black}{t} & \textcolor{black}{5.0e-12} & \textcolor{black}{2.4e-05}\tabularnewline
\hline
\textcolor{black}{Wilcoxon} & \textcolor{black}{3.9e-13} & \textcolor{black}{1.3e-05}\tabularnewline
\hline
\end{tabular}\textcolor{black}{~}%
\begin{tabular}{|c|c|c|}
\hline
\texttt{\textcolor{black}{EURGBP}} & \textcolor{black}{RF} & \textcolor{black}{logistic}\tabularnewline
\hline
\hline
\textcolor{black}{Chou Chu} & \textcolor{black}{8.2e-05} & \textcolor{black}{3.6e-04}\tabularnewline
\hline
\textcolor{black}{t} & \textcolor{black}{4.7e-03} & \textcolor{black}{1.8e-04}\tabularnewline
\hline
\textcolor{black}{Wilcoxon} & \textcolor{black}{5.0e-05} & \textcolor{black}{4.2e-05}\tabularnewline
\hline
\end{tabular}\textcolor{black}{~}%
\begin{tabular}{|c|c|c|}
\hline
\texttt{\textcolor{black}{USDJPY}} & \textcolor{black}{RF} & \textcolor{black}{logistic}\tabularnewline
\hline
\hline
\textcolor{black}{Chou Chu} & \textcolor{black}{4.8e-05} & \textcolor{black}{7.9e-01}\tabularnewline
\hline
\textcolor{black}{t} & \textcolor{black}{4.7e-03} & \textcolor{black}{3.9e-01}\tabularnewline
\hline
\textcolor{black}{Wilcoxon} & \textcolor{black}{4.0e-06} & \textcolor{black}{1.7e-03}\tabularnewline
\hline
\end{tabular}

\textcolor{black}{\caption{P-values of various tests for positive performance for logistic regression
and random forests (RF) on SQ2014-6 dataset. \label{tab:P-values-of-various}}
}
\end{table}

\end{document}